\documentclass[a4paper,11pt]{article}
\pdfoutput=1 

\usepackage{jheppub} 

\usepackage[T1]{fontenc}
\usepackage{multirow}
\usepackage{wasysym}

\allowdisplaybreaks
 
\definecolor{nicered}{rgb}{0.7,0.1,0.1}
\definecolor{nicegreen}{rgb}{0.1,0.5,0.1}
\definecolor{niceblue}{rgb}{0.0,0.1,0.7}
\hypersetup{colorlinks,citecolor=niceblue,linkcolor=niceblue,urlcolor=niceblue}

\def \bm#1{\mbox{\boldmath$#1$\unboldmath}}
\def \beq{\begin{equation}}
\def \eeq{\end{equation}}
\def \bea{\begin{eqnarray}}
\def \eea{\end{eqnarray}}

\title{QED effects in inclusive semi-leptonic $\bm{B}$ decays}

\author[]{Dante Bigi,} 

\author[a]{Marzia Bordone,} 

\author[b,c,d]{Paolo Gambino,} 

\author[c]{\\Ulrich Haisch}

\author[e]{and Andrea Piccione}

\affiliation[a]{Theoretical Physics Department, CERN, \\ 1211 Geneva 23, Switzerland}
\affiliation[b]{Dipartimento di Fisica, Universit{\`a} di Torino \& INFN, Sezione di Torino,\\ Via Pietro Giuria 1, I-10125 Turin, Italy}
\affiliation[c]{Max Planck Institute for Physics, \\ F{\"o}hringer Ring 6, 80805 M{\"u}nchen, Germany}
\affiliation[d]{Technische Universit{\"a}t M{\"u}nchen, Physik-Department, \\ James-Franck-Strasse 1, 85748 Garching, Germany}
\affiliation[e]{IIS G. Plana, Piazza Generale di Robilant 5, I-10141 Turin, Italy}

\emailAdd{bigid@iol.it,marzia.bordone@cern.ch,gambino@to.infn.it, haisch@mpp.mpg.de,andrea.piccione@plana.edu.it}

\abstract{We analyse in detail the QED corrections to the total decay width and the moments of the electron energy spectrum of the inclusive semi-leptonic $B \to X_c e \nu$ decay. Our~calculation includes short-distance electroweak~corrections, the complete~${\cal O}(\alpha)$ partonic terms and leading-logarithmic QED effects up to ${\cal O}(\Lambda^3_{\rm QCD}/m_b^3)$. A comprehensive numerical comparison of our results against those obtained with the Monte~Carlo~(MC) tool PHOTOS is presented. While the comparison indicates good overall agreement, our computation contains QED effects not included in PHOTOS and should therefore better describe photon radiation to $B \to X_c e \nu$ as measured by the $B$-factories. Our~calculations represent the first steps in the construction of a fully differential higher-order QED MC generator for inclusive semi-leptonic $B$ decays.}

\preprint{CERN-TH-2023-145, MPP-2023-186}

\begin{document} 
\maketitle
\flushbottom

\renewcommand{\arraystretch}{1.25}

\section{Introduction}
\label{sec:introduction}

After the calculation of the ${\cal O}(\alpha_s^3)$ corrections to the total semi-leptonic decay width~\cite{Fael:2020tow}, the inclusive determination of the Cabibbo–Kobayashi–Maskawa~(CKM) matrix element $|V_{cb}|$ has an uncertainty of about $1.2\%$~\cite{Bordone:2021oof}. At this level of precision a careful study of QED and electroweak~(EW) effects is imperative, not only for the inclusive semi-leptonic decay width but also for the moments of the kinematic distributions that enter the Heavy Quark Expansion (HQE) global fits of $B\to X_c \ell \nu$. At present, the inclusive extraction of $|V_{cb}|$ includes only the short-distance EW logarithmic corrections derived in the classic paper by Sirlin~\cite{Sirlin:1981ie}, while photon radiation is generally subtracted from the measured moments in the experimental analyses. The subtraction is performed using the Monte Carlo~(MC) code PHOTOS~\cite{Barberio:1993qi} which includes soft and collinear photon radiation from final-state charged leptons and hadrons with logarithmic accuracy. It however lacks the interference between initial- and final-state photons, hard and structure-dependent radiative effects as well as virtual corrections, which are sometimes enhanced by a factor $\pi^2$.

The study of inclusive semi-leptonic $B \to X_c \ell \nu$ decays is based on an Operator Product Expansion (OPE) that holds for sufficiently inclusive observables. In particular, the theoretical predictions for the total decay width and the first few moments of the lepton energy and the hadronic invariant mass can be formulated as a double-series expansion in the strong coupling constant $\alpha_s$ and the ratio $\Lambda_{\rm QCD}/m_b$ involving the QCD scale and the bottom-quark mass. The power corrections start at quadratic order and are parametrised in terms of $B$-meson matrix elements of local operators. Information on these non-perturbative matrix elements is obtained from the normalised moments, and it is then employed in the calculation of the total decay width from which $|V_{cb}|$ is extracted (cf.~for instance \cite{Bordone:2021oof,Bernlochner:2022ucr} for~recent analyses). At the~$B$-factories the moments have been measured by applying a lower cut on the lepton energy $E_\ell$ or the invariant mass of the final-state lepton pair $q^2$.

When QED interactions enter the game, things become more involved. The first important observation in this context concerns the definition of the final-state observables. If~only~QCD is considered, the leptonic and hadronic final states are completely independent and cleanly separate. Gluon radiation is fully included in the hadronic final state so that the hadronic invariant mass squared $M_X^2$ and the hadronic energy $E_X$ are defined in a completely inclusive way. This is consistently done in the theoretical calculations of the inclusive semi-leptonic decay width~\cite{Fael:2020tow} and the moments~\cite{Pak:2008qt,Biswas:2009rb,Fael:2022frj}, as well as in the corresponding experimental measurements. Photon radiation, however, affects both charged leptons and hadrons making a completely inclusive definition of their respective properties impossible. Moreover, the measurements of $E_\ell$ and $q^2$ always depend on the details of the experimental analysis. At~the~$B$-factories, for example, the momentum of an electron and hence its energy~$E_e$ is measured most precisely in the drift chambers from the curvature of its trajectory in~the~magnetic field~\cite{Belle:2000cnh,BaBar:2001yhh}. The reconstruction of the true value of $E_e$ involves modelling~the photon radiation due to the interaction with the magnetic field and the material in the detector components~\cite{Brun:1994aa,GEANT4:2002zbu} as well as bremsstrahlungs recovery~\cite{Belle:2000cnh,BaBar:2001yhh}. Given the detector geometry and the way electrons are reconstructed, most of the prompt photons emitted by the electron in the form of final-state radiation however go undetected and are not corrected for in the $E_e$ measurement. We will therefore work under the assumption that the true electron energy determined by the $B$-factory experiments corresponds to the energy before the electron enters the detector, and after all final-state radiation associated to the hard process has taken place. As a consequence, all these photons are effectively summed over in the experimental measurement which can be qualified as totally inclusive with respect to photons. While this is certainly an approximation, we believe it is closest to the experimental setup. The same assumption is also adopted in the precision~QED calculations of the electron energy spectrum in muon decay~\cite{vanRitbergen:1999fi,Arbuzov:2002pp,Arbuzov:2002cn,Arbuzov:2002rp,Anastasiou:2005pn,Pak:2008qt} when compared to the experimental measurements like those performed at MuLan~\cite{MuLan:2010shf,MuLan:2012sih}. 

A second, related observation concerns the OPE, which in the presence of QED interactions is no longer valid in general. While it still holds in the case of the total decay width, the~OPE fails for quantities that describe exclusive properties of the leptonic final state (such as~$E_\ell$ or~$q^2$) or the hadronic final state (such as $M_X^2$ or $E_X$). This is because a photon can connect the charged lepton with the hadronic initial or final states. In the context of our work, the breakdown of the OPE is signalled by the presence of collinear logarithms of the form $\alpha/\pi \ln \hspace{-0.25mm} \left (m_b^2 /m_e^2 \right )$ that appear in the electron energy spectrum and its moments. These logarithmically enhanced QED effects arise already at leading power in the HQE,~i.e.~the partonic level, and are expected to provide the bulk of the complete~${\cal O}(\alpha)$ contributions. In the absence of an OPE, we generally expect power corrections to start already at ${\cal O} (\alpha \hspace{0.125mm} \Lambda_{\rm QCD}/m_b)$. However, the logarithmically enhanced corrections related to photons collinear to the electron represent an exception: since they factorise, they can be computed using the structure function method directly from the lowest order, and this holds for the power-suppressed contributions as well.

In this article, we focus our attention on the QED corrections to the total decay width of $B \to X_c e \nu$, the electron energy spectrum and its moments. These observables play a crucial role in the HQE global fits that are used to determine $|V_{cb}|$. Our aim is to validate the subtraction procedure employed by the experiments and to improve it by calculating QED effects not included in PHOTOS. To this purpose, we identify two types of large QED corrections, namely collinear terms proportional to $\ln \hspace{-0.25mm} \left (m_b^2 /m_e^2 \right )$ and threshold corrections enhanced by a factor~$\pi^2$. For the electron energy spectrum of $b \to c e \nu$ we describe the analytic calculation of these contributions in Section~\ref{sec:LL} and Section~\ref{sec:Pi2effects}, respectively. In~Section~\ref{sec:full} we then calculate the complete ${\cal O} (\alpha)$ corrections to the electron energy spectrum and the total decay width in~$b \to c e \nu$. A comparison of the obtained results shows that the logarithmically enhanced ($\pi^2$-enhanced) terms indeed provide the dominant part of the full~${\cal O} (\alpha)$ corrections in the case of the electron energy spectrum (total decay width). In~our comparison with experiment presented in~Section~\ref{sec:comparison}, we therefore include the full short-distance EW~corrections (beyond the logarithm computed in~\cite{Sirlin:1981ie}), the complete ${\cal O}(\alpha)$ terms, and the leading-logarithmic~(LL) QED effects in the power corrections of ${\cal O}(\Lambda^2_{\rm QCD}/m_b^2)$ and ${\cal O}(\Lambda^3_{\rm QCD}/m_b^3)$. Section~\ref{sec:conclusions} contains our conclusions and an outlook, while lengthy analytic expressions and technical details are relegated to a few appendices.

\section{LL effects in the electron energy spectrum}
\label{sec:LL}

In this section, we describe the calculation of the~LL~QED corrections~to the electron energy spectrum of the $B\to X_c e \nu$ transition. These corrections are proportional to~logarithms of the ratio of the bottom-quark and the electron mass. We~will first detail the general formalism and then apply it to the partonic~case. The same formalism can also be used to compute the~LL~QED corrections to the power-suppressed contributions in the electron energy moments of $B\to X_c e \nu$. The analytic expressions for these corrections at ${\cal O} ( \Lambda_{\rm QCD}^2/m_b^2 )$ and ${\cal O} ( \Lambda_{\rm QCD}^3/m_b^3 )$ are relegated to~Appendix~\ref{app:LLpower}, while in Appendix~\ref{app:NLLpartonic} we extend the formalism presented in this section to the next-to-leading-logarithmic~(NLL) order and apply it to the leading-power computation of $B\to X_c e \nu$.

The~LL~QED corrections~to the electron energy spectrum of $B\to X_c e \nu$ can be easily computed with the structure function method~\cite{Lipatov:1974qm,Altarelli:1977zs}. Further details on this approach can be found for instance in~\cite{Arbuzov:2002pp,Arbuzov:2002cn,Arbuzov:2002rp}. 
The main ingredient to calculate these effects is the leading-order~(LO) electron-electron splitting function 
\beq \label{eq:Pee0}
P_{ee}^{(0)} (z) = \left [ \frac{1+z^2}{1-z} \right ] _+ = \lim_{\Delta \, \to \, 0} \left [ \frac{1+z^2}{1-z} \, \theta ( 1- z -\Delta ) + \left ( \frac{3}{2} + 2 \ln \Delta \right ) \delta (1 - z ) \right ] \,, 
\eeq
which is a kind of plus distribution and encodes the~LL~QED correction to the probability of finding an electron with energy fraction $z$ in the initial electron. Here $\theta ( z )$ and $\delta( z )$ are the Heaviside step function and the Dirac delta function, respectively. The LO electron-electron splitting function is normalised such that 
\beq \label{eq:normPee0}
\int_0^1 \! dz \, P_{ee}^{(0)} (z) = 0 \,,
\eeq
which guarantees that in the total decay width $\Gamma$ logarithmic corrections are not present as required by the Kinoshita-Lee-Nauenberg~(KLN) theorem~\cite{Kinoshita:1962ur,Lee:1964is}. In practice, the~LL~corrections to the electron energy spectrum are obtained by evaluating the following convolution 
\beq \label{eq:convolutionLL}
\left ( \frac{d \Gamma}{dy} \right )^{(1)}_{\rm LL} = \frac{\alpha}{2\pi} \, \bar L_{b/e} \int_y^{1 - \rho} \, \frac{dx}{x} \, P_{ee}^{(0)} \left (\frac{y}{x} \right ) \, \left ( \frac{d \Gamma}{dx} \right )^{(0)} \,,
\eeq
where $y = 2 E_e/m_b$ with $E_e$ the electron energy and $\rho = m_c^2/m_b^2$ with $m_b$ and $m_c$ the bottom- and charm-quark mass, respectively. We have furthermore introduced the abbreviation 
\beq \label{eq:Lij}
\bar L_{b/e} = \ln \bigg( \frac{m_b^2}{m_e^2} \bigg ) -1 \,. 
\eeq
and $\left ( d \Gamma/dy \right )^{(0)}$ denotes the electron energy spectrum calculated at LO assuming a massless~electron. 

To give a simple example of the general formalism let us apply~(\ref{eq:convolutionLL}) to the case of the partonic electron energy spectrum. In this case one has 
\beq \label{eq:EespectrumLOpartonic}
\left (\frac{d \Gamma}{dy} \right )^{(0)}= \Gamma^{(0)} \, f^{(0)} (y) \,, \qquad \Gamma^{(0)} = \frac{G_F^2 |V_{cb}|^2}{192 \pi^2} \hspace{0.25mm} m_b^5 \,, 
\eeq
where $G_F = 1.1663787 \cdot 10^{-5} \, {\rm GeV}^{-2}$ is the Fermi constant as extracted from muon decay, $V_{cb}$ is the relevant CKM matrix element and~\cite{Manohar:2000dt} 
\beq \label{eq:f0partonic}
f^{(0)} (y) = \Bigg [ \, 2 \left (3-2 y \right ) y^2 -6 y^2 \rho -\frac{6 y^2 \rho ^2}{(1-y)^2} + \frac{2 \left (3-y \right ) y^2 \rho ^3 }{(1-y)^3} \, \Bigg ] \; \theta ( 1 - y - \rho ) \,.
\eeq
Inserting the latter result into~(\ref{eq:convolutionLL}) and combining it with~(\ref{eq:EespectrumLOpartonic}) we obtain 
\beq \label{eq:fpartonicLL}
\frac{d \Gamma}{dy}= \Gamma^{(0)} \, f (y) \,, \qquad f(y) = f^{(0)} (y) + \frac{\alpha}{2\pi} \, \bar L_{ b/e} \, f^{(1)}_{\rm LL} (y) \,,
\eeq
where
\bea \label{eq:f1partonic}
\begin{split}
f^{(1)}_{\rm LL} (y) & = \Bigg \{ \, \frac{5}{3} + \frac{4}{3} \left(3 -6 y+ 4 y^2 \right) y +4 \left (3-2 y \right ) y^2 \ln \left(\frac{1 -y - \rho }{y}\right) \\[2mm] 
& \phantom{xxx} - \left [ 9 + 3y -12 y^2 + 8 y^3+12 \left (1-y \right ) y^2 \ln \left(\frac{1-y-\rho}{y}\right) \right ] \frac{\rho}{1-y} \\[2mm] 
& \phantom{xxx} + \left [ 9 + 2 y^3 + 6 \left(1 - 2 y - y^2\right) \ln \left(\frac{1-y}{\rho }\right) -12\hspace{0.25mm} y^2 \ln \left(\frac{1-y-\rho}{y}\right) \right ] \frac{\rho^2}{(1-y)^2} \hspace{6mm} \\[2mm]
& \phantom{xxx} - \left [ \frac{5}{3} + y + 4 y^2 - \frac{2 }{3} y^3 + 2 \left(1 - 3 y - 3 y^2 + y^3 \right) \ln \left(\frac{1-y}{\rho }\right) \right. \\[2mm]
& \phantom{xxxxxi} \left. -4 \left (3-y \right ) y^2 \ln \left(\frac{1-y-\rho}{y}\right) \right ] \frac{\rho^3}{(1-y)^3} \Bigg \} \; \theta ( 1 - y - \rho ) \,.
\end{split}
\eea
Notice that in the limit $\rho \to 0$ our result for $f^{(1)}_{\rm LL} (y)$ agrees with~(12)~of~\cite{Arbuzov:2002pp} which contains the~LL~QED terms for muon decay. This agreement is expected because in the fully massless case the electron energy spectrum of $b \to u e \nu$ and $\mu \to e \nu \nu$ are the same. 

The total decay width for $b \to c e \nu$ is obtained by integrating~(\ref{eq:fpartonicLL}) over $y \in [0, 1-\rho]$. Using the relations 
\beq \label{eq:fintegralspartonic} 
\int_0^{1-\rho} \! dy \, f^{(0)} (y) = g(\rho) \,, \qquad \int_0^{1-\rho} \! dy \, f^{(1)}_{\rm LL} (y) = 0 \,, 
\eeq
where
\beq \label{eq:g} 
 g(\rho) = 1 - 8 \rho - 12 \rho ^2 \ln \rho + 8 \rho ^3 -\rho ^4 \,,
\eeq
one finds 
\beq \label{eq:Gammapartial} 
\Gamma = \Gamma^{(0)} g(\rho) \,, 
\eeq
which implies that the total decay width of $b \to c e \nu$ does not receive~LL~QED effects. Notice~that this~cancellation of mass singularities is required by the KLN theorem. In~fact, the same cancellation occurs for the~LL and the~NLL~QED corrections calculated in Appendix~\ref{app:LLpower} and Appendix~\ref{app:NLLpartonic}, respectively. This feature is related to the existence of an OPE at~${\cal O} (\alpha)$ for the total semi-leptonic decay width and represents a non-trivial cross-check of our calculation of logarithmically enhanced QED corrections to the electron energy spectrum of~$B \to X_c e \nu$. 

\begin{figure}[t!]
\begin{center}
\includegraphics[width=0.6\textwidth]{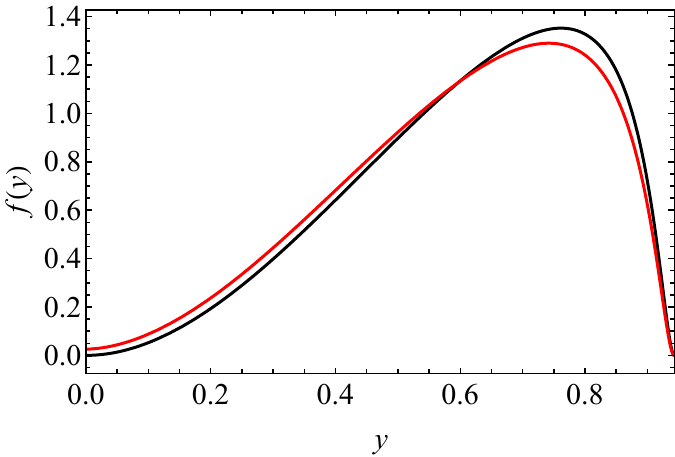}
\end{center}
\vspace{-4mm} 
\caption{\label{fig:LLparton} The red curve corresponds to $f(y)$ defined in~(\ref{eq:fpartonicLL}) while the black curve represents the LO contribution $f^{(0)}(y)$ as given in~(\ref{eq:f0partonic}). A kinetic~bottom- and $\overline{\rm MS}$ charm-quark mass is employed and final states containing electrons are considered.}
\end{figure}

While the absence of logarithmically enhanced QED corrections in the total decay width of $B \to X_c e \nu$ is guaranteed by the KLN theorem, such effects appear in general in other~observables. The impact of the~LL~QED corrections to the electron energy spectrum in the partonic $b \to c e \nu$ transition is illustrated in~Figure~\ref{fig:LLparton}. The shown results have been obtained for a kinetic bottom-quark mass of $m_b^{\rm kin} (1 \, {\rm GeV}) = 4.57 \, {\rm GeV}$ and a $\overline{\rm MS}$ charm-quark mass of $\overline{m}_c (2 \, {\rm GeV}) = 1.09 \, {\rm GeV}$, corresponding to the central values of the recent HQE global fit~\cite{Bordone:2021oof}. We~furthermore have employed $m_e = 511 \, {\rm keV}$ and $\alpha (m_b) = 1/133$. The same input will be used hereafter. From the figure it is evident that the~LL~QED corrections change the shape of the electron energy spectrum of $b \to c e \nu$ in a non-trivial way. In the tail region,~i.e.~for~$y \to 0$, the spectrum receives a positive shift while close to the endpoint,~i.e.~for~$y \to 1-\rho$, the~LL~QED corrections tend to zero. This~can be understood analytically from the following limiting behaviours:
\beq \label{eq:f1partoniclimits}
\lim_{y \, \to \, 0} f^{(1)}_{\rm LL} (y) = \frac{5}{3} -9 \rho +9 \rho ^2 -\frac{5 }{3} \rho ^3 -2 \left (3-\rho \right ) \rho ^2 \ln \rho \,, \qquad \lim_{y \, \to \,1-\rho} f^{(1)}_{\rm LL} (y) = 0 \,.
\eeq
In the peak region,~i.e.~in the vicinity of $y = 1 - \sqrt{\rho}$, the electron energy distribution is reduced by the~LL~QED effects, as expected from the second relation in~(\ref{eq:fintegralspartonic}) and~(\ref{eq:f1partoniclimits}). We add that the approximation of the logarithmically enhanced QED corrections to the electron energy spectrum that has been proposed in~\cite{Atwood:1989em} does not accurately capture the functional form of the exact expression~$f^{(1)}_{\rm LL} (y)$. In particular, the tail is not described correctly by the exponentiated radiation-damping factor introduced in~\cite{Atwood:1989em} because this approximation misses hard-collinear photon contributions that populate the low-energy spectrum. 

\section{Threshold corrections in the total decay width} 
\label{sec:Pi2effects}

In this section, we explain how to compute a certain class of $\pi^2$-enhanced QED corrections first identified in~\cite{Ginsberg:1968pz,Atwood:1989em} in the case of exclusive semi-leptonic $K$ and $B$ decays. In the modern literature discussing QED corrections to meson decays these terms are often referred to as Coulomb corrections (see for instance~\cite{Isidori:2007zt,Kubis:2010mp,deBoer:2018ipi,Cali:2019nwp,Mishra:2020orb,Bansal:2021oon}). For what concerns the $b \to c e \nu$ process these terms arise from the virtual corrections where the photon is exchanged between the charm-quark and the electron line in the final state. The corresponding Feynman diagram is shown on the right in~Figure~\ref{fig:virtuals}. Interfering the relevant virtual~matrix~element with the Born-level contribution and normalising the result to the Born-level matrix element squared,~we~find 
\beq \label{eq:Deltace}
\begin{split}
\Delta_{c e} & = \frac{2 \hspace{0.125mm} {\rm Re} \left ( {\cal M}_{\rm Born} \, {\cal M}_{\text{virtual}}^{ce \hspace{0.25mm} \ast} \right )}{|{\cal M}_{\rm Born}|^2} \\[2mm] 
& = -\frac{\alpha}{\pi} \hspace{0.5mm} Q_c \hspace{0.125mm} Q_e \hspace{0.5mm} {\rm Re} \, \Bigg \{ 1 + \frac{1}{q^2 - m_c^2 - m_e^2} \bigg [ \left ( q^2 - m_c^2 \right ) B_0 (m_c^2, 0, m_c^2) \\[2mm] 
& \hspace{3.5cm} + \left ( q^2 - m_e^2 \right ) B_0 (m_e^2, 0, m_e^2) - \left ( m_c^2 + m_e^2 \right ) B_0 (q^2, m_c^2, m_e^2) \bigg ] \\[2mm] 
& \hspace{2.5cm} \phantom{xi} - \left ( q^2 - m_c^2 - m_e^2 \right ) C_0 ( m_c^2, q^2, m_e^2, 0, m_c^2, m_e^2 ) \Bigg \} \,.
\end{split}
\eeq
Here $Q_c = 2/3$ and $Q_e = -1$ are the electric charges of the charm quark and the electron in units of $e$, while $q^2 = \left ( p_b - p_\nu \right )^2 = \left ( p_c + p_e \right )^2$ with $p_b$, $p_c$, $p_e$ and $p_\nu$ denoting the four-momentum of the bottom quark, charm quark, electron and neutrino, respectively. The correction~(\ref{eq:Deltace}) has been expressed in terms of the following one-loop Passarino-Veltman~(PV) integrals:
\beq \label{eq:PVintegrals}
\begin{split}
B_0 \big (p_1^2, m_0^2, m_1^2 \big) & = \frac{\mu^{4-d}}{i \hspace{0.125mm} \pi^{d/2} \hspace{0.25mm} r_\Gamma} \int \! \frac{d^d l}{\Pi_{i=0,1} \hspace{0.25mm} P(l+p_i, m_i)} \,, \\[2mm]
C_0 \big (p_1^2, (p_1 - p_2)^2, p_2^2, m_0^2, m_1^2, m_2^2 \big ) & = \frac{\mu^{4-d}}{i \hspace{0.125mm} \pi^{d/2} \hspace{0.25mm} r_\Gamma} \int \! \frac{d^d l}{\Pi_{i=0,1,2} \hspace{0.25mm} P(l+p_i, m_i)} \,. 
\end{split}
\eeq
Here $\mu$ is the renormalisation scale that keeps track of the correct dimension of the integrals in $d = 4 - 2 \epsilon$ space-time dimensions, $r_\Gamma = \Gamma^2 (1-\epsilon) \hspace{0.25mm} \Gamma (1+ \epsilon)/\Gamma(1 -2\epsilon)$ with $\Gamma(z)$ denoting the Euler gamma function, $P(k,m) = k^2-m^2$ and $p_0 = 0$. The definitions~(\ref{eq:PVintegrals}) resemble those of the {\tt LoopTools} package~\cite{Hahn:1998yk}. 

The $\pi^2$-enhanced QED terms now arise from the result~(\ref{eq:Deltace}) in the threshold region, specifically as $q^2 \to (m_c + m_e)^2_+$, where the subscript $+$ denotes that the limit is approached from above. In this limit the three-point one-loop PV integral $C_0 ( m_c^2, q^2, m_e^2, 0, m_c^2, m_e^2 )$ develops a singularity at threshold. Specifically, above the threshold, one has:
\beq \label{eq:ReC0}
 {\rm Re} \left ( C_0 ( m_c^2, q^2, m_e^2, 0, m_c^2, m_e^2 ) \right ) = -\frac{ \pi^2}{2 \hspace{0.25mm} \beta_{ce}} \hspace{0.5mm} \frac{1}{p_c \cdot p_e} \, \Big [ 1 + {\cal O} ( \beta_{ce}) \Big ]  \,, 
\eeq
with 
\beq \label{eq:betace}
\beta_{ce} = \sqrt{1 - \frac{m_c^2 \hspace{0.5mm} m_e^2}{(p_c \cdot p_e)^2}} \,, 
\eeq 
the relative velocity between the charm quark and the electron in the rest frame of either particle.  Notice that the observed~$1/\beta_{ce}$ factor signals the presence of a Coulomb~pole that describes the universal long-distance electromagnetic interactions between pairs of charged non-relativistic final-state particles, while the symbol ${\cal O}( \beta_{ce})$ in~(\ref{eq:ReC0}) represents terms that are non-singular. In the limit of vanishing electron mass, one has~$\beta_{ce} \to 1$ and the $\pi^2$ terms correspond to the discontinuity of $\Delta_{ce}$ around the threshold:
\beq \label{eq:discDeltace}
\text{disc} \, \Delta_{c e} = -\pi^2 \hspace{1mm} \frac{\alpha}{\pi} \hspace{0.5mm} Q_c \hspace{0.125mm} Q_e = \frac{2 \pi \alpha}{3} \,.
\eeq

To obtain the $\pi^2$-enhanced QED correction to the total decay width of $b \to c e \nu$ one now has to integrate~(\ref{eq:discDeltace}) over the relevant three-particle phase space. Since in the massless electron limit, $\text{disc} \, \Delta_{c e}$ is independent of kinematics this integration is however trivial. Instead of~(\ref {eq:Gammapartial}) one therefore obtains
\beq \label{eq:GammapartialPi2}
\Gamma = \Gamma^{(0)} g(\rho) \left ( 1 + \frac{2 \pi \alpha}{3} \right ) = \Gamma^{(0)} g(\rho) \, \Big ( 1 + 1.57\% \Big ) \,, 
\eeq
where the explicit form of the function $g(\rho)$ can be found in~(\ref {eq:g}), which is also an excellent approximation to the massive electron case. Notice that we do not resum the Coulomb corrections as done for instance in~\cite{Isidori:2007zt,deBoer:2018ipi,Cali:2019nwp} since in our case the associated Sommerfeld enhancement~\cite{Sommerfeld:1931qaf} is insignificant. Beyond the leading power term computed above, one expects power corrections enhanced by $\pi^2$, starting at ${\cal O}(\Lambda_{\rm QCD}^2/m_b^2)$ for the total semi-leptonic width. These power corrections are sensitive to the spectator quark and may lead to an observable difference between the $B^0$ and $B^+$ semi-leptonic decay widths.
 
\section{Complete $\bm{{\cal O} (\alpha)}$ calculation} 
\label{sec:full}

In the following, we describe the different ingredients of our calculation of the ${\cal O} (\alpha)$ corrections to $b \to c e \nu$. We first provide a comprehensive discussion of the anatomy of the computation, the virtual and real corrections and their combination to obtain ultraviolate~(UV) and infrared~(IR)~finite results. While our approach is general and can be used to calculate the ${\cal O} (\alpha)$ corrections to any observable in $b \to c e \nu$, we present in this article only numerical results for the total decay width and the electron energy spectrum. These~results are then compared to the~LL and threshold corrections computed earlier. Finally, the relevant quantities used in the experimental analyses are introduced and a first numerical analysis is performed. 

\subsection{The calculation in a nutshell} 
\label{sec:nutshell}

The calculation of semi-leptonic $B \to X_c e \nu$ decay properties is characterised by a large hierarchy of scales, i.e.~$M_W^2 \gg m_b^2$, and is most conveniently carried out in the framework of an effective field theory~(EFT). If power-suppressed terms of ${\cal O} (m_b^2/M_W^2)$ are neglected, the relevant interactions are encoded by the weak Hamiltonian 
\beq \label{eq:Hweak}
 {\cal H} = -\frac{4 \hspace{0.125mm} G_F}{\sqrt{2}} \, C(\mu) \hspace{0.25mm} Q \,, \qquad Q = \big ( \bar c \hspace{0.5mm} \gamma_\mu P_L \hspace{0.25mm} b \big ) \big ( \bar e \hspace{0.5mm}\gamma^\mu P_L \hspace{0.25mm} \nu \big ) \,, 
\end{equation}
where $P_L$ projects onto fermionic fields with left-handed chirality. The Wilson coefficient~$C(\mu)$ contains the short-distance contributions from scales above $\mu$. In the case of~$B \to X_c e \nu$ and at leading power in $\Lambda_{\rm QCD}/m_b$, the matrix element of the dimension-six operator $Q$ is computed using perturbation theory for the semi-leptonic decay of a free bottom quark. In QCD the hadronic current present in $Q$ is a (partially) conserved current and therefore the anomalous dimension of $Q$ vanishes. However, if photonic corrections are considered, $Q$ develops a non-zero anomalous dimension and as a result its Wilson coefficient depends non-trivially on the renormalisation scale $\mu$, but no additional operator appears. While the~LL~dependence on $\mu$ has been computed more than 40 years ago by~Sirlin in~\cite{Sirlin:1981ie}, the complete~${\cal O} (\alpha)$ matching corrections to~(\ref{eq:Hweak}) have only been found rather recently in~\cite{Brod:2008ss,Gorbahn:2022rgl}. We perform an independent calculation, whose basic steps are outlined in Appendix~\ref{app:EWcorrections}, and~obtain
\beq \label{eq:CEW}
C (\mu) = 1 + \frac{\alpha}{2 \pi} \left [ \hspace{0.25mm} \ln \left ( \frac{M_Z^2}{\mu^2} \right ) - \frac{11}{6} \hspace{0.25mm} \right ] \,,
\eeq
which agrees with~\cite{Brod:2008ss,Gorbahn:2022rgl}. Identifying the renormalisation scale $\mu$ with $m_b=m_b^{\rm kin} (1 \, {\rm GeV})$ and using $M_Z = 91.1876 \, {\rm GeV}$ the EW~correction to the Wilson coefficient amounts to a positive shit of about $0.5\%$. We note that the rational term in~(\ref{eq:CEW}) is renormalisation scheme dependent and that the quoted value corresponds to the use of dimensional regularisation in $d = 4 - 2 \epsilon$ space-time dimensions with a naive anti-commuting $\gamma_5$~(NDR). At ${\cal O} (\alpha)$ the scale and scheme dependence of $C(\mu)$ is cancelled by that of the~${\cal O} (\alpha)$ corrections to the matrix element~$\left \langle Q (\mu ) \right \rangle$ computed in the EFT using the same renormalisation prescription. The basic steps of this computation are detailed in the~following. 

We begin the calculation of the ${\cal O} (\alpha)$ corrections to the matrix element of the four-fermion operator~$Q$ by parametrising the differential decay width of $b \to ce \nu$ including ${\cal O} (\alpha)$ effects as follows 
\beq \label{eq:dGamma}
 d\Gamma =\frac{\big | C (\mu) \big |^2}{4 \hspace{0.125mm} m_b} \Big [ B \hspace{0.25mm} d\phi_3 + \frac{\alpha}{\pi} \, \big (V \hspace{0.125mm} d\phi_3+ R \hspace{0.25mm} d\phi_4 \big )\Big] \,.
\eeq
Here $B$, $V$ and $R$ encode the Born-level, virtual and real corrections, respectively, $d\phi_3$ and $d\phi_4$ represent the relevant three- and four-body phase space and $C(\mu)$ is the Wilson coefficient given in~(\ref{eq:CEW}). In terms of the dimensionless kinematic parameters
\beq \label{eq:yij} 
y_{ij} = \frac{2 p_i \cdot p_j}{m_b^2} \,. 
\eeq
the Born-level contribution is given by 
\beq \label{eq:B}
B = N_B \hspace{0.5mm} y_{b\nu} \hspace{0.5mm} y_{ce} \,, \qquad N_B = 32 \hspace{0.25mm} G_F^2 \hspace{0.25mm} |V_{cb}|^2 \hspace{0.5mm} m_b^4 \,.
\eeq

In order to regulate the UV divergences that appear in $V$ we use the NDR scheme. The~finite parts of $V$ will depend on this scheme choice, but as already mentioned this scheme dependence will cancel against the scheme dependence of the Wilson~coefficient~(\ref{eq:CEW}). The same statement is true for the scale dependence. The logarithmic IR divergences associated to soft photon emissions are isolated and regulated by using a fictitious photon mass~$m_\gamma$. These IR-singular terms appear in both the virtual and the real ${\cal O} (\alpha)$ contributions but have to cancel in their combination making the final result manifestly independent of the photon mass in the $m_\gamma \to 0$ limit. This regulator independence provides an important check of the numerical implementation of our calculation. For what concerns collinear divergences, they are naturally regulated by the electron mass, which we always~keep~finite. 

\subsubsection{Virtual corrections}
\label{sec:virtuals}

Examples of Feynman diagrams that describe the virtual ${\cal O} (\alpha)$ corrections to $b \to c e \nu$ are displayed in~Figure~\ref{fig:virtuals}. External leg corrections are not shown in the figure but included in the results given below. We decompose the virtual contributions as follows
\beq \label{eq:Vdecomposition}
V = V_{\rm UV} + V_{\rm IR} + V_{\rm fin} \,.
\eeq 
where
\begin{align}
V_{\rm UV} & = \frac{B}{4} \left ( -Q_b^2 - Q_c^2 - Q_e^2 + 2 \hspace{0.125mm} Q_b \hspace{0.25mm} Q_c + 2 \hspace{0.125mm} Q_b \hspace{0.25mm} Q_e - 8 \hspace{0.125mm} Q_c \hspace{0.25mm} Q_e \right ) \left [ \frac{1}{\epsilon} + \ln \left ( \frac{\mu^2}{m_b^2} \right ) \right ] \,, \label{eq:VUV} \\[4mm]
 V_{\rm IR } & = \frac{B}{2} \left ( Q_b^2 \hspace{0.25mm} J_b + Q_c^2 \hspace{0.25mm} J_c + Q_e^2 \hspace{0.25mm} J_e + 2 \hspace{0.25mm} Q_b \hspace{0.25mm} Q_c \hspace{0.25mm} I_{bc}^1 + 2 \hspace{0.25mm} Q_b \hspace{0.25mm} Q_e \hspace{0.25mm} I_{be}^1 + 2 \hspace{0.25mm} Q_c \hspace{0.25mm} Q_e \hspace{0.25mm} I_{ce}^1 \right ) \,, \label{eq:VIR} \\[4mm]
 V_{\rm fin} & = N_B \, \Bigg \{ \Bigg [ \left \{ \frac{Q_b^2}{4} \left ( 1 + 2 \hspace{0.125mm} I_b^0 \right ) + \left ( b \rightarrow c \right ) + \left ( b \rightarrow e \right ) \right \} \nonumber \\[2mm]
 & \hspace{1.75cm} - \left \{ Q_b \hspace{0.25mm} Q_c \left ( \frac{5}{4} + \frac{I^0_{bc}}{2} - m_b^2 \left ( I_{bc}^x + \rho \hspace{0.25mm} I_{bc}^y \right ) \right ) + \left ( c \rightarrow e, \rho \rightarrow r \right ) \right \} \nonumber \\[2mm]
 & \hspace{1.75cm}+ Q_c \hspace{0.25mm} Q_e \left ( \frac{3}{2} + 2 \hspace{0.125mm} I^0_{ce} - m_b^2 \left ( \rho \hspace{0.25mm} I_{ce}^x + r \hspace{0.25mm} I_{ce}^y \right ) \right ) \Bigg ] \, y_{b\nu} \hspace{0.5mm} y_{ce} \nonumber \\[2mm]
& \hspace{1.5cm} + \Bigg \{ Q_b \hspace{0.25mm} Q_c \, \bigg [ \, \frac{1}{4} \left ( y_{be} \hspace{0.25mm} y_{c\nu} - y_{bc} \hspace{0.25mm} y_{e\nu} \right ) + m_b^2 \hspace{0.5mm} y_{ce} \left ( y_{bc} \hspace{0.25mm} y_{b\nu} - y_{c\nu} \right ) I_{bc}^x \nonumber \\[-4mm] \label{eq:Vfin} \\[0mm] 
& \hspace{3.25cm} + m_b^2 \hspace{0.5mm} y_{b\nu} \left ( y_{bc} \hspace{0.25mm} y_{ce} - \rho \hspace{0.25mm} y_{be} \right ) I_{bc}^y \nonumber \\[2mm]
& \hspace{3.25cm} - \frac{m_b^2}{2} \hspace{0.5mm} \Big( y_{bc} \left ( y_{b\nu} \hspace{0.25mm} y_{ce} + y_{be} \hspace{0.25mm} y_{c\nu} \right ) - y_{e\nu} \hspace{0.25mm} y_{bc}^2 \nonumber \\[2mm] 
& \hspace{4.35cm} - 2 \hspace{0.5mm} \big ( y_{ce} \hspace{0.25mm} y_{c\nu} + \rho \left ( y_{be} \hspace{0.25mm} y_{b\nu} - 2 \hspace{0.25mm} y_{e\nu} \right ) \big ) \hspace{0.25mm} I_{bc}^{xy} \Big ) \bigg ] \nonumber \\[2mm] 
 & \hspace{2.25cm} + \left ( c \leftrightarrow e, \rho \rightarrow r \right ) \Bigg \} 
+ Q_c \hspace{0.25mm} Q_e \hspace{0.5mm} m_b^2 \hspace{0.5mm} y_{b\nu} \left (y_{ce}^2 - 2 \hspace{0.125mm} \rho\hspace{0.25mm} r \right ) \left ( I_{ce}^x + I_{ce}^y \right ) \Bigg \} \,. \nonumber 
 \end{align}
Here $Q_b = -1/3$, $Q_c=2/3$ and $Q_e=-1$ denote the electric charges of the particle involved and we have introduced the abbreviation $r = m_e^2/m_b^2$. The explicit expressions for the integrals $J_i$, $I_i^0$, $I_{ij}^0$, $I_{ij}^x$, $I_{ij}^y$, $I_{ij}^{xy}$ and $I_{ij}^1$ can be found in Appendix~\ref{app:loopintegrals}. Notice that both $V_{\rm UV}$ and $V_{\rm IR}$ are proportional to the Born-level contribution~(\ref{eq:B}). This is a consequence of the well-known factorisation properties of UV and IR divergences. In contrast, only parts of the finite contributions~$V_{\rm fin}$ factorise. To illustrate this property we have decomposed $V_{\rm fin}$ into a term proportional to~$y_{b\nu} \hspace{0.5mm} y_{ce}$ and terms that are not directly proportional to the Born kinematics. We finally notice that the $1/\epsilon$ poles present in $V_{\rm UV}$ are cancelled by the $\overline{\rm MS}$ renormalisation of the four-fermion operator introduced in~(\ref{eq:Hweak}). 

\begin{figure}[t!]
\begin{center}
\includegraphics[width=0.85\textwidth]{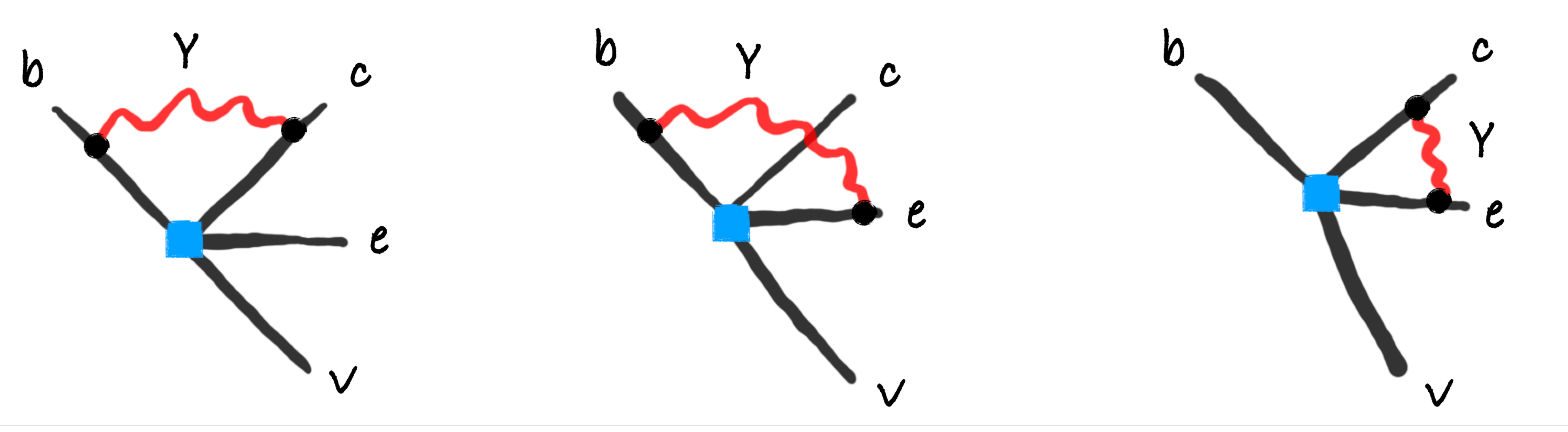}
\end{center}
\vspace{-6mm} 
\caption{\label{fig:virtuals} Virtual QED corrections to the $b \to c e \nu$ transition. The blue squares indicate an insertion of the four-fermion operator $Q$ introduced in~(\ref{eq:Hweak}) that induces the decay.}
\end{figure}
 
\subsubsection{Real corrections}
\label{sec:reals}

To determine the real contributions to the differential decay width~(\ref{eq:dGamma}) requires a treatment of the four-body phase space. In our work, we divide $d\phi_4$ into hard and soft regions using the improved phase space slicing method introduced in~\cite{Kilgore:1996sq}. One has schematically
\beq \label{eq:PSS}
\begin{split}
R \hspace{0.5mm} d\phi_4 & = \left [ R \left (1-\theta_\gamma \right ) + R \hspace{0.5mm} \theta_\gamma \right ] d\phi_4 \\[2mm] 
& = R_{\rm hard} \left (1-\theta_\gamma \right ) d\phi_4 + \theta_\gamma \, \big [ T_1(\theta_\gamma ) + T_2(\theta_\gamma ) + T_3(\theta_\gamma ) \big ] \,,
\end{split}
\eeq
where $R$ is the exact real contribution to~(\ref{eq:dGamma}) and the slicing of the phase space is indicated by~$\theta_\gamma$. In our computation, we employ $\theta_\gamma = \theta ( E_{\rm min} - E_\gamma)$ which evaluates to $0$ in the hard region, i.e.~for~$E_\gamma > E_{\rm min}$, while it returns $1$ in the soft region, i.e.~for~$E_\gamma < E_{\rm min}$. Here~$E_\gamma$~denotes the energy of the photon in the bottom-quark rest frame and $E_{\rm min}$ is a resolution parameter that determines whether the photon is resolved or unresolved. Notice that since the first term in the second line of~(\ref{eq:PSS}) encodes hard photon emission the photon mass can be set to zero when evaluating the corresponding matrix element squared. The subtraction terms in~(\ref{eq:PSS}) can be written as 
\beq \label{eq:T1T2T3}
T_1(\theta_\gamma) = R_{\rm soft} \hspace{0.5mm} d\phi^{\rm soft}_4 \,, \quad 
T_2(\theta_\gamma) = \left ( R - R_{\rm soft} \right ) d\phi_4 \,, \quad 
T_3(\theta_\gamma) = R_{\rm soft} \hspace{0.5mm} \big ( d\phi_4 - d\phi^{\rm soft}_4 \big ) \,, 
\eeq
with
\beq \label{eq:dphi4soft}
d\phi^{\rm soft}_4 = d\phi_3 \hspace{0.5mm} d\phi_\gamma \,, \qquad \quad d\phi_\gamma = \frac{1}{8\pi^2}\sqrt{E_\gamma^2-m_\gamma^2} \hspace{0.75mm} dE_\gamma \hspace{0.5mm} d\cos\theta\,.
\eeq
Here $\theta$ is the angle that the photon makes with the $z$ direction and $d\phi_3$ denotes the three-body phase space of the Born-level process $b\to c e \nu$. 
Notice that $T_1(\theta_\gamma)$ contains the soft divergences needed to cancel the singularities of the virtual contribution, while $T_2(\theta_\gamma)$ and $T_3(\theta_\gamma)$ are finite and vanish as the domain of support of $\theta_\gamma$ is taken to zero~\cite{Eynck:2001en}. In our case this means that for sufficiently small $E_{\rm min}$ only the term $T_1(\theta_\gamma)$ needs to be included in the calculation. 

\begin{figure}[t!]
\begin{center}
\includegraphics[width=0.85\textwidth]{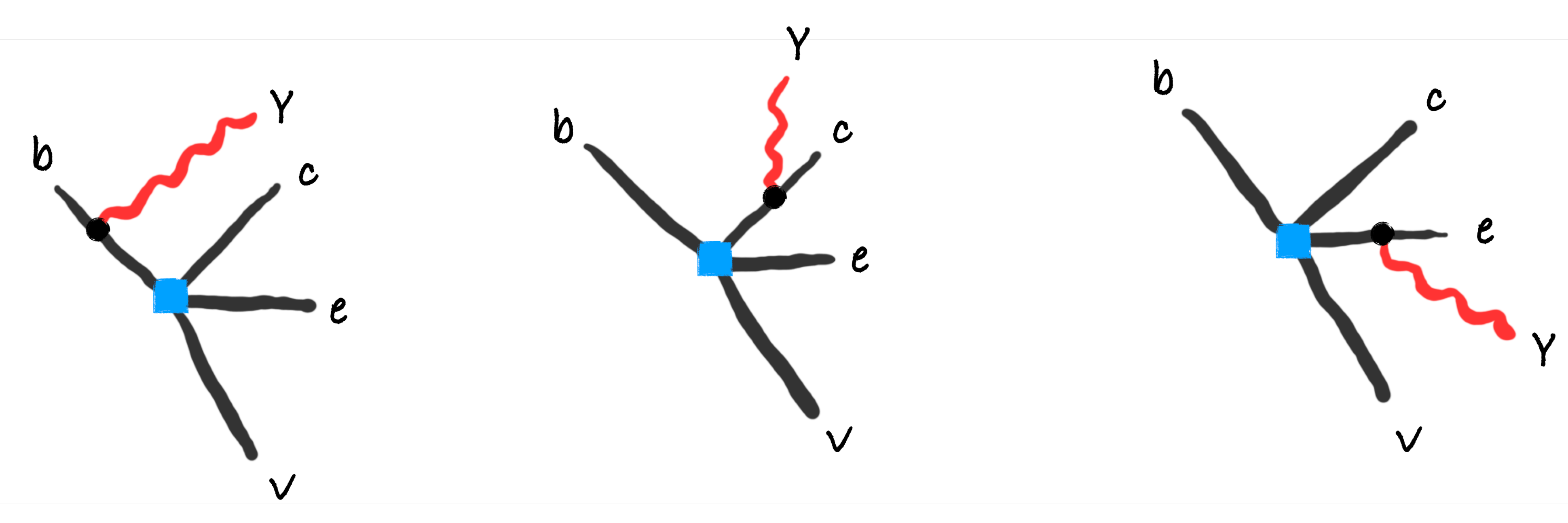}
\end{center}
\vspace{-6mm} 
\caption{\label{fig:reals} Real emission QED contributions to the $b \to c e \nu$ process. The colour coding resembles that used in~Figure~\ref{fig:virtuals}.}
\end{figure}

From the above discussion it follows that in order to determine the real ${\cal O} (\alpha)$ corrections to the partonic $b \to c e \nu$ process we need to calculate $R_{\rm hard}$ and $R_{\rm soft}$. The relevant Feynman~graphs are depicted in~Figure~\ref{fig:reals}. For the hard real contribution we obtain 
\bea \label{eq:Rhard}
\begin{split}
R_{\rm hard} & = \frac{16 \hspace{0.125mm} \pi^2 N_B}{m_b^2} \, \Bigg \{ \frac{Q_b^2 \hspace{0.5mm} y_{ce} \big ( y_{b\gamma} \hspace{0.5mm} y_{\nu \gamma} - 2 \left ( y_{b \nu} - y_{\nu \gamma} \right ) \big ) }{2 \hspace{0.25mm} y_{b\gamma}^2} + \frac{Q_c^2 \hspace{0.5mm} y_{b\nu} \big ( y_{c\gamma} \hspace{0.5mm} y_{e \gamma} - 2 \hspace{0.125mm} \rho \left ( y_{c e} + y_{e \gamma} \right ) \big ) }{2 \hspace{0.25mm} y_{c\gamma}^2} \\[2mm]
& \hspace{2.05cm} + \frac{Q_e^2 \hspace{0.5mm} y_{b\nu} \big ( y_{c\gamma} \hspace{0.5mm} y_{e \gamma} - 2 \hspace{0.25mm} r \left ( y_{c e} + y_{c \gamma} \right ) \big ) }{2 \hspace{0.25mm} y_{e\gamma}^2} - Q_b \hspace{0.5mm} Q_c \left [ \, \frac{ y_{b\nu} \left ( y_{be}+ y_{ce} \right )}{2 \hspace{0.25mm} y_{b\gamma}} \right. \\[2mm] 
& \hspace{2.55cm} \left. - \frac{ y_{ce} \left ( y_{b\nu}+ y_{c\nu} \right )}{2 \hspace{0.25mm} y_{c\gamma}} - \frac{ y_{bc} \hspace{0.25mm}\big ( y_{b\nu} \left (2 y_{ce}+ y_{e\gamma})- y_{ce} \hspace{0.25mm} y_{\nu\gamma} \right ) \big ) }{2 \hspace{0.25mm} y_{b\gamma} \hspace{0.25mm} y_{c\gamma}} \right ] \\[2mm]
& \hspace{2.05cm} - Q_b \hspace{0.5mm} Q_e \left [ \, \frac{ y_{b\nu} \left ( y_{bc}+ y_{ce} \right )}{2 \hspace{0.25mm} y_{b\gamma}} - \frac{ y_{ce} \left ( y_{b\nu}+ y_{e\nu} \right )}{2 \hspace{0.25mm} y_{e\gamma}} \right . \\[2mm] 
& \hspace{3.5cm} \left. - \frac{ y_{be} \hspace{0.25mm}\big ( y_{b\nu} \left (2 y_{ce}+ y_{c\gamma})- 2 \hspace{0.25mm} y_{ce} \hspace{0.25mm} y_{\nu\gamma} \right ) \big ) }{2 \hspace{0.25mm}y_{b\gamma} \hspace{0.25mm} y_{c\gamma}} \right ] \\[2mm]
& \hspace{2.05cm} - \frac{Q_c \hspace{0.5mm} Q_e \hspace{0.5mm} y_{b\nu} \hspace{0.5mm} \big ( \left (y_{ce} + y_{c\gamma} \right ) \left (y_{ce} + y_{e\gamma} \right ) - \rho \hspace{0.25mm} y_{e\gamma} - r \hspace{0.25mm} y_{c\gamma} \big ) }{y_{c\gamma} \hspace{0.25mm} y_{e\gamma}} \Bigg \} \,.
\end{split}
\eea 

In the case of the soft real contribution that appears in $T_1(\theta_\gamma)$ we instead find 
\bea \label{eq:Rsoft}
\begin{split}
R_{\rm soft} & = -\frac{16 \hspace{0.125mm} \pi^2 B}{m_b^2} \left [ \frac{Q_b^2}{\left ( \lambda - y_{b \gamma} \right )^2 } + \frac{Q_c^2 \hspace{0.5mm} \rho }{\left ( \lambda + y_{c \gamma} \right )^2} + \frac{Q_e^2 \hspace{0.5mm} r }{\left ( \lambda + y_{e \gamma} \right )^2} \right . \\[2mm] 
& \left. \hspace{2.1cm} + \frac{Q_b \hspace{0.25mm} Q_c \hspace{0.5mm} y_{bc} }{\left ( \lambda - y_{b \gamma} \right ) \left ( \lambda + y_{c \gamma} \right ) } + \frac{Q_b \hspace{0.25mm} Q_e \hspace{0.5mm} y_{be} }{\left ( \lambda - y_{b \gamma} \right ) \left ( \lambda + y_{e \gamma} \right ) } + \frac{Q_c \hspace{0.25mm} Q_e \hspace{0.5mm} y_{ce} }{\left ( \lambda + y_{c \gamma} \right ) \left ( \lambda + y_{e \gamma} \right ) } \right ] \,, \hspace{6mm}
\end{split}
\eea
where $\lambda = m_\gamma^2/m_b^2$. This result is simple enough that one can integrate it over the $d\phi_\gamma$ part of the soft phase space~(\ref {eq:dphi4soft}) and obtain an analytic result. We~find 
\beq \label{eq:Rsoftint}
\begin{split} 
\int\! \theta_\gamma \hspace{0.25mm} R_{\rm soft} \hspace{0.25mm} d\phi_\gamma & = \frac{B}{2} \hspace{0.5mm} \Big[ \hspace{0.25mm} -Q_b^2 \hspace{0.5mm} I_R(p_b)-Q_c^2 \hspace{0.5mm} I_R(p_c)-Q_e^2 \hspace{0.5mm} I_R(p_e) \\[2mm] 
& \hspace{1.175cm} -Q_b \hspace{0.5mm} Q_c \hspace{0.5mm} I_R(p_b,p_c)-Q_b \hspace{0.5mm} Q_e \hspace{0.5mm} I_R(p_b ,p_e)+Q_c \hspace{0.5mm} Q_e \hspace{0.5mm} I_R(p_c,p_e) \hspace{0.25mm} \Big]\,.
\end{split}
\eeq
The soft integrals $I_R(p_i)$ and $I_R(p_i,p_j)$ are given in Appendix~\ref{app:softintegrals}. Note that~(\ref{eq:Rsoft}) and~(\ref{eq:Rsoftint}) are both proportional to the Born-level contribution~(\ref{eq:B}). This is again a result of the factorisation of IR physics and a necessary condition so that soft singularities in~$R_{\rm soft}$ can cancel similar contribution in the virtual contributions~$V_{\rm IR}$. 

\subsubsection{Implementation and validation}
\label{sec:MC}

\begin{figure}[t!]
\begin{center}
\includegraphics[width=0.6\textwidth]{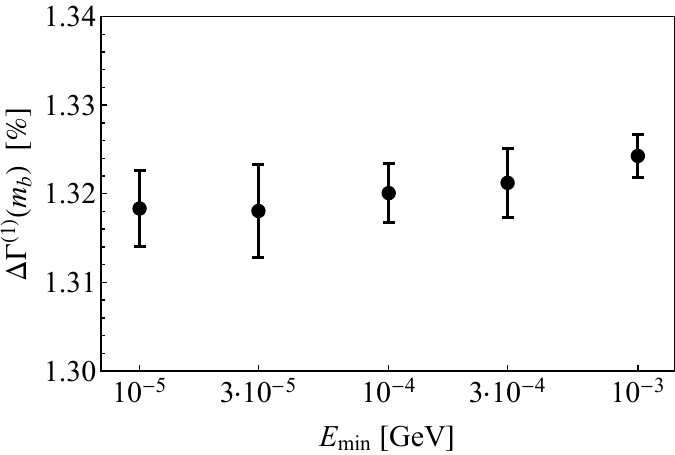}
\end{center}
\vspace{-4mm} 
\caption{\label{fig:CUBA} Dependence of the complete ${\cal O}(\alpha)$ correction to the total decay width,~i.e.~$\Delta \Gamma^{(1)}(m_b)$ in~(\ref{eq:DeltaGamma}), on~$E_\mathrm{min}$. The error bars represent the numerical integration error provided by CUBA.}
\end{figure}

In the following, we briefly discuss the numerical implementation of the virtual and real corrections, their combination and the validation of our MC code. The three- and four-particle phase spaces $d\phi_3$ and $d\phi_4$ play an important role in the calculation of the differential decay width~(\ref{eq:dGamma}). We adopt the four-body phase-space parameterisations given in~\cite{Asatrian:2012tp}. The multi-dimensional sampling necessary to integrate the virtual and real contribution over their respective phase spaces is performed with the help of the~CUBA library~\cite{Hahn:2004fe}. To~validate that soft singularities cancel between~(\ref {eq:VIR}) and~(\ref {eq:Rsoftint}) and to check the irrelevance of the subtraction terms $T_2 (\theta_\gamma)$ and $T_3 (\theta_\gamma)$ in~(\ref {eq:PSS}), we have computed the electron energy spectrum and the total decay width for few different values of the resolution parameter in the range~$E_{\rm min} \in [10^{-5}, 10^{-3}] \, {\rm GeV}$. Within the statistical uncertainties given by CUBA we find agreement between the different results. In the case of the complete ${\cal O} (\alpha)$ correction to the total decay width this feature is illustrated in~Figure~\ref{fig:CUBA}. Hereafter we employ the value $E_{\rm min} = 10^{-4} \, {\rm GeV}$ as our standard choice. We have furthermore numerically verified that the total decay width of $b \to c \ell \nu$ does not contain logarithmic mass singularities by employing $m_\ell = 511 \, {\rm keV}$ or $m_\ell = 106 \, {\rm MeV}$. These two choices correspond to the case of an electron or a muon in the final state. As a final cross-check of our MC implementation we have computed the ${\cal O} (\alpha)$ corrections to the total decay width of $\mu \to e \nu \nu$ and the ${\cal O} (\alpha_s)$ corrections to the total decay width of $b \to c e \nu$, reproducing the classic results~\cite{Berman:1958ti,Kinoshita:1958ru,Nir:1989rm} with a precision far better than $1\%$. 

\subsection{Numerical results} 
\label{sec:numerics}

\begin{figure}[t!]
\begin{center}
\includegraphics[width=0.6\textwidth]{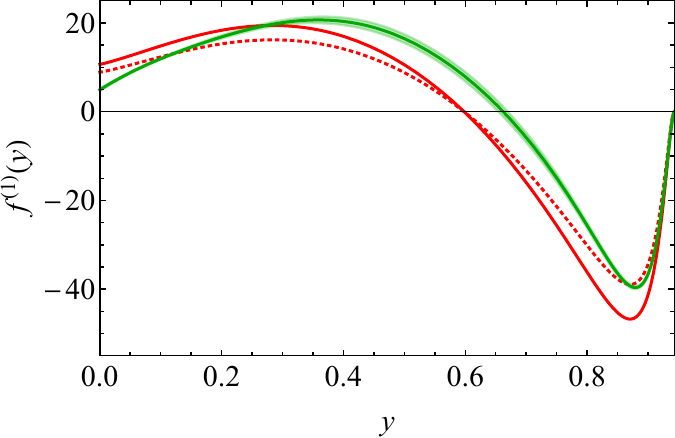}
\end{center}
\vspace{-4mm} 
\caption{\label{fig:fullparton} Comparison of the complete ${\cal O}(\alpha)$ corrections~(\ref{eq:Delta1full}) to the electron energy spectrum~(green~curve) in $b \to c e \nu$ and the corresponding~LL approximation~(red~curve). In the former case also the uncertainty of our numerical phase-space integration is indicated (green band). The~LL approximation using $\bar L_{c/e}$ instead of $\bar L_{b/e}$ is displayed~as well~(dotted red curve). See the main text for additional details. }
\end{figure}

In Figure~\ref{fig:fullparton} we display the complete~${\cal O}(\alpha)$ corrections~(green curve and band) and the corresponding~LL approximation~(red~curve) computed in Section~\ref{sec:LL}. The green curve corresponds to an interpolation obtained by considering 40 different bins that cover the full physical region of $y \in [2 \hspace{0.25mm} \sqrt{r}, 1-\rho + r]$, while the green band reflects the associated numerical integration uncertainties. Relative to the total ${\cal O}(\alpha)$ corrections these uncertainties typically amount to around $1\%$, except close to the zero of the depicted green curve. For the purpose of this comparison, we have factored the Wilson coefficient out and set the renormalisation scale $\mu$ equal to $m_b=m_b^{\rm kin} (1 \, {\rm GeV})$. We observe a relatively good agreement between the~LL terms and the complete ${\cal O}(\alpha)$ corrections to the electron energy spectrum of the partonic $b \to c e \nu$ transition, especially in the hard part of the spectrum, where the~LL approximation is expected to work best, and where the differences amount to around $10\%$ to $20\%$.
Writing 
\beq \label{eq:Delta1full} 
f^{(1)} (y) = \frac{\bar L_{ b/e}}{2} \, f^{(1)}_{\mathrm{LL}} (y) + \Delta f^{(1)} (y) \,, 
\eeq 
with $f^{(1)}_{\mathrm{LL}} (y)$ given in~(\ref {eq:f1partonic}), we can use our numerical results for $f^{(1)} (y)$ to obtain a simple approximate expression for $\Delta f^{(1)} (y)$. Employing $\rho = 0.057$ and $r = 1.25 \cdot 10^{-8}$ and identifying again the renormalisation scale $\mu$ with $m_b=m_b^{\rm kin} (1 \, {\rm GeV})$, we find 
\begin{equation}
\label{eq:Deltaf1}
\begin{aligned}
\Delta f^{(1)} (y) & = 
\Bigg [
 -2.04264+119.012 \hspace{0.5mm} y-476.678 \hspace{0.5mm} y^2+2034.14 \hspace{0.5mm} y^3 \\[2mm]
& \phantom{xxx} -4402.22 \hspace{0.5mm} y^4+4505.93 \hspace{0.5mm} y^5-1807.38 \hspace{0.5mm} y^6\\[2mm]
& \phantom{xxx} -66.8251\left (y-y_{\rm max} \right ) \ln \left (y_{\rm max} -y \right ) \, \Bigg ] \, \theta ( y_{\rm max}-y) \,,
\end{aligned}
\end{equation}
where $y_{\rm max}=1-\rho+r$. This formula encodes the exact non-LL terms for the input parameters listed above with a relative accuracy of better than $1\%$. 
It is worth noting that in Section~\ref{sec:LL} we have used $m_b$ as the hard scale in the logarithm $\bar L_{b/e}$ as defined in~(\ref{eq:Lij}). This is a somewhat arbitrary choice because the hard scale is in fact of the order of the energy released,~i.e.~of~${\cal O} (m_b-m_c)$, and using a scale lower than $m_b$ in the~LL~QED effects might thus be more appropriate. To investigate this aspect, we also display in~Figure~\ref{fig:fullparton} the electron energy spectrum obtained using $\bar L_{c/e}$ instead of $\bar L_{b/e}$ in the~LL~QED prediction~(dotted~red~curve). We observe a better agreement near the endpoint but not elsewhere, suggesting that the terms beyond the~LLs cannot be accounted for by a rescaling. Hereafter we hence evaluate all~LL~QED corrections with our standard choice~$\bar L_{b/e}$.

By direct integration over the full phase space, we also obtain a value of the ${\cal O} (\alpha)$ effects in the total decay width of the partonic $b \to c e \nu$ process, 
\beq \label{eq:Gammapartialfull}
\Gamma = \Gamma^{(0)} g(\rho) \, \big |C(\mu) \big |^2 \; \Big [ \, 1 + \Delta\Gamma^{(1)}(\mu) \, \Big ] \, ,
\eeq
where $\Gamma^{(0)}$ and $g(\rho)$ are defined in~(\ref {eq:EespectrumLOpartonic}) and~(\ref{eq:g}), respectively. The correction $\Delta \Gamma^{(1)}(\mu)$ represents the ${\cal O} (\alpha)$ contribution to the matrix element of the operator introduced in~(\ref{eq:Hweak}) evaluated at the scale $\mu$. For the input parameters used before, we find
\beq \label{eq:DeltaGamma}
\Delta\Gamma^{(1)}(\mu)= \frac{\alpha}{\pi} \left [ \hspace{0.25mm} \ln \left ( \frac{\mu^2}{m_b^2} \right ) + 5.516 \hspace{0.25mm} (14) \hspace{0.25mm} \right ] \,,
\eeq
where the coefficient of the logarithm is exact while the quoted numerical coefficient has as indicated an uncertainty of around $0.3\%$ which is associated to our MC phase-space integration. Combining~(\ref{eq:CEW}),~(\ref{eq:Gammapartialfull}) and~(\ref{eq:DeltaGamma}), one finds to ${\cal O} (\alpha)$ that 
\beq \label{eq:GammaOverGammaLO}
\begin{split}
\frac{\Gamma}{\Gamma^{(0)} g(\rho)} & = 1 + \frac{\alpha}{\pi} \left [ \hspace{0.25mm} \ln \left ( \frac{M_Z^2}{m_b^2} \right ) - \frac{11}{6} + 5.516 \hspace{0.25mm} (14) \hspace{0.25mm} \right ] \\[4mm] 
& = 1 + 1.43\% - 0.44\% + 1.32\% = 1 + 2.31 \% \,, 
\end{split}
\eeq
where in the second line we have dropped the quoted uncertainty but given the numerical results of the individual ${\cal O} (\alpha)$ terms as well as their sum. The first observation to make is that the renormalisation scale dependence has cancelled between the ${\cal O} (\alpha)$ corrections to the Wilson coefficient and the virtual contributions to the matrix element $\big($cf.~(\ref{eq:CEW}) and~(\ref{eq:VUV})$\big)$ leaving behind the EW logarithm first computed in~\cite{Sirlin:1981ie}. In fact, it is interesting to note that this logarithm represents about~$60\%$ of the total ${\cal O} (\alpha)$ correction in~(\ref{eq:GammaOverGammaLO}). Comparing the result~(\ref{eq:GammapartialPi2}) with~(\ref{eq:GammaOverGammaLO}) one furthermore observes that the $\pi^2$-enhanced terms calculated in Section~\ref{sec:Pi2effects} provide about~$120\%$ of~$\Delta \Gamma^{(1)}(m_b)$,~i.e.~the complete ${\cal O} (\alpha)$ contribution to the matrix element of~(\ref{eq:Hweak}). Hence, the complete ${\cal O} (\alpha)$ correction to the total decay width of $b \to ce\nu$ is well approximated by the sum of the EW logarithm and the $\pi^2$-enhanced threshold effects, which are both scale- and scheme-independent.

The relevant quantities used in the experimental analyses are the branching ratio of $B \to X_c e \nu$, the electron energy spectrum and its moments with a lower cut $E_{\rm cut}$ on the electron energy in the rest frame of the decaying $B$ meson. A quantity employed in the extrapolation of the semi-leptonic branching ratio with a cut $E_e > E_{\rm cut}$ to the total semi-leptonic branching ratio is
\beq \label{eq:Rstar}
R^\ast (E_{\rm cut})= \frac{\displaystyle \int_{E_{\rm cut}}^{E_{\rm max}} d E_e \, \frac{d \Gamma}{d E_e} }{\displaystyle \int_{m_e}^{E_{\rm max}} d E_e\, \frac{d \Gamma}{d E_e}} \,, 
\eeq
where $E_{\rm max} = m_b/2 \left (1 - \rho + r \right )$ denotes the maximum of the electron energy available in the partonic $b \to c e \nu$ transition. The moments of the electron energy spectrum are defined~as 
\beq \label{eq:defmoments}
\left \langle E_e^n \right \rangle_{E_e> E_{\rm cut}} = \frac{\displaystyle \int_{E_{\rm cut}}^{E_{\rm max}} d E_e \, E_e^n \, \frac{d \Gamma}{d E_e} }{\displaystyle \int_{E_{\rm cut}}^{E_{\rm max}} d E_e\, \frac{d \Gamma}{d E_e}} \,.
\eeq
In~the $B$-factory analyses~\cite{BaBar:2004bij,Belle:2006kgy,BaBar:2009zpz} the central moments of the electron energy spectrum are considered. The first three central moments are given~by 
\beq \label{eq:defli}
\begin{split}
\ell_1 (E_{\rm cut} ) & = \left \langle E_e \right \rangle_{E_e > E_{\rm cut}} \,, \\[3mm]
\ell_2 (E_{\rm cut} ) & = \Big \langle \big ( E - \langle E_e \rangle \big )^2 \Big \rangle _{E_e > E_{\rm cut}} \,, \\[2mm]
\ell_3 (E_{\rm cut} ) & = \Big \langle \big ( E - \langle E_e \rangle \big )^3 \Big \rangle _{E_e > E_{\rm cut}} \,.
\end{split}
\eeq

\begin{table}[t!]
\begin{center}
\vspace{-4mm}
\begin{tabular}{|c|c|c|c|c|}
\hline 
$E_{\rm cut}$ & $R^{\ast \hspace{0.25mm} {\rm LO}}$ & $R^\ast$& $\Delta R^{\ast \hspace{0.25mm} {\rm LL}}$ & $\Delta R^{\ast \hspace{0.25mm} (1)}$\\
\hline \hline
$0.6$ & $0.9556$ & $0.9516$ & $-0.0150$ & $-0.0120 \hspace{0.25mm} (3)$ \\
$0.8$ & $0.9005$ & $0.8919$ & $-0.0211$ & $-0.0176 \hspace{0.25mm} (3)$ \\
$1.0$ & $0.8171$ & $0.8017$ & $-0.0266$ & $-0.0230 \hspace{0.25mm} (1)$ \\
$1.2$ & $0.7041$ & $0.6798$ & $-0.0304$ & $-0.0270 \hspace{0.25mm} (1)$ \\
$1.5$ & $0.4824$ & $0.4431$ & $-0.0306$ & $-0.0282 \hspace{0.25mm} (1) $ \\
\hline \hline
$E_{\rm cut}$ & $\ell_{1}^{\rm LO}$ & $\ell_1$ & $\Delta \ell_{1}^{\rm LL}$ & $\Delta \ell_{1}^{(1)}$\\
\hline \hline
$m_e$ & $1.4187$ & $1.3867$ & $-0.0389$ & $-0.0342 \hspace{0.25mm} (1)$ \\
$0.6$ & $1.4639$ & $1.4345$ & $-0.0229$& $-0.0219 \hspace{0.25mm} (5)$ \\
$0.8$ & $1.5101$ & $1.4831$ & $-0.0181$ & $-0.0176 \hspace{0.25mm} (4)$ \\
$1.0$ & $1.5718$ & $1.5478$ & $-0.0138$ & $-0.0136 \hspace{0.25mm} (3)$ \\
$1.2$ & $1.6469$ & $1.6272$ & $-0.0102$ & $-0.0101 \hspace{0.25mm} (2)$ \\
$1.5$ & $1.7803$ & $1.7721$ & $-0.0058$ & $-0.0059 \hspace{0.25mm} (1)$ \\
\hline \hline
$E_{\rm cut}$ & $\ell_{2}^{\rm LO}$ & $\ell_2$ & $\Delta \ell_{2}^{\rm LL}$ & $\Delta \ell_{2}^{(1)}$\\
\hline \hline
$m_e$ & $0.1832$ & $0.1836$ & $+0.0163$ & $+0.0122 \hspace{0.25mm} (2)$ \\
$0.6$ & $0.1449$ & $0.1452$ & $+0.0024$ & $+0.0019 \hspace{0.25mm} (2)$ \\
$0.8$ & $0.1165$ & $0.1171$ & $+0.0006$ & $+0.0004 \hspace{0.25mm} (1)$ \\
$1.0$ & $0.0870$ & $0.0883$ & $-0.00024$ & $-0.00033 \hspace{0.25mm} (7)$ \\
$1.2$ & $0.0597$ & $0.0619$ & $-0.00048$ & $-0.00054 \hspace{0.25mm} (3)$ \\
$1.5$ & $0.0272$ & $0.0308$ & $-0.00035$ & $-0.00038 \hspace{0.25mm} (1)$ \\
\hline \hline
$E_{\rm cut}$ & $\ell_{3}^{\rm LO}$ & $\ell_3$ & $\Delta \ell_{3}^{\rm LL}$ & $\Delta \ell_{3}^{(1)}$\\
\hline \hline
$m_e$ & $-0.0416$ & $-0.0325$ & $-0.0065$ & $-0.0036 \hspace{0.25mm} (2)$ \\
$0.6$ & $-0.0184$ & $-0.0108$ & $+0.00238$ & $+0.00236 \hspace{0.25mm} (8)$ \\
$0.8$ & $-0.0090$ & $-0.0027$ & $+0.00186$ & $+0.00182 \hspace{0.25mm} (5)$ \\
$1.0$ & $-0.0031$ & $\phantom{+}0.0017$ & $+0.00114$ & $+0.00112 \hspace{0.25mm} (3)$ \\
$1.2$ & $-0.00029$ & $\phantom{+}0.0032$ & $+0.00058$ & $+0.00057 \hspace{0.25mm} (2)$ \\
$1.5$ & $\phantom{+}0.00057$ & $\phantom{+}0.0023$ & $+0.000136$ & $+0.000137 \hspace{0.25mm} (5)$ \\
\hline
\end{tabular}
\end{center}
\vspace{-2mm} 
\caption{\label{tab:section4} Comparison of the~LL and the complete ${\cal O} (\alpha)$ contributions to the observables defined in~(\ref{eq:Rstar}) and~(\ref{eq:defli}) for different values of the lower cut $E_{\rm cut}$ on the electron energy. All quantities are given in units of ${\rm GeV}$ to the appropriate power. The second column shows the LO~results while the third column corresponds to the state-of-the-art QCD predictions used in the recent HQE global~fit~\cite{Bordone:2021oof}. The fourth and fifth columns present the absolute shifts due to the~LL and the complete~${\cal O} (\alpha)$ corrections, respectively. Only the leading power terms are included in these~${\cal O} (\alpha)$~predictions. Additional explanation can be found in the main text.}
\end{table}

Notice that all the observables introduced in~(\ref{eq:Rstar}),~(\ref{eq:defmoments}) and~(\ref{eq:defli}) are ratios and as such they are insensitive to the overall normalisation. This means in particular that they neither depend on the CKM matrix element $V_{cb}$ nor on the Wilson coefficient $C(\mu)$. They are also renormalisation scale independent and, to an excellent approximation, insensitive to the $\pi^2$-enhanced threshold correction. Since they do not cancel out like in the total decay width, the~LL contributions should therefore represent the leading source of QED corrections to~(\ref{eq:Rstar}),~(\ref{eq:defmoments}) and~(\ref{eq:defli}), and it can be expected that in the case of the higher moments these corrections may be sizeable. To illustrate the latter point let us consider the central moments defined in (\ref{eq:defli}) in the absence of a cut on the electron energy,~i.e.~$E_e > m_e$. Relative to the LO results we find that the~LL~QED corrections to the first three central moments~are
\beq \label{eq:deltaliLL}
\begin{split}
\delta \ell_{1}^{\rm LL} (m_e) & = -\frac{2}{3} \, \frac{\alpha}{\pi} \, \bar L_{b,e} = -2.74\% \,, \\[2mm]
\delta \ell_{2}^{\rm LL} (m_e) & = 2.16 \, \frac{\alpha}{\pi} \, \bar L_{b,e} = +8.9\% \,, \\[2mm]
\delta \ell_{3}^{\rm LL} (m_e) & = 3.78 \, \frac{\alpha}{\pi} \, \bar L_{b,e} = +15.6\% \,. 
\end{split}
\eeq
Notice that the result for $\delta \ell_{1}^{\rm LL} (m_e)$ is independent of $\rho$ because, in the absence of a lower cut on the electron energy, the moments of a convolution factorise and the~LL correction to the first moment is proportional to the lowest order result. The results for $\delta \ell_{2}^{\rm LL} (m_e)$ and~$\delta \ell_{3}^{\rm LL} (m_e)$ instead depend on $\rho$ and we have employed the input parameters discussed before in our numerics. For the corresponding complete ${\cal O} (\alpha)$ corrections we instead obtain the relative~shifts
\beq \label{eq:deltalialpha}
\begin{split}
\delta \ell_{1}^{(1)} (m_e) & = -10.08 \hspace{0.25mm} (3) \hspace{0.25mm} \, \frac{\alpha}{\pi} = -2.412 \hspace{0.25mm} (7) \% \,, \\[2mm]
\delta \ell_{2}^{(1)} (m_e) & = 27.8 \hspace{0.25mm} (5) \hspace{0.25mm} \, \frac{\alpha}{\pi} = +6.7 \hspace{0.25mm} (1) \% \,, \\[2mm]
\delta \ell_{3}^{(1)} (m_e) & = 36 \hspace{0.25mm} (2) \, \frac{\alpha}{\pi} = +8.7 \hspace{0.25mm} (5) \% \,. 
\end{split}
\eeq
As indicated these predictions have relative uncertainties due to our numerical phase-space integration of around $0.3\%$, $1.8\%$ and $5.6\%$, respectively. From~(\ref{eq:deltaliLL}) and~(\ref{eq:deltalialpha}) one observes that while for the first central moment the~LL and the full ${\cal O} (\alpha)$ predictions agree quite well, in the case of the second and third central moments the complete ${\cal O} (\alpha)$ results are markedly smaller than their~LL counterparts. This is because in the higher central moments there are strong cancellations among the various contributing linear moments, which subdominant contributions tend to disrupt, leading to larger deviations from the~LL approximation. 

A comprehensive comparison of the~LL and the complete ${\cal O} (\alpha)$ contributions to the observables defined in~(\ref{eq:Rstar}) and~(\ref{eq:defli}) is presented in~Table~\ref{tab:section4}. The numbers given in the table show that the LL approximation captures the bulk of the full ${\cal O}(\alpha)$ corrections to $R^\ast (E_{\rm cut})$, $\ell_1 (E_{\rm cut})$, $\ell_2 (E_{\rm cut})$ and $\ell_3 (E_{\rm cut})$. In fact, the LL approximation tends to improve with increasing $E_{\rm cut}$ values. Also notice that the considered QED corrections are comparable in magnitude to the higher-order QCD and power corrections that have been included in the recent HQE global~fit~\cite{Bordone:2021oof}. We add that the latter feature probably only applies to $R^\ast (E_{\rm cut})$ and to the moments of the electron energy spectrum~(\ref{eq:defli}), where perturbative QCD and power corrections are relatively suppressed. 

\section{Comparison with BaBar implementation of QED effects} 
\label{sec:comparison}

As already explained in the introduction, both BaBar and Belle subtract photon radiation associated to charged final-state particles using PHOTOS, because such effects were not included in the HQE theoretical predictions that were available at that time. In fact, the BaBar~papers~\cite{BaBar:2004bij,BaBar:2009zpz} report explicitly the size of the PHOTOS correction that is applied for each observable. In the publication~\cite{BaBar:2009zpz}, the BaBar collaboration updated its earlier measurement~\cite{BaBar:2004bij} using more recent values of the branching ratios of some of the background processes. This more recent paper reports the size of the PHOTOS corrections only for $E_{\rm cut}= \{ 0.6, 1.5 \} \, {\rm GeV}$, finding results which are very close to those given previously in~\cite{BaBar:2004bij}. Since the earlier work~\cite{BaBar:2004bij} provides numbers for five different cuts $E_{\rm cut} = \{0.6,0.8,1.0,1.2,1.5\} \, {\rm GeV}$ we will use them for our comparison of the branching ratio of $B \to X_c e \nu$ $\big($${\rm BR}_{\rm incl} (E_{\rm cut})$$\big)$ and the first three central moments. 

To obtain QED correction factors for the distributions of the inclusive $B \to X_c e \nu$ decay, the BaBar collaboration has generated spectra for various exclusive decay channels such as $B \to De \nu$, $B \to D^\ast e \nu$ and $B \to D^{\ast \ast} e \nu$ without and with up to two photon emissions utilising PHOTOS, appropriately weighting each exclusive sample. BaBar then subtracts the obtained corrections from the measured or uncorrected values of ${\rm BR}_{\rm incl} (E_{\rm cut})$, $\ell_1(E_{\rm cut})$, $\ell_2(E_{\rm cut})$ and $\ell_3(E_{\rm cut})$ to derive the corrected or QCD values of the branching ratio and the central~moments. Notice that the procedure that BaBar uses to estimate the size of the QED corrections to $B \to X_c e \nu$ is only able to capture logarithmically enhanced effects associated to the soft and collinear photonic corrections to the matrix element of the operator~$Q$ in specific channels. It instead misses the interference between initial- and final-state photons, structure-dependent contributions, EW corrections that are associated to short-distance physics encoded in the Wilson coefficient, as well as hard photon emission and virtual photon corrections such as the $\pi^2$-enhanced threshold corrections discussed in~Section~\ref{sec:Pi2effects}.

Any comparison with the QED correction factors obtained by BaBar requires a choice of how to split the complete QED corrections to the differential decay width~(\ref{eq:dGamma}) into an effective coefficient~$A_{\rm EW}$ and a function $f (y)$ that describes the normalised electron energy spectrum in $B \to X_c e \nu$. We choose 
the following decomposition
\beq \label{eq:Eespectrumfinal}
\frac{d \Gamma}{dy}= \Gamma^{(0)} \, A_{\rm EW} \, f (y) \,, \qquad A_{\rm EW} = 1 + \frac{\alpha}{\pi} \hspace{0.25mm} \ln \left ( \frac{M_Z^2}{m_b^2} \right ) = 1.0143 \,, 
\eeq
with 
\bea \label{eq:fall}
\begin{split}
 f(y) & = f^{(0)} (y) + \frac{\alpha}{\pi} \, \left [ f^{(1)} (y) - \frac{11}{6} \hspace{0.25mm} \right ] + \left ( \frac{\alpha}{2\pi} \right )^2 \frac{\bar L_{ b/e}^2}{2} \, f^{(2)}_{\rm NLL} (y) \\[2mm] 
& + \sum_{i = \pi, G} \frac{\mu_i^2}{m_b^2} \left [ f^{(0)}_i (y) + \frac{\alpha}{2\pi} \, \bar L_{ b/e} \, f^{(1)}_{i, {\rm LL}} (y) \right ] + \sum_{j = D, SL} \frac{\rho_j^3}{m_b^3} \left [ f^{(0)}_j(y) + \frac{\alpha}{2\pi} \, \bar L_{ b/e} \, f^{(1)}_{j, {\rm LL}} (y) \right ] \,. \hspace{4mm} 
\end{split}
\eea
Here $\Gamma^{(0)}$ is given in~(\ref{eq:EespectrumLOpartonic}) and the logarithmic ${\cal O} (\alpha)$ correction in $A_{\rm EW}$ is the scheme-independent part of the Wilson coefficient~(\ref{eq:CEW}) with the renormalisation scale $\mu$ set to~$m_b$. The~numerical value of $A_{\rm EW}$ quoted in~(\ref{eq:Eespectrumfinal}) corresponds to $m_b^{\rm kin} (1 \, {\rm GeV}) = 4.57 \, {\rm GeV}$ and $M_Z = 91.1876 \, {\rm GeV}$. It is the value of~$A_{\rm EW}$ that is typically used in the literature (see~for example~\cite{Bordone:2021oof}). The functions $f^{(0)} (y)$ and $f^{(1)} (y)$ are given in~(\ref {eq:f0partonic}) and~(\ref {eq:Deltaf1}), respectively, and the term $-11/6$ is the scheme-dependent part of the Wilson coefficient~(\ref{eq:CEW}). Since it is contained in $f(y)$ the function is formally scheme independent up to ${\cal O} (\alpha)$. The function~$f^{(2)}_{\rm NLL} (y)$ describes the NLL~QED effects to the partonic electron energy spectrum and is evaluated in~Appendix~\ref{app:NLLpartonic}, while the contributions in the second line contain the LL~QED contributions to the power corrections that are detailed in Appendix~\ref{app:LLpower}. In our numerical study, we will use the values $\mu_\pi^2 = 0.477 \,{\rm GeV}^2$, $\mu_G^2 = 0.306 \,{\rm GeV}^2$, $\rho_D^3 = 0.185 \,{\rm GeV}^3$ and $\rho_{SL}^3 = -0.130 \, {\rm GeV}^3$, corresponding to the central values of the HQE parameters obtained in the recent global fit~\cite{Bordone:2021oof}. 

\begin{figure}[t!]
\begin{center}
\includegraphics[width=0.6\textwidth]{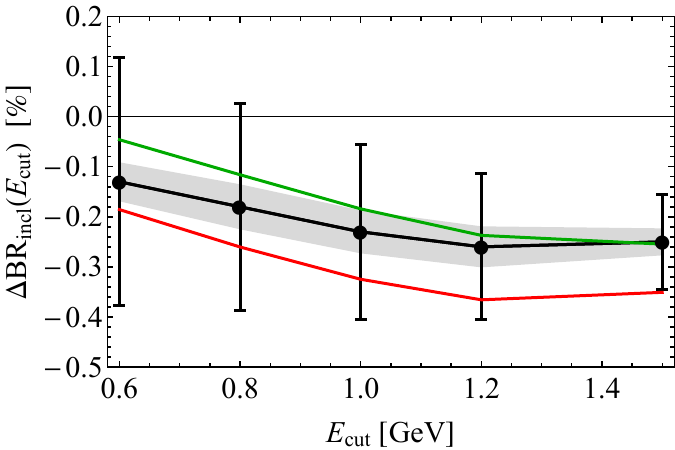}
\end{center}
\vspace{-4mm} 
\caption{\label{fig:DeltaBRincl} Comparison of the absolute shift of the QED corrections to ${\rm BR}_{\rm incl} (E_{\rm cut})$ as a function of the lower cut $E_{\rm cut}$ on the electron energy. The black curve corresponds to the correction obtained by BaBar in~\cite{BaBar:2004bij} using PHOTOS, while the red (green) curve corresponds to our QED prediction including the LL terms (all QED corrections) given in~(\ref {eq:fall}). The grey band represents the systematic uncertainty on the PHOTOS bremsstrahlungs corrections that BaBar quotes in~\cite{BaBar:2004bij}, while the black error bars correspond to the total uncertainties of the QED corrected BaBar results. For further explanations consult the main text.}
\end{figure}

In~Figure~\ref{fig:DeltaBRincl} we present a comparison of the absolute shift of the QED corrections to the inclusive branching ratio of $B \to X_c e \nu$ imposing a lower cut $E_e > E_{\rm cut}$ on the electron energy. The~black curve corresponds to the correction obtained by BaBar in~\cite{BaBar:2004bij} using PHOTOS and the grey band represents the associated systematic uncertainties. The~black error bars furthermore correspond to the total uncertainties of the QED corrected measurements. Our~LL~prediction contains all terms in~(\ref {eq:fall}) that are proportional to $\alpha/\pi \hspace{0.25mm} \bar L_{b/e}$ and is indicated by the red~line. The~green curve instead corresponds to our best QED prediction considering all contributions to~(\ref {eq:fall}). We~observe that the BaBar results and our LL predictions show relative deviations of around $40\%$. This is an expected feature because PHOTOS as any other QED parton shower MC should capture the bulk of the logarithmically enhanced photonic effects, even though it works with point-like hadrons. However, the absence of virtual effects and in particular of the $\pi^2$-enhanced threshold corrections in PHOTOS explains why our full QED predictions are visibly above the BaBar results. Also notice that for $E_{\rm cut} \lesssim 1.0 \, {\rm GeV}$ the differences between the BaBar numbers and our best QED predictions are larger than the systematic uncertainties on the PHOTOS bremsstrahlungs corrections that BaBar provides in the publication~\cite{BaBar:2004bij}. The~found differences are however always safely within the total experimental uncertainties. 

In Table~\ref{tab:DeltaBRincl} we present a more detailed breakdown of the individual relative QED corrections to~${\rm BR}_{\rm incl} (E_{\rm cut})$. All~theory predictions are normalised to the state-of-the-art QCD predictions of the branching ratio that have been obtained recently in~\cite{Bordone:2021oof}. One first observes that compared to the partonic LL~QED effects the corresponding NLL corrections are very small. Similar~observations have been made in the case of muon decay in~\cite{Arbuzov:2002pp,Arbuzov:2002cn,Arbuzov:2002rp}. Likewise, the power-suppressed LL~QED corrections of ${\cal O} (\Lambda_{\rm QCD}^2/m_b^2)$ and ${\cal O} (\Lambda_{\rm QCD}^3/m_b^3)$ have only a minor impact on ${\rm BR}_{\rm incl} (E_{\rm cut})$. In~fact, the calculated ${\cal O} (\Lambda_{\rm QCD}^2/m_b^2)$ and ${\cal O} (\Lambda_{\rm QCD}^3/m_b^3)$ effects slightly improve the agreement with the BaBar corrections. The non-logarithmic~${\cal O} (\alpha)$ effects calculated in Section~\ref{sec:full} and encoded by~$\Delta f^{(1)} (y)$ are instead numerically quite relevant and tend to reduce the partonic LL~QED effects in magnitude. Notice that this reduction would be larger by around~$0.4\%$ if the constant $-11/6$ had been included in $A_{\rm EW}$ and not in~$f(y)$ $\big($cf.~(\ref{eq:Eespectrumfinal}) and~(\ref{eq:fall})$\big)$. As~a result when using our best QED calculation to correct the BaBar measurements we obtain ${\rm BR}_{\rm incl} (E_{\rm cut})$ values that are on average larger by about~$0.2\sigma$ than the QED corrected values for ${\rm BR}_{\rm incl} (E_{\rm cut})$ given in~\cite{BaBar:2004bij}.

\begin{table}[t!]
\begin{center}
\vspace{4mm}
\begin{tabular}{|c|c|c|c|c|c|c|c|}
\hline
$E_{\rm cut}$ & $\delta {\rm BR}^{\rm BaBar}_{\rm incl}$ & $\delta {\rm BR}^{\rm LL}_{\rm incl}$ & $\delta {\rm BR}^{\rm NLL}_{\rm incl}$ & $\delta {\rm BR}^{\alpha}_{\rm incl}$ & $\delta {\rm BR}^{1/m_b^2}_{\rm incl}$ & $\delta {\rm BR}_{\rm incl}$ & $\sigma$ \\
\hline \hline
$0.6$ & $-1.26\%$ & $-1.92\%$ & $-1.95\%$ & $-0.54\%$ & $-0.50\%$ & $-0.45\%$ & $+0.34$ \\
$0.8$ & $-1.87\%$ & $-2.88\%$ & $-2.91\%$ & $-1.36\%$ & $-1.29\%$ & $-1.22\%$ & $+0.30$ \\
$1.0$ & $-2.66\%$ & $-4.03\%$ & $-4.04\%$ & $-2.38\%$ & $-2.26\%$ & $-2.15\%$ & $+0.25$ \\
$1.2$ & $-3.56\%$ & $-5.43\%$ & $-5.41\%$ & $-3.65\%$ & $-3.43\%$ & $-3.27\%$ & $+0.14$ \\
$1.5$ & $-5.22\%$ & $-8.41\%$ & $-8.26\%$ & $-6.37\%$ & $-5.73\%$ & $-5.39\%$ & $-0.09$ \\
\hline
\end{tabular}
\end{center}
\vspace{-2mm} 
\caption{\label{tab:DeltaBRincl} Relative size of the QED corrections to ${\rm BR}_{\rm incl} (E_{\rm cut})$. The values of $E_{\rm cut}$ are given in units of ${\rm GeV}$. The entries in the column $\delta {\rm BR}^{\rm BaBar}_{\rm incl}$ are the corrections obtained by BaBar in~\cite{BaBar:2004bij}, while the numbers for $\delta {\rm BR}^{\rm LL}_{\rm incl}$, $\delta {\rm BR}^{\rm NLL}_{\rm incl}$ and $\delta {\rm BR}^{\alpha}_{\rm incl}$ successively include the LL, NLL and complete ${\cal O}(\alpha)$ corrections to the $b \to ce\nu$ branching ratio. The $\delta {\rm BR}^{1/m_b^2}_{\rm incl}$ numbers include all partonic QED effects as well as the LL~QED corrections to the ${\cal O} (\Lambda_{\rm QCD}^2/m_b^2)$ power corrections. The entries in the column $\delta {\rm BR}_{\rm incl}$ represent our best predictions and include besides all partonic QED effects the power-suppressed LL~QED corrections up to ${\cal O} (\Lambda_{\rm QCD}^3/m_b^3)$ $\big($see~(\ref{eq:fall})$\big)$. The relative shifts in standard deviations~($\sigma$) that we obtain when using our best QED calculation to correct the BaBar measurements are given in the last column. See~main text for additional details.}
\end{table}

\begin{figure}[t!]
\begin{center}
\includegraphics[width=0.6\textwidth]{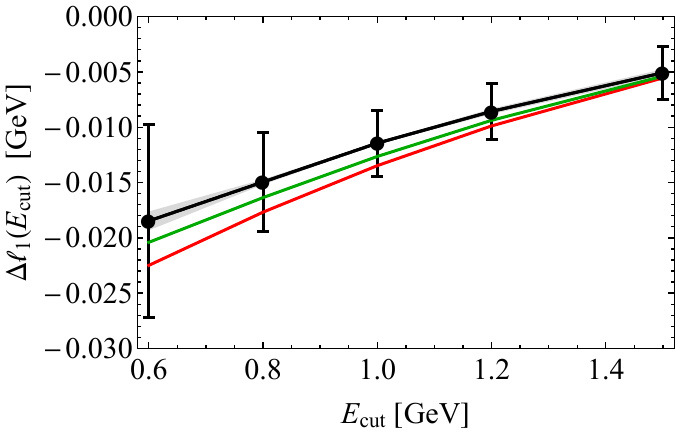}

\vspace{4mm}

\includegraphics[width=0.6\textwidth]{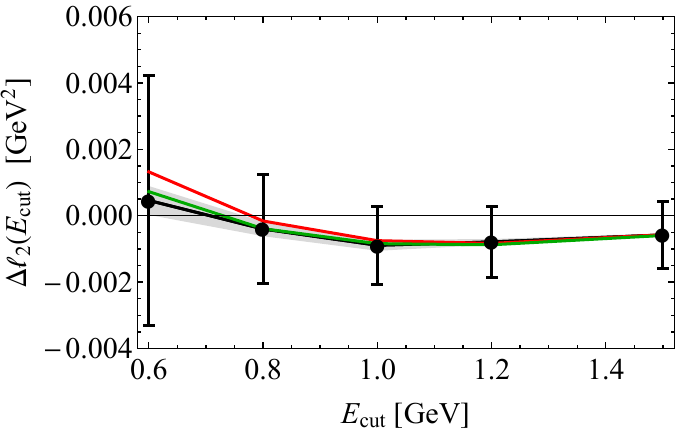}

\vspace{4mm}

\includegraphics[width=0.6\textwidth]{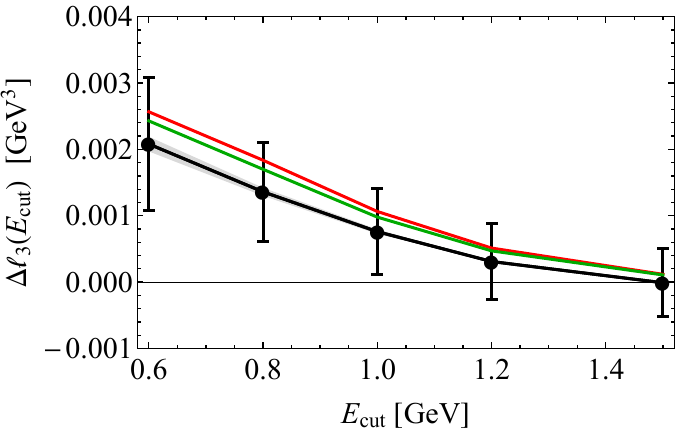}

\end{center}
\vspace{-4mm} 
\caption{\label{fig:Deltalisex} Comparison of the absolute shift of the QED corrections to~$\ell_1 (E_{\rm cut})$,~$\ell_2 (E_{\rm cut})$ and~$\ell_3 (E_{\rm cut})$ as a function of the lower cut $E_{\rm cut}$ on the electron energy. The colour coding resembles that used in~Figure~\ref{fig:DeltaBRincl}.}
\end{figure}

The absolute shift of the QED corrections to~$\ell_1 (E_{\rm cut})$,~$\ell_2 (E_{\rm cut})$ and~$\ell_3 (E_{\rm cut})$ is shown in the three panels in Figure~\ref{fig:Deltalisex}. In order to not spoil the strong cancellations between the quantum corrections to the numerator and the denominator that enter the normalised central moments~\cite{Biswas:2009rb,Gambino:2011cq} we perform a double-series expansion in $\alpha$ and $\Lambda_{\rm QCD}/m_b$ when calculating the ratios~(\ref{eq:defli}). In this expansion we keep all the terms up to the order indicated by the superscript following the notation introduced in~(\ref{eq:fall}). We~add that we have verified that the expanded and unexpanded results of the central moments are numerically quite close together. The~black curves correspond to the~QED~corrections estimated by~BaBar in~\cite{BaBar:2004bij} with the help of PHOTOS, while the red (green) lines represent our LL $\big($full~${\cal O} (\alpha)$$\big)$ predictions. The grey bands represent the systematic uncertainties that are associated to the experimental subtraction procedure of QED corrections performed in~\cite{BaBar:2004bij}, while the black error bars correspond to the total uncertainties of the BaBar measurements. From all three plots it is evident that the~LL~QED corrections describe the~BaBar corrections pretty well and that the numerical impact of the non-LL~${\cal O} (\alpha)$ corrections is notably smaller in the case of $\ell_1 (E_{\rm cut})$,~$\ell_2 (E_{\rm cut})$ and~$\ell_3 (E_{\rm cut})$ than for ${\rm BR}_{\rm incl} (E_{\rm cut})$. Still the inclusion of the term $\alpha/\pi \left ( \Delta f^{(1)} (y) - 11/6 \right )$ in the calculation of the central moments in general improves the agreement between the BaBar and our QED corrections. Also notice that in the case of $\ell_2 (E_{\rm cut})$ the differences between the BaBar numbers and our best QED predictions are within the systematic uncertainty band of the PHOTOS bremsstrahlungs corrections, while this is not the case for $\ell_1 (E_{\rm cut})$ and $\ell_3 (E_{\rm cut})$. Given that the systematic uncertainties associated to the subtraction of QED effects are always a subdominant component in the total experimental uncertainties, our absolute shifts $\Delta \ell_1 (E_{\rm cut})$, $\Delta \ell_2 (E_{\rm cut})$ and $\Delta \ell_3 (E_{\rm cut})$ are, however, always fully compatible with the combined errors quoted by BaBar. 

\begin{table}[t!]
\begin{center}
\vspace{4mm}
\begin{tabular}{|c|c|c|c|c|c|c|c|}
\hline
$E_{\rm cut}$ & $\delta \ell_1^{\rm BaBar}$ & $\delta \ell_1^{\rm LL}$ & $\delta \ell_1^{\rm NLL}$ & $\delta \ell_1^{\alpha}$ & $\delta \ell_1^{1/m_b^2}$ & $\delta \ell_1$ & $\sigma$ \\
\hline \hline
$0.6$ & $-1.29\%$ & $-1.60\%$ & $-1.58\%$ & $-1.48\%$ & $-1.45\%$ & $-1.42\%$ & $-0.22$ \\
$0.8$ & $-1.01\%$ & $-1.22\%$ & $-1.20\%$ & $-1.16\%$ & $-1.13\%$ & $-1.10\%$ & $-0.31$ \\
$1.0$ & $-0.74\%$ & $-0.89\%$ & $-0.88\%$ & $-0.87\%$ & $-0.84\%$ & $-0.82\%$ & $-0.40$ \\
$1.2$ & $-0.53\%$ & $-0.63\%$ & $-0.62\%$ & $-0.62\%$ & $-0.59\%$ & $-0.58\%$ & $-0.32$ \\
$1.5$ & $-0.29\%$ & $-0.33\%$ & $-0.32\%$ & $-0.34\%$ & $-0.31\%$ & $-0.30\%$ & $-0.13$ \\
\hline \hline
$E_{\rm cut}$ & $\delta \ell_2^{\rm BaBar}$ & $\delta \ell_2^{\rm LL}$ & $\delta \ell_2^{\rm NLL}$ & $\delta \ell_2^{\alpha}$ & $\delta \ell_2^{1/m_b^2}$ & $\delta \ell_2$ & $\sigma$ \\
\hline \hline
$0.6$ & $+0.31\%$ & $+1.65\%$ & $+1.43\%$ & $+0.91\%$ & $+0.48\%$ & $+0.50\%$ & $+0.07$ \\
$0.8$ & $-0.34\%$ & $+0.50\%$ & $+0.34\%$ & $+0.04\%$ & $-0.40\%$ & $-0.33\%$ & $+0.01$ \\
$1.0$ & $-1.00\%$ & $-0.27\%$ & $-0.38\%$ & $-0.60\%$ & $-1.08\%$ & $-0.95\%$ & $+0.04$ \\
$1.2$ & $-1.27\%$ & $-0.78\%$ & $-0.85\%$ & $-1.05\%$ & $-1.60\%$ & $-1.42\%$ & $-0.08$ \\
$1.5$ & $-1.91\%$ & $-1.15\%$ & $-1.18\%$ & $-1.40\%$ & $-2.24\%$ & $-1.93\%$& $-0.01$ \\
\hline \hline
$E_{\rm cut}$ & $\delta \ell_3^{\rm BaBar}$ & $\delta \ell_3^{\rm LL}$ & $\delta \ell_3^{\rm NLL}$ & $\delta \ell_3^{\alpha}$ & $\delta \ell_3^{1/m_b^2}$ & $\delta \ell_3$ & $\sigma$ \\
\hline \hline
$0.6$ & $-17.3\%$ & $-22.1\%$ & $-23.1\%$ & $-22.7\%$ & $-22.6\%$ & $-22.5\%$ & $+0.35$ \\
$0.8$ & $-42.8\%$ & $-68.9\%$ & $-69.2\%$ & $-66.7\%$ & $-62.9\%$ & $-62.9\%$ & $+0.45$ \\
$1.0$ & $+63.9\%$ & $+67.3\%$ &$+66.3\%$ & $+63.9\%$ & $+56.2\%$ & $+57.6\%$ & $+0.34$ \\
$1.2$ & $+11.7\%$ & $+18.1\%$ & $+17.6\%$ & $+17.1\%$ & $+13.3\%$ & $+14.7\%$ & $+0.28$ \\
$1.5$ & $-0.47\%$ & $+5.92\%$ & $+5.69\%$ & $+5.61\%$ & $+2.10\%$ & $+4.69\%$& $+0.23$ \\
\hline
\end{tabular}
\end{center}
\vspace{-2mm} 
\caption{\label{tab:Deltalisex} Relative size of the QED corrections to the first three central moments of the electron energy spectrum~(\ref{eq:defli}). The notation resembles that used in~Table~\ref{tab:DeltaBRincl}.}
\end{table}

In Table~\ref{tab:Deltalisex} we present a complete breakdown of the individual relative QED corrections to $\ell_1 (E_{\rm cut})$,~$\ell_2 (E_{\rm cut})$ and~$\ell_3 (E_{\rm cut})$. All~theory predictions are again normalised to the state-of-the-art QCD results for the central moments~\cite{Bordone:2021oof}. One first notices that while in the case of $\ell_1 (E_{\rm cut})$ and $\ell_2 (E_{\rm cut})$ the relative corrections are at the level of $1\%$ or below, for $\ell_3 (E_{\rm cut})$ the $\delta \ell_3 (E_{\rm cut})$ values typically exceed $10\%$. This feature can be traced back to the smallness of the QCD prediction for $\ell_3 (E_{\rm cut})$ (cf.~Table~\ref{tab:section4}) with $\ell_3 ( 0.9 \, {\rm GeV}) \simeq 0$ leading to a sign flip of $\delta \ell_3 (E_{\rm cut})$ between $E_{\rm cut} = 0.8 \, {\rm GeV}$ and $E_{\rm cut} = 1.0 \, {\rm GeV}$. In the case of $\ell_1 (E_{\rm cut})$, one furthermore sees that successively including the NLL, the complete ${\cal O}(\alpha)$ corrections as well as the LL~QED terms of the ${\cal O} (\Lambda_{\rm QCD}^2/m_b^2)$ and ${\cal O} (\Lambda_{\rm QCD}^3/m_b^3)$ power corrections steadily improves the agreement between our and the BaBar numbers. As a result, our best QED predictions differ from the BaBar results by no more than $0.13\%$ for $\ell_1 (E_{\rm cut})$. In the case of $\ell_2 (E_{\rm cut})$ and $\ell_3 (E_{\rm cut})$, we find that the LL~QED corrections of ${\cal O} (\Lambda_{\rm QCD}^2/m_b^2)$ and ${\cal O} (\Lambda_{\rm QCD}^3/m_b^3)$ are larger than the partonic~${\cal O} (\alpha)$ corrections, and that the inclusion of the former contributions in many cases help to improve the agreement between our and the PHOTOS calculation of QED effect. These~features are particularly apparent in the case of~$\ell_2 (E_{\rm cut})$. Numerically, we obtain differences below $0.19\%$ and below~$21\%$ for the second and third central moment, respectively. Taking into account the total experimental uncertainties the found deviations however amount to always less than~$0.5\sigma$.

\section{Conclusions and outlook} 
\label{sec:conclusions}

In order to match the increasing experimental accuracy in semi-leptonic, radiative and rare $B$ decays, a reliable assessment of EW and QED corrections is desirable in general, in some cases even mandatory. In recent years, such calculations have received considerable theoretical attention \cite{deBoer:2018ipi,Cali:2019nwp,Mishra:2020orb,Bansal:2021oon,Gambino:2000fz,Gambino:2001au,Bobeth:2003at,Huber:2005ig,Huber:2007vv,Bernlochner:2010fc,Bobeth:2013uxa,Huber:2015sra,Bordone:2016gaq,Beneke:2017vpq,Beneke:2019slt,Beneke:2020vnb,Isidori:2020acz,Beneke:2021jhp,Beneke:2021pkl,Beneke:2022msp,Herren:2022spb,Isidori:2022bzw,Cornella:2022ubo,Huang:2023nli,Choudhury:2023uhw,Gorbahn:2022rgl}. This article represents a first step towards obtaining precision predictions for inclusive semi-leptonic $B \to X_c \ell \nu$ decays including EW and QED corrections that can be directly confronted with existing and future $B$-factory measurements. To~this purpose, we have identified two types of large QED corrections, namely collinear terms proportional to $\ln \hspace{-0.25mm} \left (m_b^2 /m_\ell^2 \right )$ and threshold corrections enhanced by a factor $\pi^2$. In~order to show that these terms indeed provide the dominant part of the QED corrections, we have calculated the complete~${\cal O} (\alpha)$ corrections to the electron energy spectrum and the total decay width of the partonic process $b \to ce\nu$. The relevant virtual and real matrix elements have been calculated analytically while the phase-space integration is performed numerically, using an improved phase space slicing method to handle soft singularities. This~in~principle allows for a fully-differential calculation of any IR-safe observable in $b \to ce\nu$. Motivated by the observation that the logarithmically enhanced corrections represent the dominant~${\cal O}(\alpha)$~effects in the electron energy spectrum of~$b \to ce\nu$, we have furthermore calculated the NLL~QED corrections in the partonic case and the LL~QED corrections in the case of the power corrections of ${\cal O}(\Lambda^2_{\rm QCD}/m_b^2)$ and ${\cal O}(\Lambda^3_{\rm QCD}/m_b^3)$ to the first few moments of the spectrum.These~calculations can be carried out analytically using splitting functions. We have also completed the computation of the short-distance Wilson coefficient at ${\cal O}(\alpha)$ finding agreement with~\cite{Brod:2008ss,Gorbahn:2022rgl}. 

 Including all the above radiative corrections, we have then compared our results against those obtained by BaBar in~\cite{BaBar:2004bij} using the PHOTOS~package~(see~\cite{Bernlochner:2010fc,Cali:2019nwp,Bobeth:2013uxa,Huber:2015sra,Isidori:2022bzw} for similar comparisons). This~MC~code is the standard tool used in the relevant experimental $B$-factory analyses to subtract photon radiation from the measured distributions. While PHOTOS includes the real and virtual QED corrections that are needed to obtain a logarithmically accurate description of the soft-photon region of the phase space, effects such as the interference between initial- and final-state photons, structure-dependent contributions, EW corrections that are associated to short-distance physics, as well as hard photon emission and threshold corrections enhanced by $\pi^2$ are not incorporated in the generator. Our~comprehensive comparison to PHOTOS shows that in the case of the branching ratio the missing pieces lead to a systematic shift upwards by around~$0.8\%$ for low cuts on the electron energy. For~what concerns the first and second central moment of the electron energy spectrum, we find differences of below~$0.13\%$ and below $0.19\%$, respectively. In~the case of the third central moment of the electron energy spectrum, the maximal deviations between our and the PHOTOS results can reach up to around~$21\%$. We~also find that the differences between the BaBar numbers and our best QED predictions are often larger than the systematic uncertainties of the PHOTOS bremsstrahlungs corrections provided in~\cite{BaBar:2004bij}. Taking into account the total experimental uncertainties the found deviations however amount to always less than $0.5\sigma$. Interestingly, the inclusion of power corrections tends to improve the agreement between our and the PHOTOS calculation of QED effects in the branching ratio and the electron energy moments. Since our computation contains QED effects not included in PHOTOS, we believe that our calculation provides a better description of QED radiation in $B \to X_c e \nu$ than~PHOTOS. 

We add that the QED corrections to the total decay width and the moments of the electron energy spectrum calculated here have subsequently been included in a full HQE global fit~\cite{Finauri:2023kte}. There it has been found that the correction $\delta {\rm BR}_{\rm incl}$ at the lowest $E_{\rm cut}$ value leads to the largest shift of $-0.4\%$ in $|V_{cb}|$, as one would naively expect from our discussion in~Section~\ref{sec:comparison}. The~QED corrections to the moments of the electron energy spectrum instead amount to an effect of $+0.2\%$. Applying our new results for the QED corrections to the electron energy spectrum and its moments to the BaBar data~\cite{BaBar:2004bij,BaBar:2009zpz}, the~article~\cite{Finauri:2023kte} finds therefore a total modification of around~$-0.2\%$ in $|V_{cb}|$ compared to the inclusive determination in~\cite{Bordone:2021oof}. Notice that the HQE global fit~\cite{Finauri:2023kte} also includes the Belle data~\cite{Belle:2006kgy}. The~latter measurement subtracts the QED effects in a way that is similar to what is done by BaBar, but the paper~\cite{Belle:2006kgy} unfortunately does not provide the size of the subtraction, so that one cannot improve the treatment of QED effects for what concerns the Belle~data. 

The calculations and studies in this work can be extended in various ways. In the case of the inclusive semi-leptonic $B \to X_c \ell \nu$ decays, it would be worthwhile to generalise the computations presented here to the moments of the hadronic and leptonic invariant mass distribution, since these observables play an important role in the HQE global fits~\cite{Bordone:2021oof,Bernlochner:2022ucr} as well. Similar statements apply to the forward-backward asymmetry~\cite{Herren:2022spb,Turczyk:2016kjf} that has not been measured at BaBar and Belle, but should be accessible at Belle~II. The~employed calculation~techniques can also be applied to exclusive semi-leptonic $B$ decays. While the impact of QED radiation in~$B \to D \ell \nu$ has been studied by several groups~\cite{deBoer:2018ipi,Cali:2019nwp,Bansal:2021oon}, including both logarithmic corrections and $\pi^2$-enhanced threshold effects, in the case of~$B \to D^\ast \ell \nu$ the existing analysis~\cite{Bernlochner:2010fc} takes into account only logarithmically enhanced contributions. Improved calculations of $B \to D^{(\ast)} \ell \nu$ are also motivated by the measurements of the lepton flavour universality ratios $R(D)$ and $R(D^\ast)$ that show long-standing deviations from their respective SM predictions at the level of $3\sigma$~\cite{HeavyFlavorAveragingGroup:2022wzx}. In summary, our calculation and its planned extension to exclusive semi-leptonic $b \to c \ell \nu$ decays sets the theoretical foundations to build a new MC generator specifically designed to be used by the $B$-factory collaborations. Such~a tool will be~a crucial ingredient in any future extraction of $|V_{cb}|$ that aims at a precision close to~$1\%$.

\acknowledgments{We are grateful to Martin~Beneke, Florian~Bernlochner, Rhorry~Gauld, Martin~Gorbahn, Gino~Isidori, Martin~Jung, Marcello~Rotondo and Christoph Schwanda for useful discussions and explanations. We thank Gael Finauri for identifying the incorrect overall factor in~(\ref{eq:ReC0}) in the original version of our work. The research of PG is supported in part by the Italian Ministry of Research (MIUR) under grant PRIN 20172LNEEZ. He also received partial support from the Excellence Cluster ORIGINS which is funded by the Deutsche Forschungsgemeinschaft (DFG, German Research Foundation) under Germany's Excellence Strategy -- EXC-2094 -- 390783311 as well as from the Sino-German Collaborative Research Center TRR110 ``Symmetries and the Emergence of Structure in QCD'' (DFG Project-ID 196253076, NSFC Grant No. 12070131001, - TRR 110).}

\begin{appendix}

\section{LL effects beyond leading power} 
\label{app:LLpower}

In this appendix, we give the analytic results for LL~QED corrections to the electron energy spectrum beyond leading power. At the LO the ${\cal O} (\Lambda_{\rm QCD}^2/m_b^2)$ and ${\cal O} (\Lambda_{\rm QCD}^3/m_b^3)$ corrections to the electron energy spectrum can be parameterised as 
\beq \label{eq:EePC}
\frac{d\Gamma}{dy}= \Gamma^{(0)} \left [ f^{(0)}(y) + \frac{\mu_\pi^2}{m_b^2} \, f^{(0)}_\pi (y) + \frac{\mu_G^2}{m_b^2} \, f^{(0)}_G (y) + \frac{\rho_D^3}{m_b^3} \, f^{(0)}_D (y) + \frac{\rho_{LS}^3}{m_b^3} \, f^{(0)}_{LS} (y) \right ] \,, 
\eeq
where the expression for $\Gamma^{(0)}$ and $f^{(0)}(y)$ can be found in~(\ref {eq:EespectrumLOpartonic}) and~(\ref {eq:f0partonic}), respectively. In~the case of a massless electron, the remaining LO~functions appearing in~(\ref{eq:EePC}) are~\cite{Bigi:1993fe,Blok:1993va,Manohar:1993qn,Voloshin:1994cy,Gremm:1996df}
\begin{align} \label{eq:f0power}
f^{(0)}_\pi (y) & = \left [ -\frac{10}{3} \hspace{0.5mm} y^3- \frac{2 \left (5 - 2 y \right ) y^3 \rho ^2}{\left (1 - y \right)^4} + \frac{4 \left(10 -5 y + y^2 \right) y^3 \rho ^3}{3 \left (1-y \right )^5} \right ] \theta ( 1 - y - \rho ) \,, \\[4mm]
f^{(0)}_G (y) & = \left [ \frac{2}{3} \left (6 + 5 y \right ) y^2 -\frac{4 \left (3 - 2 y \right ) y^2 \rho }{\left (1-y \right)^2} \right. \nonumber \\[-2mm] \\[-2mm]
& \left . \phantom{xxx} - \frac{6 \left (2 - y \right ) y^2 \rho ^2}{\left (1 - y \right )^3} + \frac{10 \left( 6 -4 y+ y^2\right) y^2 \rho ^3 }{3 \left ( 1 - y \right )^4} \right ] \theta ( 1 - y - \rho ) \,, \nonumber \\[4mm]
f^{(0)}_D (y) & = \left [ -\frac{2 \left( 24 - 11 y- y^2 \right) y^2}{3 \left (1 - y \right )} + \frac{16 \left( 3 -3 y+ y^2 \right) y^2 \rho }{3 \left ( 1 - y \right )^3} \right . \nonumber \\[2mm]
& \left . \phantom{xxx} - \frac{2 \left( 24 -5 y -11 y^2 + 4 y^3 \right) y^2 \rho ^2 }{3 \left ( 1 - y \right )^5} + \frac{8 \left ( 6 - y \right ) y^2 \rho ^3 }{3 \left (1 - y \right )^6} \right ] \theta ( 1 - y - \rho ) \\[2mm]
& \phantom{xx} -\frac{2 \left (1 + \rho \right )^2 \left (1 - \rho \right )^4 }{3 \rho ^2} \hspace{0.5mm} \delta ( 1 - y - \rho ) \,, \nonumber \\[4mm]
f^{(0)}_{SL} (y) & = \left [ -\frac{2}{3}\hspace{0.5mm} y^3 +\frac{2 \left(12 -13 y + 4 y^2\right) y^2 \rho ^2 }{ \left (1-y \right)^4} \right . \nonumber \\[-2mm] \\[-2mm]
& \left. \phantom{xxx} -\frac{8 \left( 9 -10 y + 5 y^2 - y^3 \right) y^2 \rho ^3 }{3 \left (1 -y \right )^5} \right ] \theta ( 1 - y - \rho ) \,. \nonumber 
\end{align}
Notice that the function $f^{(0)}_D (y)$ has a singularity at the endpoint of the spectrum but the moments of the spectrum are well-defined.

By inserting the above results into~(\ref{eq:convolutionLL}) it is straightforward to derive the LL~QED corrections to the electron energy spectrum at ${\cal O} ( \Lambda_{\rm QCD}^2/m_b^2 )$ and ${\cal O} ( \Lambda_{\rm QCD}^3/m_b^3 )$. We obtain 
\begin{align} \label{eq:LLs}
f_{\pi, {\rm LL}}^{(1)}(y) & = \Bigg \{ \left [ -\frac{2}{9} \left(5 - y + 23 y^2 - 74 y^3 + 70 y^4 - 20 y^5\right) \right. \nonumber \\[2mm]
& \phantom{xxxxxi} \left. \hspace{-6mm} -\frac{20}{3} \left (1-y \right )^2 y^3 \ln \left(\frac{1 - y -\rho}{y}\right) \right ] \frac{1}{ \left (1-y \right )^2} \nonumber \\[2mm] 
& \phantom{xxx} \hspace{-6mm} + \frac{2}{3} \left(9 - 18 y + 12 y^2 - 28 y^3 + 29 y^4 - 10 y^5 \right) \frac{\rho}{\left (1-y \right)^3} \nonumber \\[2mm] 
& \phantom{xxx} \hspace{-6mm} - \left [ \frac{1}{3} \left(18 - 42 y - 30 y^2 - 38 y^3 + 4 y^4 + y^5 \right) \right. \nonumber \\[2mm]
& \phantom{xxxxxi} \hspace{-6mm} + \hspace{0.5mm} 4 \left ( 1 - 4 y + 6 y^2 + y^3 - y^4 \right ) \ln \left(\frac{1-y}{\rho }\right) \\[2mm] 
& \phantom{xxxxxi} \left. \hspace{-6mm} + \hspace{0.5mm} 4 \left (5 - 2 y \right ) y^3 \ln \left(\frac{1-y-\rho}{y}\right) \right ] \frac{\rho^2}{\left (1-y \right )^4} \nonumber \\[2mm]
& \phantom{xxx} \hspace{-6mm} + \left [ \frac{2}{9} \left ( 5 - 10 y - 58 y^2 - 47 y^3 + 16 y^4 - 2 y^5 \right ) \right. \nonumber \\[2mm]
& \phantom{xxxxxi} \hspace{-6mm} + \hspace{0.5mm} \frac{4}{3} \left( 1 - 5 y + 10 y^2 + 10 y^3 - 5 y^4 + y^5 \right) \ln \left(\frac{1-y}{\rho }\right) \nonumber \\[2mm]
& \phantom{xxxxxi} \left. \hspace{-6mm} + \hspace{0.5mm} \frac{8}{3} \left(10 -5 y + y^2 \right) y^3 \ln \left(\frac{1-y-\rho}{y}\right) \right ] \frac{\rho^3}{\left (1-y \right )^5}\Bigg \} \; \theta ( 1 - y - \rho ) \,, \nonumber \\[4mm]
f_{G, {\rm LL}}^{(1)}(y) & = \Bigg \{ \left [ -\frac{1}{9} \left(5 - 23 y + 24 y^2 + 100 y^3 - 40 y^4\right) \right. \nonumber \\[2mm]
& \phantom{xxxxxi} \left. \hspace{-6mm} + \hspace{0.5mm} \frac{4}{3}\left ( 6 + 5 y \right ) \left ( 1 - y \right ) y^2 \ln \left(\frac{1 - y -\rho}{y}\right) \right ] \frac{1}{ 1-y } \nonumber \\[2mm] 
& \phantom{xxx} \hspace{-6mm} + \left [ \frac{1}{3} \left( 9 - 39 y - 18 y^2 + 20 y^3 \right) \left ( 1- y \right ) - \left(4 +12 y^2 - 8 y^3\right) \ln \left(\frac{1-y}{\rho }\right) \right. \nonumber \\[2mm]
& \phantom{xxxxxi} \left. \hspace{-6mm} - \hspace{0.5mm} 8 \left ( 3 - 2 y \right ) y^2 \ln \left(\frac{1-y-\rho}{y}\right) \right ] \frac{\rho}{\left (1-y \right )^2} \nonumber \\[-2mm] \\[-2mm]
& \phantom{xxx} \hspace{-6mm} - \left [ 3 -29 y -4 y^2 +\frac{8 }{3} \hspace{0.5mm}y^3 -\frac{5 }{3} \hspace{0.5mm} y^4 -6 \left( 1 -3 y -y^2 + y^3 \right) \ln \left(\frac{1-y}{\rho }\right) \right. \nonumber \\[2mm]
& \phantom{xxxxxi} \left. \hspace{-6mm} + \hspace{0.5mm} 12 \left ( 2 - y \right ) y^2 \ln \left(\frac{1-y-\rho}{y}\right) \right ] \frac{\rho^2}{\left (1-y \right )^3 } \nonumber \\[2mm]
& \phantom{xxx} \hspace{-6mm} + \left [ \frac{5}{9} \left(1-28 y-15 y^2+2 y^3+y^4\right) - \frac{10}{3} \left( 1 - 4 y - 6 y^2 + 4 y^3 - y^4 \right) \ln \left(\frac{1-y}{\rho }\right) \right. \nonumber \\[2mm]
& \phantom{xxxxxi} \left. \hspace{-6mm} + \hspace{0.5mm} \frac{20}{3} \left(6 -4 y + y^2 \right) y^2 \ln \left(\frac{1-y-\rho}{y}\right) \right ] \frac{\rho^3}{\left (1-y \right )^4 } \Bigg \} \; \theta ( 1 - y - \rho ) \,, \nonumber \\[4mm]
f_{D, {\rm LL}}^{(1)}(y) & = \Bigg \{ \frac{2 \left ( 1 + y^2 \right )}{3 \left (1-y \right ) \rho^2} - \frac{2 \left ( 1 - 2 y - y^2 \right ) }{3 \left (1-y \right )^2 \rho} \nonumber \\[2mm]
& + \left [ \frac{2}{9} \left( 34 - 153 y + 255 y^2 - 158 y^3 + 12 y^4 + 18 y^5 - 4 y^6 \right) \right. \nonumber \\[2mm]
& \phantom{xxxxxi} \hspace{-6mm} - \hspace{0.5mm} 8 \left ( 1 + y^2 \right ) \left ( 1 - y \right )^2 \ln \left(\frac{1-y}{\rho }\right) \nonumber \\[2mm]
& \phantom{xxxxxi} \left. \hspace{-6mm} - \hspace{0.5mm} \frac{4}{3} \left (24 - 11 y - y^2 \right ) \left ( 1 - y \right )^2 y^2 \ln \left(\frac{1 - y -\rho}{y}\right) \right ] \frac{1}{ \left ( 1-y \right )^3 } \nonumber \\[2mm]
& \phantom{xxx} \hspace{-6mm} - \left [ \frac{2}{3} \left( 6 - 29 y + 110 y^2 - 112 y^3 + 28 y^4 + 3 y^5 - 2 y^6 \right) \right. \nonumber \\[2mm] 
& \phantom{xxxxxi} \hspace{-6mm} - \hspace{0.5mm} \frac{16}{3} \left( 1 + 3 y - 3 y^2 + y^3 \right) \left ( 1 - y \right ) y \ln \left(\frac{1-y}{\rho }\right) \nonumber \\[2mm]
& \phantom{xxxxxi} \left. \hspace{-6mm} - \hspace{0.5mm} \frac{32}{3} \left ( 3 - 3 y + y^2\right ) \left ( 1 - y \right ) y^2 \ln \left(\frac{1-y-\rho}{y}\right) \right ] \frac{\rho}{\left (1-y \right )^4 } \\[2mm] 
& \phantom{xxx} \hspace{-6mm} - \left [ \frac{1}{3} \left( 14 - 70 y - 152 y^2 + 86 y^3 - 32 y^4 + 11 y^5 - y^6 \right) \right. \nonumber \\[2mm]
& \phantom{xxxxxi} \hspace{-6mm}+ \hspace{0.5mm} \frac{4}{3} \left( 2 - 10 y + 44 y^2 - 25 y^3 - y^4 + 2 y^5 \right) \ln \left(\frac{1-y}{\rho }\right) \nonumber \\[2mm]
& \phantom{xxxxxi} \left. \hspace{-6mm} + \hspace{0.5mm} \frac{4}{3} \left( 24 - 5 y - 11 y^2 + 4 y^3 \right) y^2 \ln \left(\frac{1-y-\rho}{y}\right) \right ] \frac{\rho^2}{\left (1-y \right )^5 } \nonumber \\[2mm]
& \phantom{xxx} \hspace{-6mm} + \left [ \frac{2}{9} \left( 5 - 45 y - 159 y^2 + 5 y^3 + 15 y^4 - 6 y^5 + y^6 \right) + \frac{16}{3} \left ( 6 -y \right ) y^2 \ln \left(\frac{1-y}{\rho }\right) \right. \nonumber \\[2mm]
& \phantom{xxxxxi} \left. \hspace{-6mm} + \hspace{0.5mm} 
\frac{16}{3} \left ( 6 -y \right ) y^2 \ln \left(\frac{1-y-\rho}{y}\right) \right ] \frac{\rho^3}{\left (1-y \right )^6 } \Bigg \} \; \theta ( 1 - y - \rho ) \nonumber \\[2mm]
& \phantom{xxx} \hspace{-6mm} -\frac{2 \left (1 + \rho \right )^2 \left (1 - \rho \right )^4 }{3 \rho ^2} \int_y^{1-\rho} \frac{dx}{x} \, P_{ee}^{(0)} \left ( \frac{y}{x} \right ) \delta ( 1 - x - \rho ) \,, \nonumber \\[4mm]
f_{LS, {\rm LL}}^{(1)}(y) & = \Bigg \{ \left [ \frac{2}{9} \left( 8 -7 y+8 y^2 +4 y^3 -14 y^4 + 4 y^5 \right) \right. \nonumber \\[2mm]
& \phantom{xxxxxi} \left. \hspace{-6mm} - \hspace{0.5mm} \frac{4}{3} \left (1-y \right )^2 y^3 \ln \left(\frac{1 - y -\rho}{y}\right) \right ] \frac{1}{ \left ( 1-y \right )^2 } \nonumber \\[2mm]
& \phantom{xxx} \hspace{-6mm} + \frac{2}{3} \left( 9 + 15 y - 29 y^2 + 13 y^3 - 2 y^4 \right) y \, \frac{\rho}{\left (1-y \right)^3} \nonumber \\[2mm]
& \phantom{xxx} \hspace{-6mm} - \left [ \frac{1}{3} \left( 78 +48 y -58 y^2 +26 y^3 -7 y^4 \right) y + \left( 8 -32 y +20 y^3 -8 y^4 \right) \ln \left(\frac{1-y}{\rho }\right) \right. \nonumber \hspace{10mm} \\[-2mm] \\[-2mm]
& \phantom{xxxxxi} \left. \hspace{-6mm} - \hspace{0.5mm} 4 \left( 12 -13 y + 4 y^2 \right) y^2 \ln \left(\frac{1-y-\rho}{y}\right) \right ] \frac{\rho^2}{\left (1-y \right )^4 } \nonumber \\[2mm]
& \phantom{xxx} \hspace{-6mm} - \left [ \frac{2}{9} \left( 8 - 97 y - 19 y^2 + 4 y^3 + 13 y^4 - 5 y^5 \right) \right. \nonumber \\[2mm]
& \phantom{xxxxxi} \hspace{-6mm} - \hspace{0.5mm} \frac{8}{3} \left( 1 -5 y -8 y^2 + 10 y^3 -5 y^4 + y^5\right) \ln \left(\frac{1-y}{\rho }\right) \nonumber \\[2mm]
& \phantom{xxxxxi} \left. \hspace{-6mm} + \hspace{0.5mm} \frac{16}{3} \left( 9 - 10 y + 5 y^2 - y^3 \right) y^2 \ln \left(\frac{1-y-\rho}{y}\right) \right ] \frac{\rho^3}{\left (1-y \right )^5 } \Bigg \} \; \theta ( 1 - y - \rho ) \,. \nonumber
\end{align}

The result for $f_{D, {\rm LL}}^{(1)}(y)$ still contains the convolution with a delta function
\beq \label{eq:deltaD}
\delta_{D} (y) = \int_y^{1-\rho} \frac{dx}{x} \, P_{ee}^{(0)} \left ( \frac{y}{x} \right ) \delta ( 1 - x - \rho ) = \frac{P_{ee}^{(0)}\Big(\frac{y}{1-\rho}\Big)}{1-\rho} \, ,
\eeq
where $P_{ee}^{(0)}(z)$ is given in~(\ref{eq:Pee0}). The corrected spectrum $f_{D, {\rm LL}}^{(1)}(y)$ has still a singularity regulated by the plus distribution. The moments of the spectrum however are regular and can be expressed as
\beq \label{eq:twoforone}
\begin{split} 
\int_{y_{\rm cut}}^{1-\rho} dy \, y^i \, \delta_{D} (y) & = 
\Delta_{D}^i + \int_{0}^{y_{\rm cut}} dy \, y^i \left[ \frac{y^2 + \left (1 - \rho \right )^2}{ \left (y-1+\rho \right ) \left (1- \rho \right )^2} \right ] \,, 
\end{split}
\eeq
where we have written the integral over $y$ as the difference of two integrals, one ranging over $y \in [0, 1- \rho]$ and a second one ranging over $y \in [0, y_{\rm cut}]$. While the latter does not include the endpoint of the spectrum and can be trivially calculated, the first integral can be solved using the definition of the plus distribution~(\ref{eq:Pee0}) leading to
\beq \label{eq:DeltaDi}
\Delta_{D}^0 = 0 \,, \quad 
\Delta_{D}^1 = -\frac{4}{3} \left ( 1- \rho \right ) \,, \quad 
\Delta_{D}^2 = -\frac{25}{12} \left ( 1- \rho \right )^2 \,, \quad 
\Delta_{D}^3 = -\frac{157}{60} \left ( 1- \rho \right )^3 \,. \quad 
\eeq
Notice that $\Delta_{D}^0 = 0$ is a necessary condition to satisfy the KLN theorem. 

\section{NLL effects in the partonic case} 
\label{app:NLLpartonic}

The NLL~QED corrections to the electron energy spectrum in the inclusive $B \to X_c e \nu$ decay can be obtained by the following convolution
\beq \label{eq:convolutionNLO}
\left ( \frac{d \Gamma}{dy} \right )^{(2)}_{\rm NLL} = \left ( \frac{\alpha}{2\pi} \right )^2 \, \frac{\bar L_{b/e}^2}{2} \hspace{0.5mm} \left \{ \, \int_y^{1 - \rho} \, \frac{dx}{x} \hspace{0.25mm} \left [ P_{ee}^{(1)} \left (\frac{y}{x} \right ) + \frac{2}{3} \hspace{0.5mm} P_{ee}^{(0)} \left (\frac{y}{x} \right ) \right] \, \left ( \frac{d \Gamma}{dx} \right )^{(0)} \, \right \} \,.
\eeq
The expression for the LO~electron-electron splitting function $P_{ee}^{(0)}(z)$ has already been given in~(\ref{eq:Pee0}) and the next-to-leading-order~(NLO) equivalent takes the following form 
\bea \label{eq:Pee1}
\begin{split}
P_{ee}^{(1)} (z) & = \lim_{\Delta \, \to \, 0} \Bigg \{ \, 2 \left [ \frac{1 + z^2}{1 - z} \left ( 2 \ln (1- z) - \ln z + \frac{3}{2} \right ) + \frac{1+z}{2} \hspace{0.5mm} \ln z + z -1 \right ] \, \theta ( 1- z -\Delta ) \\[2mm] 
& \hspace{1.5cm} \, + \left [ \left ( \frac{3}{2} + 2 \ln \Delta \right )^2 - \frac{2}{3} \hspace{0.25mm} \pi^2 \right ] \delta (1 - z ) \, \Bigg \} \,.
\end{split}
\eea
Notice that since we are interested in the total NLL~QED effects including both real and virtual $e^+ e^-$ pairs that result from a photon splitting, we have included in~(\ref{eq:convolutionNLO}) only the non-singlet part of the pair corrections. As explained for instance in the articles~\cite{Arbuzov:2002pp,Arbuzov:2002cn,Arbuzov:2002rp}, in this way double counting of real $e^+ e^-$ pairs is avoided.

Extending~(\ref{eq:fpartonicLL}) to the NLL we then write 
\beq \label{eq:fpartonicNLL}
\frac{d \Gamma}{dy} = \Gamma^{(0)} \hspace{0.25mm} f (y) \,, \qquad f(y) = f^{(0)} (y) + \frac{\alpha}{2\pi} \, \bar L_{ b/e} \, f^{(1)}_{\rm LL} (y) + \left ( \frac{\alpha}{2\pi} \right )^2 \frac{\bar L_{ b/e}^2}{2} \, f^{(2)}_{\rm NLL} (y) \,,
\eeq
where the expression for $f^{(0)} (y)$ and $f^{(1)}_{\rm LL} (y)$ can be found in~(\ref{eq:f0partonic}) and~(\ref{eq:f1partonic}), respectively, and we decompose the function $f^{(2)}_{\rm NLL} (y)$ as follows 
\begin{align} \label{eq:NLLs}
f^{(2)}_{\rm NLL} (y) = \Phi(y) \, \theta ( 1 - y - \rho ) + \frac{2}{3} \hspace{0.25mm} f^{(1)}_{\rm LL}(y) \,.
\end{align} 
Here $\Phi(y)$ is a rather complicated analytic function involving various logarithms and dilogarithms that depend on both $y$ and $\rho$. Instead of giving the explicit form of $\Phi(y)$ we provide an approximate result. Using $\rho = 0.057$, we~obtain
\bea \label{eq:Phi}
\begin{split}
\Phi(y) & = \frac{320605}{2} + 2335446 \hspace{0.5mm} y - \frac{31925665}{4} \hspace{0.5mm} y^2 + 8569202 \hspace{0.5mm} y^3 - \frac{13352367}{4} \hspace{0.5mm} y^4 + \frac{427559}{2} \hspace{0.5mm} y^5 \\[2mm] 
& \phantom{xx} + \frac{83233}{2} \hspace{0.5mm} y^6 - \frac{5}{4} \ln y + \left ( y_{\rm max} - y \right ) \Bigg [ \frac{6927}{2} + \frac{776600}{3} \hspace{0.5mm} \left ( y_{\rm max} - y \right ) \\[2mm] 
& \hspace{0.8cm} + \frac{3229267}{2} \hspace{0.5mm} \left ( y_{\rm max} - y \right )^2 + \frac{4356808}{3} \hspace{0.5mm} \left ( y_{\rm max} - y \right )^3 \Bigg ] \ln ( y_{\rm max} - y ) \,,
\end{split}
\eea
with $y_{\rm max}= 1 - \rho = 0.943$. This approximation works to a relative accuracy of better than~$1\%$ for the considered parameters. 

\section{Virtual integrals} 
\label{app:loopintegrals}

In this appendix, we collect the analytic expressions for the integrals that appear in the virtual corrections~(\ref {eq:VIR}) and~(\ref {eq:Vfin}). Many of the integrals given below can be evaluated following the discussion in~\cite{Aquila:2005hq} but some integrals related to photon exchange between the final-state particles need a dedicated treatment. 

After introducing the Feynman parameter $x$, the diagrams describing the external leg corrections depend on 
\beq \label{eq:Mi2}
M_i^2 = x^2 \hspace{0.25mm} m_i^2 + m_\gamma^2 \left (1 - x \right )\,,
\eeq
where $m_i$ denotes the mass of the corresponding internal fermion. The loop integrals relevant for the self-energy corrections are 
\begin{align} 
J_i &= 2\int_0^1 \! dx \, x \left (1-x^2 \right )\frac{m_i^2}{M_i^2} = -\ln \left ( \frac{m_\gamma^2}{m^2_i} \right ) -1\,, \label{eq:Ji} \\[2mm] 
I^0_i &= \int_0^1 \! dx \left ( 1 - x \right ) \ln\left(\frac{M_i^2}{m_b^2}\right)= \frac{1}{2} \left [\ln \left(\frac{m_i^2}{m_b^2}\right) -3 \right ] \,, \label{eq:I0i} 
\end{align}
where the final results correspond to the lowest order in an expansion for small photon~mass.

The structure of the loop integrals that arise from the Feynman diagrams where the photon connects two different fermions are more complicated. After introducing the Feynman parameters $x$ and $y$, these integrals depend on 
\beq \label{eq:Mij2}
M_{ij}^2 = \left (x \hspace{0.25mm} p_i \pm y \hspace{0.25mm} p_j \right )^2 + m_\gamma^2 \left (1-x-y \right ) \,,
\eeq
where the index pair can take the following values $ij =\{bc, be, ce\}$ while $p_i$ and $p_j$ denote the associated four-momenta. The $\pm$ sign in the definition~(\ref{eq:Mij2}) encapsulates the different kinematic configurations of the relevant graphs. In the case of $ij =\{bc, be\}$ where the photon is exchanged between the initial and final state one has to take the $+$ sign, while for $ij = \{ce \}$ corresponding to pure final-state interactions one has to choose the $-$ sign. To write the results for the vertex integrals in a compact form, we introduce the following set of variables
\beq \label{eq:ijsdefs} 
\omega_{ij} = -2 \hspace{0.25mm} p_i \cdot p_j \,, \quad \sigma_{ij} = \sqrt{\omega_{ij}^2 - 4 \hspace{0.25mm} m_i^2 \hspace{0.125mm} m_j^2} \,, \quad z_{ij}^\pm = -\frac{\omega_{ij} \pm \sigma_{ij}}{2 m_i^2} \,, \quad \Sigma_{ij} = \omega_{ij}+m_i^2+m_j^2 \,.
\eeq

In terms of the abbreviations introduced in~(\ref{eq:ijsdefs}) we obtain 
\begin{align}
I^0_{ij} &= \int_0^1\! dx \int_0^{1-x} \! dy \, \ln \left(\frac{M_{ij}^4}{m_b^4} \right) \nonumber \\[-2mm] \label{eq:I0ij} \\[-2mm] 
 & = I^0_i + I^0_j +\frac{1}{2 \hspace{0.25mm} \Sigma_{ij} } \left [ \sigma_{ij} \ln \left ( \frac{z_{ij}^+}{z_{ij}^-} \right ) + \left (m^2_i - m^2_j \right ) \ln \left ( \frac{m_i^2}{m_j^2} \right ) \right ] \,, \nonumber \\[4mm]
 I^x_{ij} &= \int_0^1\! dx \int_0^{1-x} \! dy \, \frac{x}{M_{ij}^2} = -\frac{1}{2 \hspace{0.25mm} \Sigma_{ij} } \left [ \frac{\omega_{ij} + 2 \hspace{0.125mm} m_j^2}{\sigma_{ij}} \hspace{0.25mm} \ln \left ( \frac{z_{ij}^+}{z_{ij}^-} \right ) - \ln \left ( \frac{m_i^2}{m_j^2} \right ) \right ] \,, \label{eq:Ixij} \\[4mm]
 I^y_{ij} &= \int_0^1\! dx \int_0^{1-x} \! dy \, \frac{y}{M_{ij}^2} = -\frac{1}{2 \hspace{0.25mm} \Sigma_{ij} } \left [ \frac{\omega_{ij} + 2 \hspace{0.125mm} m_i^2}{\sigma_{ij}} \hspace{0.25mm} \ln \left ( \frac{z_{ij}^+}{z_{ij}^-} \right ) + \ln \left ( \frac{m_i^2}{m_j^2} \right ) \right ] \,, \label{eq:Iyij} \\[4mm] 
I^{xy}_{ij} &= \int_0^1\! dx \int_0^{1-x} \! dy \, \frac{x \hspace{0.125mm} y}{M_{ij}^2} \nonumber \\[-2mm] \label{eq:Ixyij} \\[-2mm] 
& = \frac{\big ( m_i^2 - m_j^2 \big )^2 - \big ( m_i^2 +m_j^2 \big ) \hspace{0.5mm} \Sigma_{ij}}{4 \hspace{0.125mm} \sigma_{ij} \hspace{0.25mm} \Sigma_{ij}^2} \, \ln \left ( \frac{z_{ij}^+}{z_{ij}^-} \right ) + \frac{m_i^2 - m_j^2}{4 \hspace{0.125mm}\Sigma_{ij}^2}\ln \left ( \frac{m_i^2}{m_j^2} \right ) -\frac{1}{2 \hspace{0.125mm}\Sigma_{ij}}\,. \nonumber
\end{align}
These results hold in the cases $ij = \{bc,be\}$ while for $ij = \{ce\}$ one has to use the replacement $\omega_{ij} \to -\omega_{ij}$ in the above expressions as well as in $z_{ij}^\pm$. Notice that we have again taken the limit $m_\gamma \to 0$ to obtain the final expressions.

The expression for the integral $I_{ij}^1$ is the most complicated one. It also depends in a non-trivial way on whether one considers $ij = \{bc,be\}$ or $ij = \{ce\}$. In the case 
 $ij = \{bc,be\}$ we obtain 
\beq\label{eq:I1ijIF}
I^1_{ij} = \omega_{ij} \int_0^1\! dx \int_0^{1-x} \! dy \, \frac{1}{M_{ij}^2} = \frac{\omega_{ij}}{2 \hspace{0.25mm} \sigma_{ij}} \left [ \ln \left( \frac{m_\gamma^2}{m_i \hspace{0.25mm} m_j} \right ) \ln \left ( \frac{z_{ij}^+}{z_{ij}^-} \right ) + 2 \hspace{0.25mm} \Delta I_{ij}^1 \right ] \,,
\eeq
with 
\beq \label{eq;DeltaIij}
\Delta I_{ij}^1 = \ln \big (z_{ij}^- \big ) \ln \left ( \frac{\sqrt{z_{ij}^-} \hspace{0.75mm} \bar z_{ij}^+}{ \sqrt{z_{ij}^+} \hspace{0.75mm} \bar z_{ij}^-} \right ) + {\rm Li}_2 \left ( 1- \frac{z_{ij}^+ \hspace{0.75mm} \bar z_{ij}^- }{ z_{ij}^- \hspace{0.75mm} \bar z_{ij}^+} \right ) - {\rm Li}_2 \left ( 1- \frac{ \bar z_{ij}^- }{ \bar z_{ij}^+ } \right ) \,.
\eeq
Here ${\rm Li}_2 (z)$ is the usual dilogarithm and we have introduced the abbreviations $\bar z_{ij}^\pm = 1- z_{ij}^\pm$. The result~(\ref{eq:I1ijIF}) again applies in the limit $m_\gamma \to 0$. 

For $ij = \{ce\}$ the integral $I_{ij}^1$ cannot be obtained by a suitable substitution from~(\ref{eq:I1ijIF}), because it develops an imaginary part in the physical region and therefore has to be computed separately. Using the techniques described in Appendix~B of~\cite{Aquila:2005hq} it is however relatively straightforward to compute the integral $I_{ij}^1$ for the case of virtual photon exchange between the charged final-state particles. We find 
\beq \label{eq:Ice1}
I_{ce}^1 = \frac{\omega_{ce}}{2 \hspace{0.125mm} \sigma_{ce}} \hspace{0.75mm} \Bigg \{ \left [ \ln \left ( \frac{y_+ \hspace{0.5mm} \bar y_- }{y_- \hspace{0.5mm} \bar y_+} \right ) - 2 \hspace{0.125mm} i \hspace{0.125mm} \pi \right ] \ln \left ( \frac{m_\gamma^2} {q^2}\right ) + \Delta I_{ce}^1 \Bigg \} \,, 
\eeq
where as before $q^2 = \left ( p_b - p_\nu \right )^2 = \left ( p_c + p_e \right )^2$ and 
\beq \label{eq:ypm}
y_{\pm} = \frac{q^2 + m_c^2 - m_e^2 \pm \sigma_{ce}}{2 \hspace{0.25mm} q^2} \,, \qquad \bar y_\pm = 1 - y_\pm \,.
\eeq
The real and imaginary parts of $\Delta I_{ce}^1$ are given by
\bea \label{eq:ReIm}
\begin{split}
{\rm Re} \left ( \Delta I_{ce}^1\right ) & = -2 \hspace{0.25mm} {\rm Li}_2 \left ( -\frac{y_-}{ y_+ - y_-} \right ) - 2 \hspace{0.25mm} {\rm Li}_2 \left ( -\frac{\bar y_+ }{ y_+ - y_-} \right ) + 2 \ln \left (y_+ - y_- \right ) \ln \left ( \frac{y_- \hspace{0.5mm} \bar y_+ }{y_+ - y_-} \right ) \\[2mm]
& \phantom{xx} + \frac{1}{2} \ln^2 \left ( \frac{\bar y_-}{\bar y_+} \right) + \frac{1}{2} \ln^2 \left ( \frac{y_-}{y_+} \right) - \ln^2 \left ( \bar y_- \right) - \ln^2 \left ( y_+ \right) + \frac{4 \hspace{0.125mm} \pi^2}{3} \,, \\[4mm] 
{\rm Im} \left ( \Delta I_{ce}^1\right ) & = 4 \hspace{0.125mm} i \hspace{0.125mm} \pi \ln \left ( y_+ - y_- \right ) \,.
\end{split}
\eea
Notice that the dilogarithms in~(\ref{eq:ReIm}) have all been chosen to be real by using the identity ${\rm Li}_2 \left (z \right ) = -{\rm Li}_2 \left (1-z \right ) - \ln \left ( z \right ) \ln \left (1-z \right) + \pi^2/6$ and that only the real part of $I_{ce}^1$ enters our calculation. Observe that the threshold corrections~(\ref{eq:discDeltace}) discussed in Section~\ref{sec:Pi2effects} can be recovered by evaluating $\lim_{\sigma_{ce} \to 0} {\rm Re} \left ( \Delta I_{ce}^1\right ) = 2 \hspace{0.25mm} \pi^2$ with $\sigma_{ce}$ defined in~(\ref{eq:ijsdefs}).

We finally remark that only the virtual integrals~(\ref {eq:Ji}),~(\ref {eq:I1ijIF}) and~(\ref{eq:Ice1}) develop logarithmic singularities for $m_\gamma \to 0$. These divergences are precisely the soft singularities which are needed to cancel the analogous divergent terms in the soft integrals to be discussed~next. 

\section{Soft integrals} 
\label{app:softintegrals}

The soft integrals that enter~(\ref{eq:Rsoftint}) are given by the following expressions
\begin{align}
I_R(p_i) & = -\ln \left ( \frac{m_\gamma^2}{4 \hspace{0.125mm} E_\mathrm{min}^2} \right)+\frac{p_i^0}{|\vec{p}_i|} \, \ln\left(\frac{p_i^0-|\vec{p}_i|}{p_i^0+|\vec{p}_i|}\right)\,, \label{eq:IR1} \\[2mm]
I_R(p_i,p_j) & = -\frac{2 p_i\cdot p_j}{\sigma_{ij}} \left [ \ln \left ( \frac{m_\gamma^2}{4 E_\mathrm{min}^2} \right) \ln \left(\frac{z^+_{ij}}{z^-_{ij}}\right) + 2 \hspace{0.125mm} \Delta I_R(p_i,p_j) \right]\,, \label{eq:IR2}
\end{align}
where
\bea \label{eq:DeltaIR}
\begin{split}
\Delta I_R (p_i,p_j) & = {\rm Li}_2 \left ( 1+ z_{ij}^- \, \frac{p_i^0+|\vec{p}_i|}{v_{ij}} \right) + {\rm Li}_2 \left ( 1+ z_{ij}^- \, \frac{p_i^0-|\vec{p}_i|}{v_{ij}} \right) - {\rm Li}_2 \left ( 1+ \frac{p_j^0+|\vec{p}_j|}{v_{ij}} \right ) \hspace{6mm} \\[2mm] 
& \phantom{xx} - {\rm Li}_2 \left ( 1+ \frac{p_j^0-|\vec{p}_j|}{v_{ij}} \right ) + \frac{1}{4} \left [ \ln^2 \Bigg( \frac{p_i^0-|\vec{p}_i|}{p_i^0+|\vec{p}_i|} \Bigg ) - \ln^2 \Bigg ( \frac{p_j^0-|\vec{p}_j|}{p_j^0+|\vec{p}_j|} \Bigg ) \right ] \,,
\end{split}
\eea
with
\beq \label{eq:vij}
v_{ij} = -\frac{1}{2} \, \frac{( z_{ij}^-)^2 \hspace{0.25mm} m_i^2 - m_j^2}{z_{ij}^- \hspace{0.5mm}p_i^0 - \hspace{0.25mm}p_j^0} \,.
\eeq
Here $p_i^0$ and $|\vec{p}_i |$ denote the energy component and the magnitude of the three-momentum of~$p_i$, respectively. Notice that both~(\ref{eq:IR1}) and~(\ref{eq:IR2}) contain soft singularities of the form $\ln \left (m_\gamma^2/(4 E_{\rm min}^2 ) \right)$ that depend on the phase-space slicing parameter $E_{\rm min}$. For sufficiently small $E_{\rm min}$ the sum over virtual and real contributions however becomes independent of this resolution parameter.

\section{EW corrections to the Wilson coefficient}
\label{app:EWcorrections}

The EW corrections to low-energy charged-current~(CC) processes have been studied long ago both for the semi-leptonic~\cite{Sirlin:1981ie} and purely leptonic case~\cite{Sirlin:1980nh}. However, in these early works only the scheme-independent logarithmic terms have been computed while scheme-dependent finite one-loop corrections have not been considered. These corrections have subsequently been calculated in the leptonic, semi-leptonic and non-leptonic case in the articles~\cite{Brod:2008ss,Gorbahn:2022rgl,Gambino:2000fz,Gambino:2001au,Hill:2019xq}. Below we present a comprehensive discussion of the calculation of~EW corrections to all three types of CC processes. This is accomplished by splitting the corrections into universal and non-universal parts and providing the explicit electric charge dependence for the latter. 

As we are interested in computing the $O(\alpha)$ contributions to the Wilson coefficient~$C(\mu)$ entering~(\ref {eq:Hweak}), we can neglect all external momenta as well as all quark and lepton masses because they are all much smaller than the weak scale. In this approximation the Born-level matrix elements of a generic~CC process is given by the matrix element of the local operator
\beq \label{eq:MBornCC}
{\cal M}_{\rm Born}^{\rm CC} = \frac{4 \hspace{0.125mm} G_F}{\sqrt{2}} \hspace{0.5mm} \langle Q_{\rm CC} \rangle \,, \qquad Q_{\rm CC} = \big ( \bar \psi_u\hspace{0.125mm} \gamma_{\mu} P_L \hspace{0.125mm} \psi_d \big ) \big( \bar \psi_{d^\prime} \hspace{0.125mm} \gamma^{\mu} P_L \hspace{0.125mm} \psi_{u^\prime} \big ) \,.
\eeq
Here we have removed possible CKM factors and $\psi_{u^{(\prime)}}$ $\big($$\psi_{d^{(\prime)}}$$\big)$ denotes the component of a fermionic $SU(2)_L$ doublet field with weak isospin $T_3 = 1/2$ ($T_3 = -1/2$). Notice that we have expressed the above Born-level result in terms of the Fermi constant $G_F$ as extracted from muon decay. In this way, as we will see in a moment, most of the EW radiative corrections cancel, including all terms that depend on the top-quark and the Higgs-boson mass. In the limit of vanishing external momenta, all ${\cal O} (G_F \hspace{0.25mm} \alpha) $ corrections are proportional to $\langle Q_{\rm CC} \rangle$ and can thus be expressed as ${\cal O} (\alpha)$ contribution to its Wilson coefficient $C_{\rm CC} (\mu)$. In other words, there is no mixing with other operators.

The virtual EW corrections to a given low-energy CC process can schematically be written as 
\beq \label{eq:MvirtualCC}
{\cal M}_{\rm virtual}^{\rm CC} = {\cal M}_{\rm Born}^{\rm CC} \, \left [ \delta_{\rm box}+\delta_{\rm vertex} + \frac{{\rm Re} \hspace{0.125mm} \big ( A_{WW}(M_W^2) \big) - A_{WW}(0)}{M_W^2} - \frac{2 \hspace{0.125mm} \delta g}{g} \right] \,, 
\eeq
where the contributions $\delta_{\rm box}$ and $\delta_{\rm vertex}$ encode the relevant box and vertex corrections, $A_{WW}(q^2)$ is the unrenormalised two-point function of the $W$ boson, and ${\delta g}$ is the counterterm of the $SU(2)_L$ coupling constant $g$, whose precise definition is irrelevant for our purposes. The $W$-boson mass is renormalised on-shell, as indicated by its counterterm ${\rm Re} \hspace{0.125mm} \big ( A_{WW}(M_W^2) \big)$. The sum of these four contributions is UV finite. We use the Feynman~gauge.

\begin{figure}[t!]
\begin{center}
\includegraphics[width=0.75\textwidth]{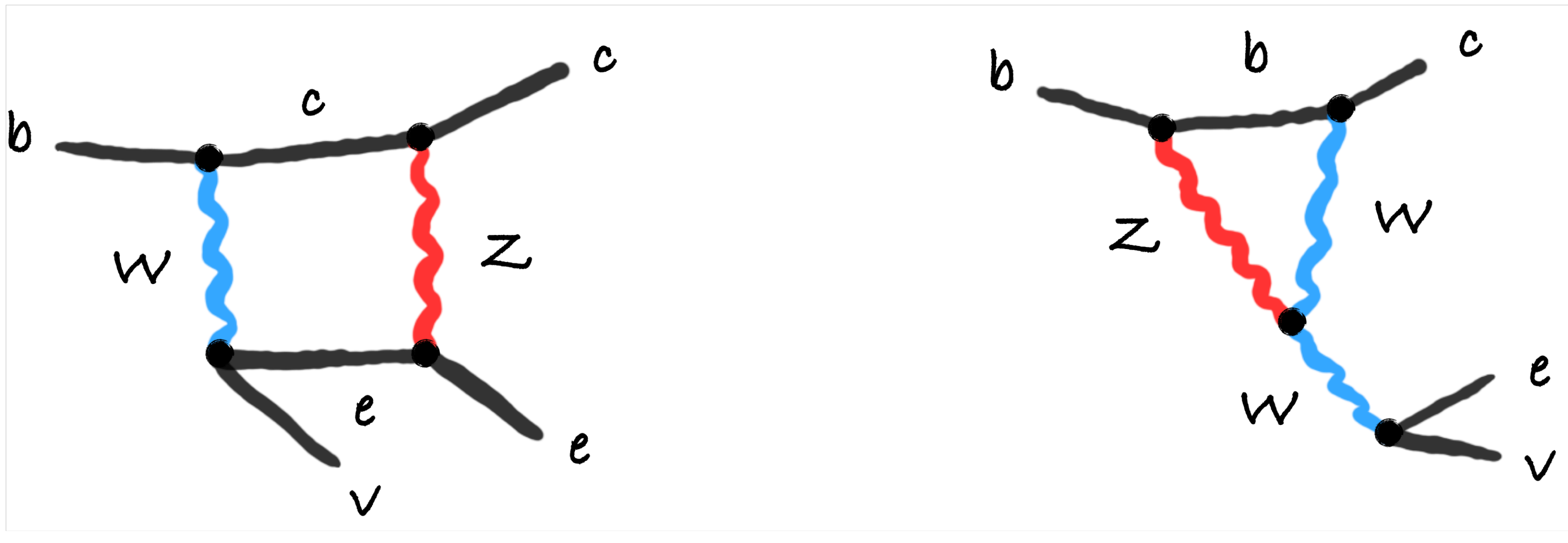}
\end{center}
\vspace{-6mm} 
\caption{\label{fig:boxtriangle} Examples of Feynman diagrams that contribute to the short-distance EW corrections encoded by the Wilson coefficient $C(\mu)$. Only contributions involving $Z$-boson exchange are shown but there are also photon loops.}
\end{figure}

In the case of the $b\to c e \nu$ transition the term $\delta_{\rm box}$ in~(\ref{eq:MvirtualCC}) receives contributions from Feynman diagrams like the one shown on the left-hand side in~Figure~\ref{fig:boxtriangle}. Using dimensional regularisation for both UV and IR singularities and keeping the electric charges $Q_{i}$ of the external fermions explicit, we obtain 
\beq \label{eq:deltabox} 
\delta_{\rm box} = \frac{g^2}{16\pi^2} \; \Bigg \{ \left [ -\frac{1}{\epsilon} + \ln \left ( \frac{M_Z^2}{\mu^2} \right ) \right ] \chi \hspace{0.25mm} s_w^2 + \zeta \hspace{0.25mm} s_w^2 - \left [ 5 - \frac{5}{2 \hspace{0.125mm} s_w^2} \right ] \ln \left ( \frac{M_Z^2}{M_W^2} \right ) \Bigg \} \,.
\eeq
Here $s_w$ denotes the sine of the weak mixing angle. The~$1/\epsilon$ pole is a IR divergence originating from the photon boxes, while the $Z$-boson boxes are finite. The process-dependent factors $\chi$ and $\zeta$ are given by 
\beq \label{eq:zetachi}
\begin{split}
\chi & = Q_{u} \big ( Q_{u^\prime} - 4 \hspace{0.125mm} Q_{d^\prime} \big ) + Q_{d} \big ( Q_{d^\prime} - 4 \hspace{0.125mm} Q_{u^\prime} \big ) = 
\begin{cases} \, 1 \,, & \text{leptonic} \,, \\
 \, 3 \,, & \text{semi-leptonic} \,, \\
 \, \frac{7}{3} \,, & \text{non-leptonic} \,, \\
 \end{cases} \\[4mm]
\zeta & = Q_{u} \left ( \frac{Q_{u^\prime}}{2} +5 \hspace{0.125mm} Q_{d^\prime} \right ) + Q_{d} \left( \frac{Q_{d^\prime}}{2} + 5 \hspace{0.125mm} Q_{u^\prime} \right ) = 
\begin{cases} \, \frac{1}{2} \,, & \text{leptonic} \,, \\[1mm]
 \, -\frac{19}{6} \,, & \text{semi-leptonic} \,, \\[1mm]
 \, -\frac{35}{18} \,, & \text{non-leptonic} \,. \\
 \end{cases} 
\end{split}
\eeq
To reduce products of strings of three Dirac matrices resulting from one-loop boxes involving the exchange of a $W$ boson and a photon, we have employed the relation 
\beq \label{eq:3to1}
\big ( \bar \psi_u \hspace{0.125mm} \gamma_{\mu} \gamma_{\nu} \gamma_{\lambda} P_L \hspace{0.125mm} \psi_d \big ) \big( \bar \psi_{d^\prime} \hspace{0.125mm} \gamma^{\mu} \gamma^{\nu} \gamma^{\lambda} P_L \hspace{0.125mm} \psi_{u^\prime} \big ) = \left (16 - 4 \hspace{0.125mm} \epsilon \right ) Q_{\rm CC} + E_{\rm CC} \,, 
\eeq
where $Q_{\rm CC}$ is the operator introduced in~(\ref{eq:MBornCC}), $E_{\rm CC}$ is a so-called evanescent operator and the particular choice of ${\cal O} (\epsilon)$ term in~(\ref{eq:3to1}) corresponds to the NDR scheme definition~\cite{Buras:1989xd}. This scheme choice only affects~$\zeta$, while~$\chi$ is unaffected and hence scheme independent. The~scheme dependence of~$\zeta$ however cancels in the final physical results, provided that all parts are calculated consistently using~(\ref{eq:3to1}). Notice that in our calculation of the matrix element of the operator $Q$ as detailed in Section~\ref{sec:nutshell} we also use the NDR scheme and as a result the final result~(\ref{eq:Eespectrumfinal}) for the electron energy spectrum is scheme independent. 

The term $\delta_{\rm vertex}$ in~(\ref{eq:MvirtualCC}) receives contributions from vertex diagrams and external leg corrections associated to wave function renormalisation~(WFR). An example of a possible vertex graph relevant in the case of $b \to c e \nu$ is depicted on the right in~Figure~\ref{fig:boxtriangle}. 
In~agreement with~\cite{Sirlin:1977sv}, we find that $\delta_{\rm vertex}$ does not depend on the fermion doublet under consideration and therefore represents a universal process-independent contribution. In the case of a quark doublet the renormalisation turns out to be more subtle. In the limit of vanishing external quark masses, one can limit oneself to the renormalisation of the left-handed quark fields. For down quarks, the WFR matrix can be decomposed as
\beq \label{eq:ZdL}
(Z_d^L)_{ij}= \delta_{ij}+ (\delta Z_d^L)_{ij} \,, \qquad (\delta Z_d^L)_{ij}=A_d\hspace{0.5mm} \delta_{ij}+ V_{ti} V_{tj}^\ast \hspace{0.5mm} B_d \,, 
\eeq
where $A_d$ contains the light-quark contribution, while $B_d$ includes the difference between the top- and the light-quark contribution that results from the Glashow-Iliopoulos-Maiani~(GIM) mechanism. The explicit form of these coefficients can be found in~(39) and~(40) of~\cite{Gambino:1998rt}. In~the case of a quark vertex like in~$b \to c e \nu$, the counterterm contribution corresponding to the renormalisation of the external bottom-quark line is proportional to the Born-level matrix element~(\ref{eq:MBornCC}) and takes the following form 
\beq \label{eq:deltactbc}
\delta_{b, \rm ct} = \sum_{q=d,s,b} \, \frac{(\delta Z_d^L)_{bq}}{2} \, V_{cq} = \sum_{q=d,s,b} \, \frac{1}{2} \, \Big [ A_d \hspace{0.5mm} \delta_{bq}+ V_{tb} V_{tq}^\ast \hspace{0.5mm} B_d \Big ] \, V_{cq} = \frac{A_d}{2} \, V_{cb} \,.
\eeq
Notice that in order to arrive at the final result we have used the unitarity of the CKM matrix,~i.e.~$\sum_{q=d,s,b} V_{iq} V_{jq}^\ast = \delta_{ij}$. Due to the GIM mechanism the counterterm~(\ref{eq:deltactbc}) hence does not receive a contribution from top-quark loops which is a necessary condition to obtain a universal vertex correction $\delta_{\rm vertex}$. For what concerns the WFR of the external charm-quark line, it proceeds in the same way as described above but in that case $B_u=0$. The~corresponding counterterm $\delta_{c, \rm ct} = A_u/2 \, V_{cb}$ therefore also receives contributions only from loops involving massless quarks. 

Since we have normalised the Born-level amplitude~(\ref{eq:MBornCC}) in terms of $G_F$, to calculate the ${\cal O} (\alpha)$ correction to the Wilson coefficient of $Q_{\rm CC}$ relevant for the semi-leptonic or non-leptonic case one has to subtract from~(\ref {eq:MvirtualCC}) the corresponding leptonic matrix element. Thereby the universal part of~(\ref {eq:deltabox}) cancels out. The same is true for the last two terms in~(\ref{eq:MvirtualCC}) as well as for the universal vertex correction $\delta_{\rm vertex}$. We add that, in principle, also the CKM matrix has to be renormalised. However, as we are assuming massless light quarks, we just need to remove the ill-defined anti-hermitian part of the WFR constants~\cite{Denner:1990yz,Gambino:1998ec}. Once~this is done, the rest of calculation is unaffected, meaning that it resembles the calculation that did not take CKM renormalisation into account. Notice also that all the residual photonic corrections to muon decay (those described by photonic corrections in the EFT) are included in the conventional definition of $G_F$~\cite{Sirlin:1980nh} and do not need to be subtracted. The final result of our full SM calculation performed at vanishing external momenta and masses is therefore 
\beq \label{eq:renSM} 
{\cal M}_{\rm virtual}^{\rm CC} = {\cal M}_{\rm Born}^{\rm CC} \; \frac{\alpha}{4\pi} \left \{ \left (\chi-1 \right ) \left[ -\frac{1}{\epsilon}+\ln \left ( \frac{M_Z^2}{\mu^2} \right) \right ] +\zeta-\frac{1}{2} \right\} \,.
\eeq

In order to obtain the Wilson coefficient we still need to match the full SM calculation with the EFT one, performed in the same scheme and at vanishing momenta and masses. As the EFT photonic loops vanish under these conditions, the only contribution from the EFT is the counterterm related to the operator $\overline{\rm MS}$ renormalisation, a pure $1/\epsilon$ pole that cancels the divergence in~(\ref{eq:renSM}). We therefore obtain the following Wilson coefficients for the CC processes of interest 
\beq \label{eq:CCC}
C_{\rm CC} (\mu) = 1 + \frac{\alpha (\mu)}{4 \pi} \begin{cases} 
\, \left [ 2 \ln \left ( \frac{M_Z^2}{\mu^2} \right ) - \frac{11}{3} \right ] \,, & \text{semi-leptonic} \,, \\[3mm]
\, \left [ \frac{4}{3} \ln \left ( \frac{M_Z^2}{\mu^2} \right ) - \frac{22}{9} \right ] \,, & \text{non-leptonic} \,,
\end{cases} 
\eeq
where as indicated the electromagnetic coupling $\alpha$ is naturally also evaluated at~$\mu$. The~first line leads to the result given in~(\ref{eq:Eespectrumfinal}) and agrees with~\cite{Brod:2008ss,Gorbahn:2022rgl}, while the second line coincides with the expression for the Wilson coefficient derived in~\cite{Gambino:2000fz,Gambino:2001au,Brod:2008ss}. 

\end{appendix}


\begin{thebibliography}{73}%
\makeatletter
\providecommand \@ifxundefined [1]{%
 \@ifx{#1\undefined}
}%
\providecommand \@ifnum [1]{%
 \ifnum #1\expandafter \@firstoftwo
 \else \expandafter \@secondoftwo
 \fi
}%
\providecommand \@ifx [1]{%
 \ifx #1\expandafter \@firstoftwo
 \else \expandafter \@secondoftwo
 \fi
}%
\providecommand \natexlab [1]{#1}%
\providecommand \enquote [1]{``#1''}%
\providecommand \bibnamefont [1]{#1}%
\providecommand \bibfnamefont [1]{#1}%
\providecommand \citenamefont [1]{#1}%
\providecommand \href@noop [0]{\@secondoftwo}%
\providecommand \href [0]{\begingroup \@sanitize@url \@href}%
\providecommand \@href[1]{\@@startlink{#1}\@@href}%
\providecommand \@@href[1]{\endgroup#1\@@endlink}%
\providecommand \@sanitize@url [0]{\catcode `\\12\catcode `\$12\catcode
 `\&12\catcode `\#12\catcode `\^12\catcode `\_12\catcode `\%12\relax}%
\providecommand \@@startlink[1]{}%
\providecommand \@@endlink[0]{}%
\providecommand \url [0]{\begingroup\@sanitize@url \@url }%
\providecommand \@url [1]{\endgroup\@href {#1}{\urlprefix }}%
\providecommand \urlprefix [0]{URL }%
\providecommand \Eprint [0]{\href }%
\providecommand \doibase [0]{http://dx.doi.org/}%
\providecommand \selectlanguage [0]{\@gobble}%
\providecommand \bibinfo [0]{\@secondoftwo}%
\providecommand \bibfield [0]{\@secondoftwo}%
\providecommand \translation [1]{[#1]}%
\providecommand \BibitemOpen [0]{}%
\providecommand \bibitemStop [0]{}%
\providecommand \bibitemNoStop [0]{.\EOS\space}%
\providecommand \EOS [0]{\spacefactor3000\relax}%
\providecommand\BibitemShut [1]{\csname bibitem#1\endcsname}%
\let\auto@bib@innerbib\@empty
\bibitem [{\citenamefont {Fael}\ \emph {et~al.}(2021)\citenamefont {Fael},
 \citenamefont {Sch\"onwald},\ and\ \citenamefont
 {Steinhauser}}]{Fael:2020tow}%
 \BibitemOpen
 \bibfield {author} {\bibinfo {author} {\bibfnamefont {M.}~\bibnamefont
 {Fael}}, \bibinfo {author} {\bibfnamefont {K.}~\bibnamefont {Sch\"onwald}}, \
 and\ \bibinfo {author} {\bibfnamefont {M.}~\bibnamefont {Steinhauser}},\
 }\href {\doibase 10.1103/PhysRevD.104.016003} {\bibfield {journal} {\bibinfo
 {journal} {Phys. Rev. D}\ }\textbf {\bibinfo {volume} {104}},\ \bibinfo
 {pages} {016003} (\bibinfo {year} {2021})},\ \Eprint
 {http://arxiv.org/abs/2011.13654} {arXiv:2011.13654 [hep-ph]}\BibitemShut 
 {NoStop}%
\bibitem [{\citenamefont {Bordone}\ \emph {et~al.}(2021)\citenamefont
 {Bordone}, \citenamefont {Capdevila},\ and\ \citenamefont
 {Gambino}}]{Bordone:2021oof}%
 \BibitemOpen
 \bibfield {author} {\bibinfo {author} {\bibfnamefont {M.}~\bibnamefont
 {Bordone}}, \bibinfo {author} {\bibfnamefont {B.}~\bibnamefont {Capdevila}},
 \ and\ \bibinfo {author} {\bibfnamefont {P.}~\bibnamefont {Gambino}},\ }\href
 {\doibase 10.1016/j.physletb.2021.136679} {\bibfield {journal} {\bibinfo
 {journal} {Phys. Lett. B}\ }\textbf {\bibinfo {volume} {822}},\ \bibinfo
 {pages} {136679} (\bibinfo {year} {2021})},\ \Eprint
 {http://arxiv.org/abs/2107.00604} {arXiv:2107.00604 [hep-ph]}\BibitemShut 
 {NoStop}%
\bibitem [{\citenamefont {Sirlin}(1982)}]{Sirlin:1981ie}%
 \BibitemOpen
 \bibfield {author} {\bibinfo {author} {\bibfnamefont {A.}~\bibnamefont
 {Sirlin}},\ }\href {\doibase 10.1016/0550-3213(82)90303-0} {\bibfield
 {journal} {\bibinfo {journal} {Nucl. Phys. B}\ }\textbf {\bibinfo {volume}
 {196}},\ \bibinfo {pages} {83} (\bibinfo {year} {1982})}\BibitemShut
 {NoStop}%
\bibitem [{\citenamefont {Barberio}\ and\ \citenamefont
 {Was}(1994)}]{Barberio:1993qi}%
 \BibitemOpen
 \bibfield {author} {\bibinfo {author} {\bibfnamefont {E.}~\bibnamefont
 {Barberio}}\ and\ \bibinfo {author} {\bibfnamefont {Z.}~\bibnamefont {Was}},\
 }\href {\doibase 10.1016/0010-4655(94)90074-4} {\bibfield {journal}
 {\bibinfo {journal} {Comput. Phys. Commun.}\ }\textbf {\bibinfo {volume}
 {79}},\ \bibinfo {pages} {291} (\bibinfo {year} {1994})}\BibitemShut
 {NoStop}%
\bibitem [{\citenamefont {Bernlochner}\ \emph {et~al.}(2022)\citenamefont
 {Bernlochner}, \citenamefont {Fael}, \citenamefont {Olschewsky},
 \citenamefont {Persson}, \citenamefont {van Tonder}, \citenamefont {Vos},\
 and\ \citenamefont {Welsch}}]{Bernlochner:2022ucr}%
 \BibitemOpen
 \bibfield {author} {\bibinfo {author} {\bibfnamefont {F.}~\bibnamefont
 {Bernlochner}}, \bibinfo {author} {\bibfnamefont {M.}~\bibnamefont {Fael}},
 \bibinfo {author} {\bibfnamefont {K.}~\bibnamefont {Olschewsky}}, \bibinfo
 {author} {\bibfnamefont {E.}~\bibnamefont {Persson}}, \bibinfo {author}
 {\bibfnamefont {R.}~\bibnamefont {van Tonder}}, \bibinfo {author}
 {\bibfnamefont {K.~K.}\ \bibnamefont {Vos}}, \ and\ \bibinfo {author}
 {\bibfnamefont {M.}~\bibnamefont {Welsch}},\ }\href {\doibase
 10.1007/JHEP10(2022)068} {\bibfield {journal} {\bibinfo {journal} {JHEP}\
 }\textbf {\bibinfo {volume} {10}},\ \bibinfo {pages} {068} (\bibinfo {year}
 {2022})},\ \Eprint {http://arxiv.org/abs/2205.10274} {arXiv:2205.10274
 [hep-ph]}\BibitemShut {NoStop}%
\bibitem [{\citenamefont {Pak}\ and\ \citenamefont
 {Czarnecki}(2008)}]{Pak:2008qt}%
 \BibitemOpen
 \bibfield {author} {\bibinfo {author} {\bibfnamefont {A.}~\bibnamefont
 {Pak}}\ and\ \bibinfo {author} {\bibfnamefont {A.}~\bibnamefont
 {Czarnecki}},\ }\href {\doibase 10.1103/PhysRevLett.100.241807} {\bibfield
 {journal} {\bibinfo {journal} {Phys. Rev. Lett.}\ }\textbf {\bibinfo
 {volume} {100}},\ \bibinfo {pages} {241807} (\bibinfo {year} {2008})},\
 \Eprint {http://arxiv.org/abs/0803.0960} {arXiv:0803.0960 [hep-ph]}
 \BibitemShut {NoStop}%
\bibitem [{\citenamefont {Biswas}\ and\ \citenamefont
 {Melnikov}(2010)}]{Biswas:2009rb}%
 \BibitemOpen
 \bibfield {author} {\bibinfo {author} {\bibfnamefont {S.}~\bibnamefont
 {Biswas}}\ and\ \bibinfo {author} {\bibfnamefont {K.}~\bibnamefont
 {Melnikov}},\ }\href {\doibase 10.1007/JHEP02(2010)089} {\bibfield {journal}
 {\bibinfo {journal} {JHEP}\ }\textbf {\bibinfo {volume} {02}},\ \bibinfo
 {pages} {089} (\bibinfo {year} {2010})},\ \Eprint
 {http://arxiv.org/abs/0911.4142} {arXiv:0911.4142 [hep-ph]}\BibitemShut 
 {NoStop}%
\bibitem [{\citenamefont {Fael}\ \emph {et~al.}(2022)\citenamefont {Fael},
 \citenamefont {Sch\"onwald},\ and\ \citenamefont
 {Steinhauser}}]{Fael:2022frj}%
 \BibitemOpen
 \bibfield {author} {\bibinfo {author} {\bibfnamefont {M.}~\bibnamefont
 {Fael}}, \bibinfo {author} {\bibfnamefont {K.}~\bibnamefont {Sch\"onwald}}, \
 and\ \bibinfo {author} {\bibfnamefont {M.}~\bibnamefont {Steinhauser}},\
 }\href {\doibase 10.1007/JHEP08(2022)039} {\bibfield {journal} {\bibinfo
 {journal} {JHEP}\ }\textbf {\bibinfo {volume} {08}},\ \bibinfo {pages} {039}
 (\bibinfo {year} {2022})},\ \Eprint {http://arxiv.org/abs/2205.03410}
 {arXiv:2205.03410 [hep-ph]}\BibitemShut {NoStop}%
\bibitem [{\citenamefont {Abashian}\ \emph {et~al.}(2002)\citenamefont
 {Abashian} \emph {et~al.}}]{Belle:2000cnh}%
 \BibitemOpen
 \bibfield {author} {\bibinfo {author} {\bibfnamefont {A.}~\bibnamefont
 {Abashian}} \emph {et~al.} (\bibinfo {collaboration} {Belle}),\ }\href
 {\doibase 10.1016/S0168-9002(01)02013-7} {\bibfield {journal} {\bibinfo
 {journal} {Nucl. Instrum. Meth. A}\ }\textbf {\bibinfo {volume} {479}},\
 \bibinfo {pages} {117} (\bibinfo {year} {2002})}\BibitemShut {NoStop}%
\bibitem [{\citenamefont {Aubert}\ \emph {et~al.}(2002)\citenamefont {Aubert}
 \emph {et~al.}}]{BaBar:2001yhh}%
 \BibitemOpen
 \bibfield {author} {\bibinfo {author} {\bibfnamefont {B.}~\bibnamefont
 {Aubert}} \emph {et~al.} (\bibinfo {collaboration} {BaBar}),\ }\href
 {\doibase 10.1016/S0168-9002(01)02012-5} {\bibfield {journal} {\bibinfo
 {journal} {Nucl. Instrum. Meth. A}\ }\textbf {\bibinfo {volume} {479}},\
 \bibinfo {pages} {1} (\bibinfo {year} {2002})},\ \Eprint
 {http://arxiv.org/abs/hep-ex/0105044} {arXiv:hep-ex/0105044}\BibitemShut 
 {NoStop}%
\bibitem [{\citenamefont {Brun}\ \emph {et~al.}(1994)\citenamefont {Brun},
 \citenamefont {Bruyant}, \citenamefont {Carminati}, \citenamefont {Giani},
 \citenamefont {Maire}, \citenamefont {McPherson}, \citenamefont {Patrick},\
 and\ \citenamefont {Urban}}]{Brun:1994aa}%
 \BibitemOpen
 \bibfield {author} {\bibinfo {author} {\bibfnamefont {R.}~\bibnamefont
 {Brun}}, \bibinfo {author} {\bibfnamefont {F.}~\bibnamefont {Bruyant}},
 \bibinfo {author} {\bibfnamefont {F.}~\bibnamefont {Carminati}}, \bibinfo
 {author} {\bibfnamefont {S.}~\bibnamefont {Giani}}, \bibinfo {author}
 {\bibfnamefont {M.}~\bibnamefont {Maire}}, \bibinfo {author} {\bibfnamefont
 {A.}~\bibnamefont {McPherson}}, \bibinfo {author} {\bibfnamefont
 {G.}~\bibnamefont {Patrick}}, \ and\ \bibinfo {author} {\bibfnamefont
 {L.}~\bibnamefont {Urban}},\ }\href {\doibase 10.17181/CERN.MUHF.DMJ1} {\
 (\bibinfo {year} {1994}),\ 10.17181/CERN.MUHF.DMJ1}\BibitemShut {NoStop}%
\bibitem [{\citenamefont {Agostinelli}\ \emph {et~al.}(2003)\citenamefont
 {Agostinelli} \emph {et~al.}}]{GEANT4:2002zbu}%
 \BibitemOpen
 \bibfield {author} {\bibinfo {author} {\bibfnamefont {S.}~\bibnamefont
 {Agostinelli}} \emph {et~al.} (\bibinfo {collaboration} {GEANT4}),\ }\href
 {\doibase 10.1016/S0168-9002(03)01368-8} {\bibfield {journal} {\bibinfo
 {journal} {Nucl. Instrum. Meth. A}\ }\textbf {\bibinfo {volume} {506}},\
 \bibinfo {pages} {250} (\bibinfo {year} {2003})}\BibitemShut {NoStop}%
\bibitem [{\citenamefont {van Ritbergen}\ and\ \citenamefont
 {Stuart}(2000)}]{vanRitbergen:1999fi}%
 \BibitemOpen
 \bibfield {author} {\bibinfo {author} {\bibfnamefont {T.}~\bibnamefont {van
 Ritbergen}}\ and\ \bibinfo {author} {\bibfnamefont {R.~G.}\ \bibnamefont
 {Stuart}},\ }\href {\doibase 10.1016/S0550-3213(99)00572-6} {\bibfield
 {journal} {\bibinfo {journal} {Nucl. Phys. B}\ }\textbf {\bibinfo {volume}
 {564}},\ \bibinfo {pages} {343} (\bibinfo {year} {2000})},\ \Eprint
 {http://arxiv.org/abs/hep-ph/9904240} {arXiv:hep-ph/9904240}\BibitemShut 
 {NoStop}%
\bibitem [{\citenamefont {Arbuzov}\ \emph {et~al.}(2002)\citenamefont
 {Arbuzov}, \citenamefont {Czarnecki},\ and\ \citenamefont
 {Gaponenko}}]{Arbuzov:2002pp}%
 \BibitemOpen
 \bibfield {author} {\bibinfo {author} {\bibfnamefont {A.}~\bibnamefont
 {Arbuzov}}, \bibinfo {author} {\bibfnamefont {A.}~\bibnamefont {Czarnecki}},
 \ and\ \bibinfo {author} {\bibfnamefont {A.}~\bibnamefont {Gaponenko}},\
 }\href {\doibase 10.1103/PhysRevD.65.113006} {\bibfield {journal} {\bibinfo
 {journal} {Phys. Rev. D}\ }\textbf {\bibinfo {volume} {65}},\ \bibinfo
 {pages} {113006} (\bibinfo {year} {2002})},\ \Eprint
 {http://arxiv.org/abs/hep-ph/0202102} {arXiv:hep-ph/0202102}\BibitemShut 
 {NoStop}%
\bibitem [{\citenamefont {Arbuzov}\ and\ \citenamefont
 {Melnikov}(2002)}]{Arbuzov:2002cn}%
 \BibitemOpen
 \bibfield {author} {\bibinfo {author} {\bibfnamefont {A.}~\bibnamefont
 {Arbuzov}}\ and\ \bibinfo {author} {\bibfnamefont {K.}~\bibnamefont
 {Melnikov}},\ }\href {\doibase 10.1103/PhysRevD.66.093003} {\bibfield
 {journal} {\bibinfo {journal} {Phys. Rev. D}\ }\textbf {\bibinfo {volume}
 {66}},\ \bibinfo {pages} {093003} (\bibinfo {year} {2002})},\ \Eprint
 {http://arxiv.org/abs/hep-ph/0205172} {arXiv:hep-ph/0205172}\BibitemShut 
 {NoStop}%
\bibitem [{\citenamefont {Arbuzov}(2003)}]{Arbuzov:2002rp}%
 \BibitemOpen
 \bibfield {author} {\bibinfo {author} {\bibfnamefont {A.}~\bibnamefont
 {Arbuzov}},\ }\href {\doibase 10.1088/1126-6708/2003/03/063} {\bibfield
 {journal} {\bibinfo {journal} {JHEP}\ }\textbf {\bibinfo {volume} {03}},\
 \bibinfo {pages} {063} (\bibinfo {year} {2003})},\ \Eprint
 {http://arxiv.org/abs/hep-ph/0206036} {arXiv:hep-ph/0206036}\BibitemShut 
 {NoStop}%
\bibitem [{\citenamefont {Anastasiou}\ \emph {et~al.}(2007)\citenamefont
 {Anastasiou}, \citenamefont {Melnikov},\ and\ \citenamefont
 {Petriello}}]{Anastasiou:2005pn}%
 \BibitemOpen
 \bibfield {author} {\bibinfo {author} {\bibfnamefont {C.}~\bibnamefont
 {Anastasiou}}, \bibinfo {author} {\bibfnamefont {K.}~\bibnamefont
 {Melnikov}}, \ and\ \bibinfo {author} {\bibfnamefont {F.}~\bibnamefont
 {Petriello}},\ }\href {\doibase 10.1088/1126-6708/2007/09/014} {\bibfield
 {journal} {\bibinfo {journal} {JHEP}\ }\textbf {\bibinfo {volume} {09}},\
 \bibinfo {pages} {014} (\bibinfo {year} {2007})},\ \Eprint
 {http://arxiv.org/abs/hep-ph/0505069} {arXiv:hep-ph/0505069}\BibitemShut 
 {NoStop}%
\bibitem [{\citenamefont {Webber}\ \emph {et~al.}(2011)\citenamefont {Webber}
 \emph {et~al.}}]{MuLan:2010shf}%
 \BibitemOpen
 \bibfield {author} {\bibinfo {author} {\bibfnamefont {D.~M.}\ \bibnamefont
 {Webber}} \emph {et~al.} (\bibinfo {collaboration} {MuLan}),\ }\href
 {\doibase 10.1103/PhysRevLett.106.079901} {\bibfield {journal} {\bibinfo
 {journal} {Phys. Rev. Lett.}\ }\textbf {\bibinfo {volume} {106}},\ \bibinfo
 {pages} {041803} (\bibinfo {year} {2011})},\ \Eprint
 {http://arxiv.org/abs/1010.0991} {arXiv:1010.0991 [hep-ex]}\BibitemShut 
 {NoStop}%
\bibitem [{\citenamefont {Tishchenko}\ \emph {et~al.}(2013)\citenamefont
 {Tishchenko} \emph {et~al.}}]{MuLan:2012sih}%
 \BibitemOpen
 \bibfield {author} {\bibinfo {author} {\bibfnamefont {V.}~\bibnamefont
 {Tishchenko}} \emph {et~al.} (\bibinfo {collaboration} {MuLan}),\ }\href
 {\doibase 10.1103/PhysRevD.87.052003} {\bibfield {journal} {\bibinfo
 {journal} {Phys. Rev. D}\ }\textbf {\bibinfo {volume} {87}},\ \bibinfo
 {pages} {052003} (\bibinfo {year} {2013})},\ \Eprint
 {http://arxiv.org/abs/1211.0960} {arXiv:1211.0960 [hep-ex]}\BibitemShut 
 {NoStop}%
\bibitem [{\citenamefont {Lipatov}(1974)}]{Lipatov:1974qm}%
 \BibitemOpen
 \bibfield {author} {\bibinfo {author} {\bibfnamefont {L.~N.}\ \bibnamefont
 {Lipatov}},\ }\href@noop {} {\bibfield {journal} {\bibinfo {journal} {Yad.
 Fiz.}\ }\textbf {\bibinfo {volume} {20}},\ \bibinfo {pages} {181} (\bibinfo
 {year} {1974})}\BibitemShut {NoStop}%
\bibitem [{\citenamefont {Altarelli}\ and\ \citenamefont
 {Parisi}(1977)}]{Altarelli:1977zs}%
 \BibitemOpen
 \bibfield {author} {\bibinfo {author} {\bibfnamefont {G.}~\bibnamefont
 {Altarelli}}\ and\ \bibinfo {author} {\bibfnamefont {G.}~\bibnamefont
 {Parisi}},\ }\href {\doibase 10.1016/0550-3213(77)90384-4} {\bibfield
 {journal} {\bibinfo {journal} {Nucl. Phys. B}\ }\textbf {\bibinfo {volume}
 {126}},\ \bibinfo {pages} {298} (\bibinfo {year} {1977})}\BibitemShut
 {NoStop}%
\bibitem [{\citenamefont {Kinoshita}(1962)}]{Kinoshita:1962ur}%
 \BibitemOpen
 \bibfield {author} {\bibinfo {author} {\bibfnamefont {T.}~\bibnamefont
 {Kinoshita}},\ }\href {\doibase 10.1063/1.1724268} {\bibfield {journal}
 {\bibinfo {journal} {J. Math. Phys.}\ }\textbf {\bibinfo {volume} {3}},\
 \bibinfo {pages} {650} (\bibinfo {year} {1962})}\BibitemShut {NoStop}%
\bibitem [{\citenamefont {Lee}\ and\ \citenamefont
 {Nauenberg}(1964)}]{Lee:1964is}%
 \BibitemOpen
 \bibfield {author} {\bibinfo {author} {\bibfnamefont {T.~D.}\ \bibnamefont
 {Lee}}\ and\ \bibinfo {author} {\bibfnamefont {M.}~\bibnamefont
 {Nauenberg}},\ }\href {\doibase 10.1103/PhysRev.133.B1549} {\bibfield
 {journal} {\bibinfo {journal} {Phys. Rev.}\ }\textbf {\bibinfo {volume}
 {133}},\ \bibinfo {pages} {B1549} (\bibinfo {year} {1964})}\BibitemShut
 {NoStop}%
\bibitem [{\citenamefont {Manohar}\ and\ \citenamefont
 {Wise}(2000)}]{Manohar:2000dt}%
 \BibitemOpen
 \bibfield {author} {\bibinfo {author} {\bibfnamefont {A.~V.}\ \bibnamefont
 {Manohar}}\ and\ \bibinfo {author} {\bibfnamefont {M.~B.}\ \bibnamefont
 {Wise}},\ }\href@noop {} {\emph {\bibinfo {title} {{Heavy quark physics}}}},\
 Vol.~\bibinfo {volume} {10}\ (\bibinfo {year} {2000})\BibitemShut {NoStop}%
\bibitem [{\citenamefont {Atwood}\ and\ \citenamefont
 {Marciano}(1990)}]{Atwood:1989em}%
 \BibitemOpen
 \bibfield {author} {\bibinfo {author} {\bibfnamefont {D.}~\bibnamefont
 {Atwood}}\ and\ \bibinfo {author} {\bibfnamefont {W.~J.}\ \bibnamefont
 {Marciano}},\ }\href {\doibase 10.1103/PhysRevD.41.1736} {\bibfield
 {journal} {\bibinfo {journal} {Phys. Rev. D}\ }\textbf {\bibinfo {volume}
 {41}},\ \bibinfo {pages} {1736} (\bibinfo {year} {1990})}\BibitemShut
 {NoStop}%
\bibitem [{\citenamefont {Ginsberg}(1968)}]{Ginsberg:1968pz}%
 \BibitemOpen
 \bibfield {author} {\bibinfo {author} {\bibfnamefont {E.~S.}\ \bibnamefont
 {Ginsberg}},\ }\href {\doibase 10.1103/PhysRev.171.1675} {\bibfield
 {journal} {\bibinfo {journal} {Phys. Rev.}\ }\textbf {\bibinfo {volume}
 {171}},\ \bibinfo {pages} {1675} (\bibinfo {year} {1968})},\ \bibinfo {note}
 {[Erratum: Phys. Rev. {\bf 174}, 2169 (1968)]}\BibitemShut {NoStop}%
\bibitem [{\citenamefont {Isidori}(2008)}]{Isidori:2007zt}%
 \BibitemOpen
 \bibfield {author} {\bibinfo {author} {\bibfnamefont {G.}~\bibnamefont
 {Isidori}},\ }\href {\doibase 10.1140/epjc/s10052-007-0487-0} {\bibfield
 {journal} {\bibinfo {journal} {Eur. Phys. J. C}\ }\textbf {\bibinfo {volume}
 {53}},\ \bibinfo {pages} {567} (\bibinfo {year} {2008})},\ \Eprint
 {http://arxiv.org/abs/0709.2439} {arXiv:0709.2439 [hep-ph]}\BibitemShut 
 {NoStop}%
\bibitem [{\citenamefont {Kubis}\ and\ \citenamefont
 {Schmidt}(2010)}]{Kubis:2010mp}%
 \BibitemOpen
 \bibfield {author} {\bibinfo {author} {\bibfnamefont {B.}~\bibnamefont
 {Kubis}}\ and\ \bibinfo {author} {\bibfnamefont {R.}~\bibnamefont
 {Schmidt}},\ }\href {\doibase 10.1140/epjc/s10052-010-1442-z} {\bibfield
 {journal} {\bibinfo {journal} {Eur. Phys. J. C}\ }\textbf {\bibinfo {volume}
 {70}},\ \bibinfo {pages} {219} (\bibinfo {year} {2010})},\ \Eprint
 {http://arxiv.org/abs/1007.1887} {arXiv:1007.1887 [hep-ph]}\BibitemShut 
 {NoStop}%
\bibitem [{\citenamefont {de~Boer}\ \emph {et~al.}(2018)\citenamefont
 {de~Boer}, \citenamefont {Kitahara},\ and\ \citenamefont
 {Nisandzic}}]{deBoer:2018ipi}%
 \BibitemOpen
 \bibfield {author} {\bibinfo {author} {\bibfnamefont {S.}~\bibnamefont
 {de~Boer}}, \bibinfo {author} {\bibfnamefont {T.}~\bibnamefont {Kitahara}}, \
 and\ \bibinfo {author} {\bibfnamefont {I.}~\bibnamefont {Nisandzic}},\ }\href
 {\doibase 10.1103/PhysRevLett.120.261804} {\bibfield {journal} {\bibinfo
 {journal} {Phys. Rev. Lett.}\ }\textbf {\bibinfo {volume} {120}},\ \bibinfo
 {pages} {261804} (\bibinfo {year} {2018})},\ \Eprint
 {http://arxiv.org/abs/1803.05881} {arXiv:1803.05881 [hep-ph]}\BibitemShut 
 {NoStop}%
\bibitem [{\citenamefont {Cal\'\i{}}\ \emph {et~al.}(2019)\citenamefont
 {Cal\'\i{}}, \citenamefont {Klaver}, \citenamefont {Rotondo},\ and\
 \citenamefont {Sciascia}}]{Cali:2019nwp}%
 \BibitemOpen
 \bibfield {author} {\bibinfo {author} {\bibfnamefont {S.}~\bibnamefont
 {Cal\'\i{}}}, \bibinfo {author} {\bibfnamefont {S.}~\bibnamefont {Klaver}},
 \bibinfo {author} {\bibfnamefont {M.}~\bibnamefont {Rotondo}}, \ and\
 \bibinfo {author} {\bibfnamefont {B.}~\bibnamefont {Sciascia}},\ }\href
 {\doibase 10.1140/epjc/s10052-019-7254-x} {\bibfield {journal} {\bibinfo
 {journal} {Eur. Phys. J. C}\ }\textbf {\bibinfo {volume} {79}},\ \bibinfo
 {pages} {744} (\bibinfo {year} {2019})},\ \Eprint
 {http://arxiv.org/abs/1905.02702} {arXiv:1905.02702 [hep-ph]}\BibitemShut 
 {NoStop}%
\bibitem [{\citenamefont {Mishra}\ and\ \citenamefont
 {Mahajan}(2021)}]{Mishra:2020orb}%
 \BibitemOpen
 \bibfield {author} {\bibinfo {author} {\bibfnamefont {D.}~\bibnamefont
 {Mishra}}\ and\ \bibinfo {author} {\bibfnamefont {N.}~\bibnamefont
 {Mahajan}},\ }\href {\doibase 10.1103/PhysRevD.103.056022} {\bibfield
 {journal} {\bibinfo {journal} {Phys. Rev. D}\ }\textbf {\bibinfo {volume}
 {103}},\ \bibinfo {pages} {056022} (\bibinfo {year} {2021})},\ \Eprint
 {http://arxiv.org/abs/2010.10853} {arXiv:2010.10853 [hep-ph]}\BibitemShut 
 {NoStop}%
\bibitem [{\citenamefont {Bansal}\ \emph {et~al.}(2022)\citenamefont {Bansal},
 \citenamefont {Mahajan},\ and\ \citenamefont {Mishra}}]{Bansal:2021oon}%
 \BibitemOpen
 \bibfield {author} {\bibinfo {author} {\bibfnamefont {A.}~\bibnamefont
 {Bansal}}, \bibinfo {author} {\bibfnamefont {N.}~\bibnamefont {Mahajan}}, \
 and\ \bibinfo {author} {\bibfnamefont {D.}~\bibnamefont {Mishra}},\ }\href
 {\doibase 10.1007/JHEP02(2022)130} {\bibfield {journal} {\bibinfo {journal}
 {JHEP}\ }\textbf {\bibinfo {volume} {02}},\ \bibinfo {pages} {130} (\bibinfo
 {year} {2022})},\ \Eprint {http://arxiv.org/abs/2112.00363} {arXiv:2112.00363
 [hep-ph]}\BibitemShut {NoStop}%
\bibitem [{\citenamefont {Hahn}\ and\ \citenamefont
 {Perez-Victoria}(1999)}]{Hahn:1998yk}%
 \BibitemOpen
 \bibfield {author} {\bibinfo {author} {\bibfnamefont {T.}~\bibnamefont
 {Hahn}}\ and\ \bibinfo {author} {\bibfnamefont {M.}~\bibnamefont
 {Perez-Victoria}},\ }\href {\doibase 10.1016/S0010-4655(98)00173-8}
 {\bibfield {journal} {\bibinfo {journal} {Comput. Phys. Commun.}\ }\textbf
 {\bibinfo {volume} {118}},\ \bibinfo {pages} {153} (\bibinfo {year}
 {1999})},\ \Eprint {http://arxiv.org/abs/hep-ph/9807565}
 {arXiv:hep-ph/9807565}\BibitemShut {NoStop}%
\bibitem [{\citenamefont {Sommerfeld}(1931)}]{Sommerfeld:1931qaf}%
 \BibitemOpen
 \bibfield {author} {\bibinfo {author} {\bibfnamefont {A.}~\bibnamefont
 {Sommerfeld}},\ }\href {\doibase 10.1002/andp.19314030302} {\bibfield
 {journal} {\bibinfo {journal} {Annalen Phys.}\ }\textbf {\bibinfo {volume}
 {403}},\ \bibinfo {pages} {257} (\bibinfo {year} {1931})}\BibitemShut
 {NoStop}%
 \bibitem [{\citenamefont {Brod}\ and\ \citenamefont
 {Gorbahn}(2008)}]{Brod:2008ss}%
 \BibitemOpen
 \bibfield {author} {\bibinfo {author} {\bibfnamefont {J.}~\bibnamefont
 {Brod}}\ and\ \bibinfo {author} {\bibfnamefont {M.}~\bibnamefont {Gorbahn}},\
 }\href {\doibase 10.1103/PhysRevD.78.034006} {\bibfield {journal} {\bibinfo
 {journal} {Phys. Rev. D}\ }\textbf {\bibinfo {volume} {78}},\ \bibinfo
 {pages} {034006} (\bibinfo {year} {2008})},\ \Eprint
 {http://arxiv.org/abs/0805.4119} {arXiv:0805.4119 [hep-ph]}\BibitemShut 
 {NoStop}%
 \bibitem [{\citenamefont {Gorbahn}\ \emph {et~al.}(2023)\citenamefont
 {Gorbahn}, \citenamefont {J\"ager}, \citenamefont {Moretti},\ and\
 \citenamefont {van~der Merwe}}]{Gorbahn:2022rgl}%
 \BibitemOpen
 \bibfield {author} {\bibinfo {author} {\bibfnamefont {M.}~\bibnamefont
 {Gorbahn}}, \bibinfo {author} {\bibfnamefont {S.}~\bibnamefont {J\"ager}},
 \bibinfo {author} {\bibfnamefont {F.}~\bibnamefont {Moretti}}, \ and\
 \bibinfo {author} {\bibfnamefont {E.}~\bibnamefont {van~der Merwe}},\ }\href
 {\doibase 10.1007/JHEP01(2023)159} {\bibfield {journal} {\bibinfo {journal}
 {JHEP}\ }\textbf {\bibinfo {volume} {01}},\ \bibinfo {pages} {159} (\bibinfo
 {year} {2023})},\ \Eprint {http://arxiv.org/abs/2209.05289} {arXiv:2209.05289
 [hep-ph]}\BibitemShut {NoStop}%
\bibitem [{\citenamefont {Kilgore}\ and\ \citenamefont
 {Giele}(1997)}]{Kilgore:1996sq}%
 \BibitemOpen
 \bibfield {author} {\bibinfo {author} {\bibfnamefont {W.~B.}\ \bibnamefont
 {Kilgore}}\ and\ \bibinfo {author} {\bibfnamefont {W.~T.}\ \bibnamefont
 {Giele}},\ }\href {\doibase 10.1103/PhysRevD.55.7183} {\bibfield {journal}
 {\bibinfo {journal} {Phys. Rev. D}\ }\textbf {\bibinfo {volume} {55}},\
 \bibinfo {pages} {7183} (\bibinfo {year} {1997})},\ \Eprint
 {http://arxiv.org/abs/hep-ph/9610433} {arXiv:hep-ph/9610433}\BibitemShut 
 {NoStop}%
\bibitem [{\citenamefont {Eynck}\ \emph {et~al.}(2002)\citenamefont {Eynck},
 \citenamefont {Laenen}, \citenamefont {Phaf},\ and\ \citenamefont
 {Weinzierl}}]{Eynck:2001en}%
 \BibitemOpen
 \bibfield {author} {\bibinfo {author} {\bibfnamefont {T.~O.}\ \bibnamefont
 {Eynck}}, \bibinfo {author} {\bibfnamefont {E.}~\bibnamefont {Laenen}},
 \bibinfo {author} {\bibfnamefont {L.}~\bibnamefont {Phaf}}, \ and\ \bibinfo
 {author} {\bibfnamefont {S.}~\bibnamefont {Weinzierl}},\ }\href {\doibase
 10.1007/s100520100868} {\bibfield {journal} {\bibinfo {journal} {Eur. Phys.
 J. C}\ }\textbf {\bibinfo {volume} {23}},\ \bibinfo {pages} {259} (\bibinfo
 {year} {2002})},\ \Eprint {http://arxiv.org/abs/hep-ph/0109246}
 {arXiv:hep-ph/0109246}\BibitemShut {NoStop}%
\bibitem [{\citenamefont {Asatrian}\ \emph {et~al.}(2012)\citenamefont
 {Asatrian}, \citenamefont {Hovhannisyan},\ and\ \citenamefont
 {Yeghiazaryan}}]{Asatrian:2012tp}%
 \BibitemOpen
 \bibfield {author} {\bibinfo {author} {\bibfnamefont {H.~M.}\ \bibnamefont
 {Asatrian}}, \bibinfo {author} {\bibfnamefont {A.}~\bibnamefont
 {Hovhannisyan}}, \ and\ \bibinfo {author} {\bibfnamefont {A.}~\bibnamefont
 {Yeghiazaryan}},\ }\href {\doibase 10.1103/PhysRevD.86.114023} {\bibfield
 {journal} {\bibinfo {journal} {Phys. Rev. D}\ }\textbf {\bibinfo {volume}
 {86}},\ \bibinfo {pages} {114023} (\bibinfo {year} {2012})},\ \Eprint
 {http://arxiv.org/abs/1210.7939} {arXiv:1210.7939 [hep-ph]}\BibitemShut 
 {NoStop}%
\bibitem [{\citenamefont {Hahn}(2005)}]{Hahn:2004fe}%
 \BibitemOpen
 \bibfield {author} {\bibinfo {author} {\bibfnamefont {T.}~\bibnamefont
 {Hahn}},\ }\href {\doibase 10.1016/j.cpc.2005.01.010} {\bibfield {journal}
 {\bibinfo {journal} {Comput. Phys. Commun.}\ }\textbf {\bibinfo {volume}
 {168}},\ \bibinfo {pages} {78} (\bibinfo {year} {2005})},\ \Eprint
 {http://arxiv.org/abs/hep-ph/0404043} {arXiv:hep-ph/0404043}\BibitemShut 
 {NoStop}%
\bibitem [{\citenamefont {Berman}(1958)}]{Berman:1958ti}%
 \BibitemOpen
 \bibfield {author} {\bibinfo {author} {\bibfnamefont {S.~M.}\ \bibnamefont
 {Berman}},\ }\href {\doibase 10.1103/PhysRev.112.267} {\bibfield {journal}
 {\bibinfo {journal} {Phys. Rev.}\ }\textbf {\bibinfo {volume} {112}},\
 \bibinfo {pages} {267} (\bibinfo {year} {1958})}\BibitemShut {NoStop}%
\bibitem [{\citenamefont {Kinoshita}\ and\ \citenamefont
 {Sirlin}(1959)}]{Kinoshita:1958ru}%
 \BibitemOpen
 \bibfield {author} {\bibinfo {author} {\bibfnamefont {T.}~\bibnamefont
 {Kinoshita}}\ and\ \bibinfo {author} {\bibfnamefont {A.}~\bibnamefont
 {Sirlin}},\ }\href {\doibase 10.1103/PhysRev.113.1652} {\bibfield {journal}
 {\bibinfo {journal} {Phys. Rev.}\ }\textbf {\bibinfo {volume} {113}},\
 \bibinfo {pages} {1652} (\bibinfo {year} {1959})}\BibitemShut {NoStop}%
\bibitem [{\citenamefont {Nir}(1989)}]{Nir:1989rm}%
 \BibitemOpen
 \bibfield {author} {\bibinfo {author} {\bibfnamefont {Y.}~\bibnamefont
 {Nir}},\ }\href {\doibase 10.1016/0370-2693(89)91495-0} {\bibfield {journal}
 {\bibinfo {journal} {Phys. Lett. B}\ }\textbf {\bibinfo {volume} {221}},\
 \bibinfo {pages} {184} (\bibinfo {year} {1989})}\BibitemShut {NoStop}%
\bibitem [{\citenamefont {Aubert}\ \emph {et~al.}(2004)\citenamefont {Aubert}
 \emph {et~al.}}]{BaBar:2004bij}%
 \BibitemOpen
 \bibfield {author} {\bibinfo {author} {\bibfnamefont {B.}~\bibnamefont
 {Aubert}} \emph {et~al.} (\bibinfo {collaboration} {BaBar}),\ }\href
 {\doibase 10.1103/PhysRevD.69.111104} {\bibfield {journal} {\bibinfo
 {journal} {Phys. Rev. D}\ }\textbf {\bibinfo {volume} {69}},\ \bibinfo
 {pages} {111104} (\bibinfo {year} {2004})},\ \Eprint
 {http://arxiv.org/abs/hep-ex/0403030} {arXiv:hep-ex/0403030}\BibitemShut 
 {NoStop}%
\bibitem [{\citenamefont {Urquijo}\ \emph {et~al.}(2007)\citenamefont {Urquijo}
 \emph {et~al.}}]{Belle:2006kgy}%
 \BibitemOpen
 \bibfield {author} {\bibinfo {author} {\bibfnamefont {P.}~\bibnamefont
 {Urquijo}} \emph {et~al.} (\bibinfo {collaboration} {Belle}),\ }\href
 {\doibase 10.1103/PhysRevD.75.032001} {\bibfield {journal} {\bibinfo
 {journal} {Phys. Rev. D}\ }\textbf {\bibinfo {volume} {75}},\ \bibinfo
 {pages} {032001} (\bibinfo {year} {2007})},\ \Eprint
 {http://arxiv.org/abs/hep-ex/0610012} {arXiv:hep-ex/0610012}\BibitemShut 
 {NoStop}%
\bibitem [{\citenamefont {Aubert}\ \emph {et~al.}(2010)\citenamefont {Aubert}
 \emph {et~al.}}]{BaBar:2009zpz}%
 \BibitemOpen
 \bibfield {author} {\bibinfo {author} {\bibfnamefont {B.}~\bibnamefont
 {Aubert}} \emph {et~al.} (\bibinfo {collaboration} {BaBar}),\ }\href
 {\doibase 10.1103/PhysRevD.81.032003} {\bibfield {journal} {\bibinfo
 {journal} {Phys. Rev. D}\ }\textbf {\bibinfo {volume} {81}},\ \bibinfo
 {pages} {032003} (\bibinfo {year} {2010})},\ \Eprint
 {http://arxiv.org/abs/0908.0415} {arXiv:0908.0415 [hep-ex]}\BibitemShut 
 {NoStop}%
 \bibitem [{\citenamefont {Gambino}(2011)}]{Gambino:2011cq}%
 \BibitemOpen
 \bibfield {author} {\bibinfo {author} {\bibfnamefont {P.}~\bibnamefont
 {Gambino}},\ }\href {\doibase 10.1007/JHEP09(2011)055} {\bibfield {journal}
 {\bibinfo {journal} {JHEP}\ }\textbf {\bibinfo {volume} {09}},\ \bibinfo
 {pages} {055} (\bibinfo {year} {2011})},\ \Eprint
 {http://arxiv.org/abs/1107.3100} {arXiv:1107.3100 [hep-ph]}\BibitemShut
 {NoStop}%
\bibitem [{\citenamefont {Gambino}\ and\ \citenamefont
 {Haisch}(2000)}]{Gambino:2000fz}%
 \BibitemOpen
 \bibfield {author} {\bibinfo {author} {\bibfnamefont {P.}~\bibnamefont
 {Gambino}}\ and\ \bibinfo {author} {\bibfnamefont {U.}~\bibnamefont
 {Haisch}},\ }\href {\doibase 10.1088/1126-6708/2000/09/001} {\bibfield
 {journal} {\bibinfo {journal} {JHEP}\ }\textbf {\bibinfo {volume} {09}},\
 \bibinfo {pages} {001} (\bibinfo {year} {2000})},\ \Eprint
 {http://arxiv.org/abs/hep-ph/0007259} {arXiv:hep-ph/0007259}\BibitemShut 
 {NoStop}%
\bibitem [{\citenamefont {Gambino}\ and\ \citenamefont
 {Haisch}(2001)}]{Gambino:2001au}%
 \BibitemOpen
 \bibfield {author} {\bibinfo {author} {\bibfnamefont {P.}~\bibnamefont
 {Gambino}}\ and\ \bibinfo {author} {\bibfnamefont {U.}~\bibnamefont
 {Haisch}},\ }\href {\doibase 10.1088/1126-6708/2001/10/020} {\bibfield
 {journal} {\bibinfo {journal} {JHEP}\ }\textbf {\bibinfo {volume} {10}},\
 \bibinfo {pages} {020} (\bibinfo {year} {2001})},\ \Eprint
 {http://arxiv.org/abs/hep-ph/0109058} {arXiv:hep-ph/0109058}\BibitemShut 
 {NoStop}%
\bibitem [{\citenamefont {Bobeth}\ \emph {et~al.}(2004)\citenamefont {Bobeth},
 \citenamefont {Gambino}, \citenamefont {Gorbahn},\ and\ \citenamefont
 {Haisch}}]{Bobeth:2003at}%
 \BibitemOpen
 \bibfield {author} {\bibinfo {author} {\bibfnamefont {C.}~\bibnamefont
 {Bobeth}}, \bibinfo {author} {\bibfnamefont {P.}~\bibnamefont {Gambino}},
 \bibinfo {author} {\bibfnamefont {M.}~\bibnamefont {Gorbahn}}, \ and\
 \bibinfo {author} {\bibfnamefont {U.}~\bibnamefont {Haisch}},\ }\href
 {\doibase 10.1088/1126-6708/2004/04/071} {\bibfield {journal} {\bibinfo
 {journal} {JHEP}\ }\textbf {\bibinfo {volume} {04}},\ \bibinfo {pages} {071}
 (\bibinfo {year} {2004})},\ \Eprint {http://arxiv.org/abs/hep-ph/0312090}
 {arXiv:hep-ph/0312090}\BibitemShut {NoStop}%
\bibitem [{\citenamefont {Huber}\ \emph {et~al.}(2006)\citenamefont {Huber},
 \citenamefont {Lunghi}, \citenamefont {Misiak},\ and\ \citenamefont
 {Wyler}}]{Huber:2005ig}%
 \BibitemOpen
 \bibfield {author} {\bibinfo {author} {\bibfnamefont {T.}~\bibnamefont
 {Huber}}, \bibinfo {author} {\bibfnamefont {E.}~\bibnamefont {Lunghi}},
 \bibinfo {author} {\bibfnamefont {M.}~\bibnamefont {Misiak}}, \ and\ \bibinfo
 {author} {\bibfnamefont {D.}~\bibnamefont {Wyler}},\ }\href {\doibase
 10.1016/j.nuclphysb.2006.01.037} {\bibfield {journal} {\bibinfo {journal}
 {Nucl. Phys. B}\ }\textbf {\bibinfo {volume} {740}},\ \bibinfo {pages} {105}
 (\bibinfo {year} {2006})},\ \Eprint {http://arxiv.org/abs/hep-ph/0512066}
 {arXiv:hep-ph/0512066}\BibitemShut {NoStop}%
\bibitem [{\citenamefont {Huber}\ \emph {et~al.}(2008)\citenamefont {Huber},
 \citenamefont {Hurth},\ and\ \citenamefont {Lunghi}}]{Huber:2007vv}%
 \BibitemOpen
 \bibfield {author} {\bibinfo {author} {\bibfnamefont {T.}~\bibnamefont
 {Huber}}, \bibinfo {author} {\bibfnamefont {T.}~\bibnamefont {Hurth}}, \ and\
 \bibinfo {author} {\bibfnamefont {E.}~\bibnamefont {Lunghi}},\ }\href
 {\doibase 10.1016/j.nuclphysb.2008.04.028} {\bibfield {journal} {\bibinfo
 {journal} {Nucl. Phys. B}\ }\textbf {\bibinfo {volume} {802}},\ \bibinfo
 {pages} {40} (\bibinfo {year} {2008})},\ \Eprint
 {http://arxiv.org/abs/0712.3009} {arXiv:0712.3009 [hep-ph]}\BibitemShut 
 {NoStop}%
\bibitem [{\citenamefont {Bernlochner}\ and\ \citenamefont
 {Sch{\"o}nherr}(2010)}]{Bernlochner:2010fc}%
 \BibitemOpen
 \bibfield {author} {\bibinfo {author} {\bibfnamefont {F.~U.}\ \bibnamefont
 {Bernlochner}}\ and\ \bibinfo {author} {\bibfnamefont {M.}~\bibnamefont
 {Sch{\"o}nherr}},\ }\href@noop {} {\ (\bibinfo {year} {2010})},\ \Eprint
 {http://arxiv.org/abs/1010.5997} {arXiv:1010.5997 [hep-ph]}\BibitemShut{NoStop}%
\bibitem [{\citenamefont {Bobeth}\ \emph {et~al.}(2014)\citenamefont {Bobeth},
 \citenamefont {Gorbahn}, \citenamefont {Hermann}, \citenamefont {Misiak},
 \citenamefont {Stamou},\ and\ \citenamefont {Steinhauser}}]{Bobeth:2013uxa}%
 \BibitemOpen
 \bibfield {author} {\bibinfo {author} {\bibfnamefont {C.}~\bibnamefont
 {Bobeth}}, \bibinfo {author} {\bibfnamefont {M.}~\bibnamefont {Gorbahn}},
 \bibinfo {author} {\bibfnamefont {T.}~\bibnamefont {Hermann}}, \bibinfo
 {author} {\bibfnamefont {M.}~\bibnamefont {Misiak}}, \bibinfo {author}
 {\bibfnamefont {E.}~\bibnamefont {Stamou}}, \ and\ \bibinfo {author}
 {\bibfnamefont {M.}~\bibnamefont {Steinhauser}},\ }\href {\doibase
 10.1103/PhysRevLett.112.101801} {\bibfield {journal} {\bibinfo {journal}
 {Phys. Rev. Lett.}\ }\textbf {\bibinfo {volume} {112}},\ \bibinfo {pages}
 {101801} (\bibinfo {year} {2014})},\ \Eprint {http://arxiv.org/abs/1311.0903}
 {arXiv:1311.0903 [hep-ph]}\BibitemShut {NoStop}%
\bibitem [{\citenamefont {Huber}\ \emph {et~al.}(2015)\citenamefont {Huber},
 \citenamefont {Hurth},\ and\ \citenamefont {Lunghi}}]{Huber:2015sra}%
 \BibitemOpen
 \bibfield {author} {\bibinfo {author} {\bibfnamefont {T.}~\bibnamefont
 {Huber}}, \bibinfo {author} {\bibfnamefont {T.}~\bibnamefont {Hurth}}, \ and\
 \bibinfo {author} {\bibfnamefont {E.}~\bibnamefont {Lunghi}},\ }\href
 {\doibase 10.1007/JHEP06(2015)176} {\bibfield {journal} {\bibinfo {journal}
 {JHEP}\ }\textbf {\bibinfo {volume} {06}},\ \bibinfo {pages} {176} (\bibinfo
 {year} {2015})},\ \Eprint {http://arxiv.org/abs/1503.04849} {arXiv:1503.04849
 [hep-ph]}\BibitemShut {NoStop}%
\bibitem [{\citenamefont {Bordone}\ \emph {et~al.}(2016)\citenamefont
 {Bordone}, \citenamefont {Isidori},\ and\ \citenamefont
 {Pattori}}]{Bordone:2016gaq}%
 \BibitemOpen
 \bibfield {author} {\bibinfo {author} {\bibfnamefont {M.}~\bibnamefont
 {Bordone}}, \bibinfo {author} {\bibfnamefont {G.}~\bibnamefont {Isidori}}, \
 and\ \bibinfo {author} {\bibfnamefont {A.}~\bibnamefont {Pattori}},\ }\href
 {\doibase 10.1140/epjc/s10052-016-4274-7} {\bibfield {journal} {\bibinfo
 {journal} {Eur. Phys. J. C}\ }\textbf {\bibinfo {volume} {76}},\ \bibinfo
 {pages} {440} (\bibinfo {year} {2016})},\ \Eprint
 {http://arxiv.org/abs/1605.07633} {arXiv:1605.07633 [hep-ph]}\BibitemShut 
 {NoStop}%
\bibitem [{\citenamefont {Beneke}\ \emph {et~al.}(2018)\citenamefont {Beneke},
 \citenamefont {Bobeth},\ and\ \citenamefont {Szafron}}]{Beneke:2017vpq}%
 \BibitemOpen
 \bibfield {author} {\bibinfo {author} {\bibfnamefont {M.}~\bibnamefont
 {Beneke}}, \bibinfo {author} {\bibfnamefont {C.}~\bibnamefont {Bobeth}}, \
 and\ \bibinfo {author} {\bibfnamefont {R.}~\bibnamefont {Szafron}},\ }\href
 {\doibase 10.1103/PhysRevLett.120.011801} {\bibfield {journal} {\bibinfo
 {journal} {Phys. Rev. Lett.}\ }\textbf {\bibinfo {volume} {120}},\ \bibinfo
 {pages} {011801} (\bibinfo {year} {2018})},\ \Eprint
 {http://arxiv.org/abs/1708.09152} {arXiv:1708.09152 [hep-ph]}\BibitemShut 
 {NoStop}%
\bibitem [{\citenamefont {Beneke}\ \emph {et~al.}(2019)\citenamefont {Beneke},
 \citenamefont {Bobeth},\ and\ \citenamefont {Szafron}}]{Beneke:2019slt}%
 \BibitemOpen
 \bibfield {author} {\bibinfo {author} {\bibfnamefont {M.}~\bibnamefont
 {Beneke}}, \bibinfo {author} {\bibfnamefont {C.}~\bibnamefont {Bobeth}}, \
 and\ \bibinfo {author} {\bibfnamefont {R.}~\bibnamefont {Szafron}},\ }\href
 {\doibase 10.1007/JHEP10(2019)232} {\bibfield {journal} {\bibinfo {journal}
 {JHEP}\ }\textbf {\bibinfo {volume} {10}},\ \bibinfo {pages} {232} (\bibinfo
 {year} {2019})},\ \bibinfo {note} {[Erratum: JHEP {\bf 11}, 099 (2022)]},\ \Eprint
 {http://arxiv.org/abs/1908.07011} {arXiv:1908.07011 [hep-ph]}\BibitemShut 
 {NoStop}%
 \bibitem [{\citenamefont {Beneke}\ \emph {et~al.}(2020)\citenamefont {Beneke},
 \citenamefont {B\"oer}, \citenamefont {Toelstede},\ and\ \citenamefont
 {Vos}}]{Beneke:2020vnb}%
 \BibitemOpen
 \bibfield {author} {\bibinfo {author} {\bibfnamefont {M.}~\bibnamefont
 {Beneke}}, \bibinfo {author} {\bibfnamefont {P.}~\bibnamefont {B\"oer}},
 \bibinfo {author} {\bibfnamefont {J.-N.}\ \bibnamefont {Toelstede}}, \ and\
 \bibinfo {author} {\bibfnamefont {K.~K.}\ \bibnamefont {Vos}},\ }\href
 {\doibase 10.1007/JHEP11(2020)081} {\bibfield {journal} {\bibinfo {journal}
 {JHEP}\ }\textbf {\bibinfo {volume} {11}},\ \bibinfo {pages} {081} (\bibinfo
 {year} {2020})},\ \Eprint {http://arxiv.org/abs/2008.10615} {arXiv:2008.10615
 [hep-ph]}\BibitemShut {NoStop}%
\bibitem [{\citenamefont {Isidori}\ \emph {et~al.}(2020)\citenamefont
 {Isidori}, \citenamefont {Nabeebaccus},\ and\ \citenamefont
 {Zwicky}}]{Isidori:2020acz}%
 \BibitemOpen
 \bibfield {author} {\bibinfo {author} {\bibfnamefont {G.}~\bibnamefont
 {Isidori}}, \bibinfo {author} {\bibfnamefont {S.}~\bibnamefont
 {Nabeebaccus}}, \ and\ \bibinfo {author} {\bibfnamefont {R.}~\bibnamefont
 {Zwicky}},\ }\href {\doibase 10.1007/JHEP12(2020)104} {\bibfield {journal}
 {\bibinfo {journal} {JHEP}\ }\textbf {\bibinfo {volume} {12}},\ \bibinfo
 {pages} {104} (\bibinfo {year} {2020})},\ \Eprint
 {http://arxiv.org/abs/2009.00929} {arXiv:2009.00929 [hep-ph]}\BibitemShut 
 {NoStop}%
 \bibitem [{\citenamefont {Beneke}\ \emph
 {et~al.}(2021{\natexlab{a}})\citenamefont {Beneke}, \citenamefont {B\"oer},
 \citenamefont {Finauri},\ and\ \citenamefont {Vos}}]{Beneke:2021jhp}%
 \BibitemOpen
 \bibfield {author} {\bibinfo {author} {\bibfnamefont {M.}~\bibnamefont
 {Beneke}}, \bibinfo {author} {\bibfnamefont {P.}~\bibnamefont {B\"oer}},
 \bibinfo {author} {\bibfnamefont {G.}~\bibnamefont {Finauri}}, \ and\
 \bibinfo {author} {\bibfnamefont {K.~K.}\ \bibnamefont {Vos}},\ }\href
 {\doibase 10.1007/JHEP10(2021)223} {\bibfield {journal} {\bibinfo {journal}
 {JHEP}\ }\textbf {\bibinfo {volume} {10}},\ \bibinfo {pages} {223} (\bibinfo
 {year} {2021}{\natexlab{a}})},\ \Eprint {http://arxiv.org/abs/2107.03819}
 {arXiv:2107.03819 [hep-ph]}\BibitemShut {NoStop}%
\bibitem [{\citenamefont {Beneke}\ \emph
 {et~al.}(2021{\natexlab{b}})\citenamefont {Beneke}, \citenamefont {B\"oer},
 \citenamefont {Toelstede},\ and\ \citenamefont {Vos}}]{Beneke:2021pkl}%
 \BibitemOpen
 \bibfield {author} {\bibinfo {author} {\bibfnamefont {M.}~\bibnamefont
 {Beneke}}, \bibinfo {author} {\bibfnamefont {P.}~\bibnamefont {B\"oer}},
 \bibinfo {author} {\bibfnamefont {J.-N.}\ \bibnamefont {Toelstede}}, \ and\
 \bibinfo {author} {\bibfnamefont {K.~K.}\ \bibnamefont {Vos}},\ }\href
 {\doibase 10.1007/JHEP11(2021)059} {\bibfield {journal} {\bibinfo {journal}
 {JHEP}\ }\textbf {\bibinfo {volume} {11}},\ \bibinfo {pages} {059} (\bibinfo
 {year} {2021}{\natexlab{b}})},\ \Eprint {http://arxiv.org/abs/2108.05589}
 {arXiv:2108.05589 [hep-ph]}\BibitemShut {NoStop}%
\bibitem [{\citenamefont {Beneke}\ \emph {et~al.}(2022)\citenamefont {Beneke},
 \citenamefont {B\"oer}, \citenamefont {Toelstede},\ and\ \citenamefont
 {Vos}}]{Beneke:2022msp}%
 \BibitemOpen
 \bibfield {author} {\bibinfo {author} {\bibfnamefont {M.}~\bibnamefont
 {Beneke}}, \bibinfo {author} {\bibfnamefont {P.}~\bibnamefont {B\"oer}},
 \bibinfo {author} {\bibfnamefont {J.-N.}\ \bibnamefont {Toelstede}}, \ and\
 \bibinfo {author} {\bibfnamefont {K.~K.}\ \bibnamefont {Vos}},\ }\href
 {\doibase 10.1007/JHEP08(2022)020} {\bibfield {journal} {\bibinfo {journal}
 {JHEP}\ }\textbf {\bibinfo {volume} {08}},\ \bibinfo {pages} {020} (\bibinfo
 {year} {2022})},\ \Eprint {http://arxiv.org/abs/2204.09091} {arXiv:2204.09091
 [hep-ph]}\BibitemShut {NoStop}%
\bibitem [{\citenamefont {Herren}(2023)}]{Herren:2022spb}%
 \BibitemOpen
 \bibfield {author} {\bibinfo {author} {\bibfnamefont {F.}~\bibnamefont
 {Herren}},\ }\href {\doibase 10.21468/SciPostPhys.14.2.020} {\bibfield
 {journal} {\bibinfo {journal} {SciPost Phys.}\ }\textbf {\bibinfo {volume}
 {14}},\ \bibinfo {pages} {020} (\bibinfo {year} {2023})},\ \Eprint
 {http://arxiv.org/abs/2205.03427} {arXiv:2205.03427 [hep-ph]}\BibitemShut
 {NoStop}%
\bibitem [{\citenamefont {Isidori}\ \emph {et~al.}(2022)\citenamefont
 {Isidori}, \citenamefont {Lancierini}, \citenamefont {Nabeebaccus},\ and\
 \citenamefont {Zwicky}}]{Isidori:2022bzw}%
 \BibitemOpen
 \bibfield {author} {\bibinfo {author} {\bibfnamefont {G.}~\bibnamefont
 {Isidori}}, \bibinfo {author} {\bibfnamefont {D.}~\bibnamefont {Lancierini}},
 \bibinfo {author} {\bibfnamefont {S.}~\bibnamefont {Nabeebaccus}}, \ and\
 \bibinfo {author} {\bibfnamefont {R.}~\bibnamefont {Zwicky}},\ }\href
 {\doibase 10.1007/JHEP10(2022)146} {\bibfield {journal} {\bibinfo {journal}
 {JHEP}\ }\textbf {\bibinfo {volume} {10}},\ \bibinfo {pages} {146} (\bibinfo
 {year} {2022})},\ \Eprint {http://arxiv.org/abs/2205.08635} {arXiv:2205.08635
 [hep-ph]}\BibitemShut {NoStop}%
 \bibitem [{\citenamefont {Cornella}\ \emph {et~al.}(2023)\citenamefont
 {Cornella}, \citenamefont {K\"onig},\ and\ \citenamefont
 {Neubert}}]{Cornella:2022ubo}%
 \BibitemOpen
 \bibfield {author} {\bibinfo {author} {\bibfnamefont {C.}~\bibnamefont
 {Cornella}}, \bibinfo {author} {\bibfnamefont {M.}~\bibnamefont {K\"onig}}, \
 and\ \bibinfo {author} {\bibfnamefont {M.}~\bibnamefont {Neubert}},\ }\href
 {\doibase 10.1103/PhysRevD.108.L031502} {\bibfield {journal} {\bibinfo
 {journal} {Phys. Rev. D}\ }\textbf {\bibinfo {volume} {108}},\ \bibinfo
 {pages} {L031502} (\bibinfo {year} {2023})},\ \Eprint
 {http://arxiv.org/abs/2212.14430} {arXiv:2212.14430 [hep-ph]}\BibitemShut
 {NoStop}%
\bibitem [{\citenamefont {Huang}\ \emph {et~al.}(2023)\citenamefont {Huang},
 \citenamefont {Shen}, \citenamefont {Zhao},\ and\ \citenamefont
 {Zhou}}]{Huang:2023nli}%
 \BibitemOpen
 \bibfield {author} {\bibinfo {author} {\bibfnamefont {Y.-K.}\ \bibnamefont
 {Huang}}, \bibinfo {author} {\bibfnamefont {Y.-L.}\ \bibnamefont {Shen}},
 \bibinfo {author} {\bibfnamefont {X.-C.}\ \bibnamefont {Zhao}}, \ and\
 \bibinfo {author} {\bibfnamefont {S.-H.}\ \bibnamefont {Zhou}},\ }\href@noop
 {} {\ (\bibinfo {year} {2023})},\ \Eprint {http://arxiv.org/abs/2301.00697}
 {arXiv:2301.00697 [hep-ph]}\BibitemShut {NoStop}%
\bibitem [{\citenamefont {Choudhury}\ \emph {et~al.}(2023)\citenamefont
 {Choudhury}, \citenamefont {Das},\ and\ \citenamefont
 {Das}}]{Choudhury:2023uhw}%
 \BibitemOpen
 \bibfield {author} {\bibinfo {author} {\bibfnamefont {D.}~\bibnamefont
 {Choudhury}}, \bibinfo {author} {\bibfnamefont {D.}~\bibnamefont {Das}}, \
 and\ \bibinfo {author} {\bibfnamefont {J.}~\bibnamefont {Das}},\ }\href@noop
 {} {\ (\bibinfo {year} {2023})},\ \Eprint {http://arxiv.org/abs/2307.07578}
 {arXiv:2307.07578 [hep-ph]}\BibitemShut {NoStop}%
\bibitem [{\citenamefont {Finauri}\ and\ \citenamefont
 {Gambino}(2023)}]{Finauri:2023kte}%
 \BibitemOpen
 \bibfield {author} {\bibinfo {author} {\bibfnamefont {G.}~\bibnamefont
 {Finauri}}\ and\ \bibinfo {author} {\bibfnamefont {P.}~\bibnamefont
 {Gambino}},\ }\href@noop {} {\ (\bibinfo {year} {2023})},\ \Eprint
 {http://arxiv.org/abs/2310.20324} {arXiv:2310.20324 [hep-ph]} \BibitemShut
 {NoStop}%
\bibitem [{\citenamefont {Turczyk}(2016)}]{Turczyk:2016kjf}%
 \BibitemOpen
 \bibfield {author} {\bibinfo {author} {\bibfnamefont {S.}~\bibnamefont
 {Turczyk}},\ }\href {\doibase 10.1007/JHEP04(2016)131} {\bibfield {journal}
 {\bibinfo {journal} {JHEP}\ }\textbf {\bibinfo {volume} {04}},\ \bibinfo
 {pages} {131} (\bibinfo {year} {2016})},\ \Eprint
 {http://arxiv.org/abs/1602.02678} {arXiv:1602.02678 [hep-ph]}\BibitemShut 
 {NoStop}%
\bibitem [{\citenamefont {Amhis}\ \emph {et~al.}(2023)\citenamefont {Amhis}
 \emph {et~al.}}]{HeavyFlavorAveragingGroup:2022wzx}%
 \BibitemOpen
 \bibfield {author} {\bibinfo {author} {\bibfnamefont {Y.~S.}\ \bibnamefont
 {Amhis}} \emph {et~al.} (\bibinfo {collaboration} {Heavy Flavor Averaging
 Group, HFLAV}),\ }\href {\doibase 10.1103/PhysRevD.107.052008} {\bibfield
 {journal} {\bibinfo {journal} {Phys. Rev. D}\ }\textbf {\bibinfo {volume}
 {107}},\ \bibinfo {pages} {052008} (\bibinfo {year} {2023})},\ \Eprint
 {http://arxiv.org/abs/2206.07501} {arXiv:2206.07501 [hep-ex]}\BibitemShut 
 {NoStop}%
\bibitem [{\citenamefont {Bigi}\ \emph {et~al.}(1993)\citenamefont {Bigi},
 \citenamefont {Shifman}, \citenamefont {Uraltsev},\ and\ \citenamefont
 {Vainshtein}}]{Bigi:1993fe}%
 \BibitemOpen
 \bibfield {author} {\bibinfo {author} {\bibfnamefont {I.~I.~Y.}\
 \bibnamefont {Bigi}}, \bibinfo {author} {\bibfnamefont {M.~A.}\ \bibnamefont
 {Shifman}}, \bibinfo {author} {\bibfnamefont {N.~G.}\ \bibnamefont
 {Uraltsev}}, \ and\ \bibinfo {author} {\bibfnamefont {A.~I.}\ \bibnamefont
 {Vainshtein}},\ }\href {\doibase 10.1103/PhysRevLett.71.496} {\bibfield
 {journal} {\bibinfo {journal} {Phys. Rev. Lett.}\ }\textbf {\bibinfo
 {volume} {71}},\ \bibinfo {pages} {496} (\bibinfo {year} {1993})},\ \Eprint
 {http://arxiv.org/abs/hep-ph/9304225} {arXiv:hep-ph/9304225}\BibitemShut 
 {NoStop}%
 \bibitem [{\citenamefont {Blok}\ \emph {et~al.}(1994)\citenamefont {Blok},
 \citenamefont {Koyrakh}, \citenamefont {Shifman},\ and\ \citenamefont
 {Vainshtein}}]{Blok:1993va}%
 \BibitemOpen
 \bibfield {author} {\bibinfo {author} {\bibfnamefont {B.}~\bibnamefont
 {Blok}}, \bibinfo {author} {\bibfnamefont {L.}~\bibnamefont {Koyrakh}},
 \bibinfo {author} {\bibfnamefont {M.~A.}\ \bibnamefont {Shifman}}, \ and\
 \bibinfo {author} {\bibfnamefont {A.~I.}\ \bibnamefont {Vainshtein}},\ }\href
 {\doibase 10.1103/PhysRevD.50.3572} {\bibfield {journal} {\bibinfo
 {journal} {Phys. Rev. D}\ }\textbf {\bibinfo {volume} {49}},\ \bibinfo
 {pages} {3356} (\bibinfo {year} {1994})},\ \bibinfo {note} {[Erratum: Phys.
 Rev. \textbf{D} 50, 3572 (1994)]},\ \Eprint {http://arxiv.org/abs/hep-ph/9307247}
 {arXiv:hep-ph/9307247}\BibitemShut {NoStop}%
\bibitem [{\citenamefont {Manohar}\ and\ \citenamefont
 {Wise}(1994)}]{Manohar:1993qn}%
 \BibitemOpen
 \bibfield {author} {\bibinfo {author} {\bibfnamefont {A.~V.}\ \bibnamefont
 {Manohar}}\ and\ \bibinfo {author} {\bibfnamefont {M.~B.}\ \bibnamefont
 {Wise}},\ }\href {\doibase 10.1103/PhysRevD.49.1310} {\bibfield {journal}
 {\bibinfo {journal} {Phys. Rev. D}\ }\textbf {\bibinfo {volume} {49}},\
 \bibinfo {pages} {1310} (\bibinfo {year} {1994})},\ \Eprint
 {http://arxiv.org/abs/hep-ph/9308246} {arXiv:hep-ph/9308246}\BibitemShut 
 {NoStop}%
\bibitem [{\citenamefont {Voloshin}(1995)}]{Voloshin:1994cy}%
 \BibitemOpen
 \bibfield {author} {\bibinfo {author} {\bibfnamefont {M.~B.}\ \bibnamefont
 {Voloshin}},\ }\href {\doibase 10.1103/PhysRevD.51.4934} {\bibfield
 {journal} {\bibinfo {journal} {Phys. Rev. D}\ }\textbf {\bibinfo {volume}
 {51}},\ \bibinfo {pages} {4934} (\bibinfo {year} {1995})},\ \Eprint
 {http://arxiv.org/abs/hep-ph/9411296} {arXiv:hep-ph/9411296}\BibitemShut 
 {NoStop}%
\bibitem [{\citenamefont {Gremm}\ and\ \citenamefont
 {Kapustin}(1997)}]{Gremm:1996df}%
 \BibitemOpen
 \bibfield {author} {\bibinfo {author} {\bibfnamefont {M.}~\bibnamefont
 {Gremm}}\ and\ \bibinfo {author} {\bibfnamefont {A.}~\bibnamefont
 {Kapustin}},\ }\href {\doibase 10.1103/PhysRevD.55.6924} {\bibfield
 {journal} {\bibinfo {journal} {Phys. Rev. D}\ }\textbf {\bibinfo {volume}
 {55}},\ \bibinfo {pages} {6924} (\bibinfo {year} {1997})},\ \Eprint
 {http://arxiv.org/abs/hep-ph/9603448} {arXiv:hep-ph/9603448}\BibitemShut 
 {NoStop}%
\bibitem [{\citenamefont {Aquila}\ \emph {et~al.}(2005)\citenamefont {Aquila},
 \citenamefont {Gambino}, \citenamefont {Ridolfi},\ and\ \citenamefont
 {Uraltsev}}]{Aquila:2005hq}%
 \BibitemOpen
 \bibfield {author} {\bibinfo {author} {\bibfnamefont {V.}~\bibnamefont
 {Aquila}}, \bibinfo {author} {\bibfnamefont {P.}~\bibnamefont {Gambino}},
 \bibinfo {author} {\bibfnamefont {G.}~\bibnamefont {Ridolfi}}, \ and\
 \bibinfo {author} {\bibfnamefont {N.}~\bibnamefont {Uraltsev}},\ }\href
 {\doibase 10.1016/j.nuclphysb.2005.04.031} {\bibfield {journal} {\bibinfo
 {journal} {Nucl. Phys. B}\ }\textbf {\bibinfo {volume} {719}},\ \bibinfo
 {pages} {77} (\bibinfo {year} {2005})},\ \Eprint
 {http://arxiv.org/abs/hep-ph/0503083} {arXiv:hep-ph/0503083}\BibitemShut 
 {NoStop}%
\bibitem [{\citenamefont {Sirlin}(1980)}]{Sirlin:1980nh}%
 \BibitemOpen
 \bibfield {author} {\bibinfo {author} {\bibfnamefont {A.}~\bibnamefont
 {Sirlin}},\ }\href {\doibase 10.1103/PhysRevD.22.971} {\bibfield {journal}
 {\bibinfo {journal} {Phys. Rev. D}\ }\textbf {\bibinfo {volume} {22}},\
 \bibinfo {pages} {971} (\bibinfo {year} {1980})}\BibitemShut {NoStop}%
\bibitem [{\citenamefont {Hill}\ and\ \citenamefont
 {Tomalak}(2020)}]{Hill:2019xq}%
 \BibitemOpen
 \bibfield {author} {\bibinfo {author} {\bibfnamefont {R.~J.}\ \bibnamefont
 {Hill}}\ and\ \bibinfo {author} {\bibfnamefont {O.}~\bibnamefont {Tomalak}},\
 }\href {\doibase 10.1016/j.physletb.2020.135466} {\bibfield {journal}
 {\bibinfo {journal} {Phys. Lett. B}\ }\textbf {\bibinfo {volume} {805}},\
 \bibinfo {pages} {135466} (\bibinfo {year} {2020})},\ \Eprint
 {http://arxiv.org/abs/1911.01493} {arXiv:1911.01493 [hep-ph]}\BibitemShut 
 {NoStop}%
\bibitem [{\citenamefont {Buras}\ and\ \citenamefont
 {Weisz}(1990)}]{Buras:1989xd}%
 \BibitemOpen
 \bibfield {author} {\bibinfo {author} {\bibfnamefont {A.~J.}\ \bibnamefont
 {Buras}}\ and\ \bibinfo {author} {\bibfnamefont {P.~H.}\ \bibnamefont
 {Weisz}},\ }\href {\doibase 10.1016/0550-3213(90)90223-Z} {\bibfield
 {journal} {\bibinfo {journal} {Nucl. Phys. B}\ }\textbf {\bibinfo {volume}
 {333}},\ \bibinfo {pages} {66} (\bibinfo {year} {1990})}\BibitemShut
 {NoStop}%
 \bibitem [{\citenamefont {Gambino}\ \emph
 {et~al.}(1999{\natexlab{a}})\citenamefont {Gambino}, \citenamefont
 {Kwiatkowski},\ and\ \citenamefont {Pott}}]{Gambino:1998rt}%
 \BibitemOpen
 \bibfield {author} {\bibinfo {author} {\bibfnamefont {P.}~\bibnamefont
 {Gambino}}, \bibinfo {author} {\bibfnamefont {A.}~\bibnamefont
 {Kwiatkowski}}, \ and\ \bibinfo {author} {\bibfnamefont {N.}~\bibnamefont
 {Pott}},\ }\href {\doibase 10.1016/S0550-3213(98)00860-8} {\bibfield
 {journal} {\bibinfo {journal} {Nucl. Phys. B}\ }\textbf {\bibinfo {volume}
 {544}},\ \bibinfo {pages} {532} (\bibinfo {year} {1999}{\natexlab{a}})},\
 \Eprint {http://arxiv.org/abs/hep-ph/9810400} {arXiv:hep-ph/9810400}\BibitemShut {NoStop}%
\bibitem [{\citenamefont {Sirlin}(1978)}]{Sirlin:1977sv}%
 \BibitemOpen
 \bibfield {author} {\bibinfo {author} {\bibfnamefont {A.}~\bibnamefont
 {Sirlin}},\ }\href {\doibase 10.1103/RevModPhys.50.573} {\bibfield {journal}
 {\bibinfo {journal} {Rev. Mod. Phys.}\ }\textbf {\bibinfo {volume} {50}},\
 \bibinfo {pages} {573} (\bibinfo {year} {1978})},\ \bibinfo {note} {[Erratum:
 Rev. Mod. Phys. {\bf 50}, 905 (1978)]}\BibitemShut {NoStop}%
\bibitem [{\citenamefont {Denner}\ and\ \citenamefont
 {Sack}(1990)}]{Denner:1990yz}%
 \BibitemOpen
 \bibfield {author} {\bibinfo {author} {\bibfnamefont {A.}~\bibnamefont
 {Denner}}\ and\ \bibinfo {author} {\bibfnamefont {T.}~\bibnamefont {Sack}},\
 }\href {\doibase 10.1016/0550-3213(90)90557-T} {\bibfield {journal}
 {\bibinfo {journal} {Nucl. Phys. B}\ }\textbf {\bibinfo {volume} {347}},\
 \bibinfo {pages} {203} (\bibinfo {year} {1990})}\BibitemShut {NoStop}%
\bibitem [{\citenamefont {Gambino}\ \emph {et~al.}(1999)\citenamefont
 {Gambino}, \citenamefont {Grassi},\ and\ \citenamefont
 {Madricardo}}]{Gambino:1998ec}%
 \BibitemOpen
 \bibfield {author} {\bibinfo {author} {\bibfnamefont {P.}~\bibnamefont
 {Gambino}}, \bibinfo {author} {\bibfnamefont {P.~A.}\ \bibnamefont {Grassi}},
 \ and\ \bibinfo {author} {\bibfnamefont {F.}~\bibnamefont {Madricardo}},\
 }\href {\doibase 10.1016/S0370-2693(99)00321-4} {\bibfield {journal}
 {\bibinfo {journal} {Phys. Lett. B}\ }\textbf {\bibinfo {volume} {454}},\
 \bibinfo {pages} {98} (\bibinfo {year} {1999})},\ \Eprint
 {http://arxiv.org/abs/hep-ph/9811470} {arXiv:hep-ph/9811470}\BibitemShut 
 {NoStop}%
\end{thebibliography}


%

\end{document}